\documentclass[a4paper,11pt]{article}
\usepackage{graphicx}
\usepackage{enumerate}
\usepackage{amssymb}
\usepackage{amsmath}
\usepackage{amsthm}
\usepackage{dsfont}
\usepackage{accents}
\usepackage{wasysym}
\usepackage{tikz}
\newcommand\tikzmark[1]{\tikz[remember picture,overlay,baseline=-1.5ex] \node (#1) {};}
\usepackage{jheppub}

\theoremstyle{plain}
\newtheorem{thm}{Theorem}[section]

\newtheorem{remark}[thm]{Remark}

\newcommand{\R}{\mathbb{R}}
\newcommand{\C}{\mathbb{C}}
\newcommand{\Z}{\mathbb{Z}}

\newcommand{\Zs}{\mathbb{Z}{\times}\mathbb{Z}}

\newcommand{\h}{\mathfrak{h}}
\newcommand{\GL}{\mathrm{GL}}
\newcommand{\gl}{\mathfrak{gl}}
\newcommand{\SL}{\mathrm{SL}}
\newcommand{\sla}{\mathfrak{sl}}
\newcommand{\SU}{\mathrm{SU}}
\newcommand{\su}{\mathfrak{su}}

\newcommand{\so}{\mathfrak{so}}
\newcommand{\U}{\mathrm{U}}
\newcommand{\ua}{\mathfrak{u}}
\newcommand{\da}{\boldsymbol{\mathtt{d}}}

\newcommand{\boldell}{\boldsymbol{\ell}}
\newcommand{\rad}{\mathop{\mathrm{rad}}\nolimits}
\newcommand{\nil}{\mathop{\mathrm{nil}}\nolimits}

\newcommand{\D}{\mathrm{D}}
\newcommand{\Real}{\mathop{\mathrm{Re}}\nolimits}
\newcommand{\Imag}{\mathop{\mathrm{Im}}\nolimits}

\newcommand{\Tr}{\mathop{\mathrm{Tr}}\nolimits}
\newcommand{\Ad}{\mathop{\mathrm{Ad}}\nolimits}
\newcommand{\ad}{\mathop{\mathrm{ad}}\nolimits}

\newcommand{\Lin}{\mathop{\mathrm{Lin}}\nolimits}

\newcommand{\M}{\mathcal{M}}
\newcommand{\A}{\mathcal{A}}
\newcommand{\G}{\mathcal{G}}
\newcommand{\bepsilon}{\boldsymbol{\epsilon}}
\newcommand{\blambda}{\boldsymbol{\lambda}}
\newcommand{\bomega}{\boldsymbol{\omega}}
\newcommand{\bL}{\mathbf{L}}
\newcommand{\bsigma}{\boldsymbol{\sigma}}
\newcommand{\bdelta}{\boldsymbol{\delta}}
\newcommand{\bg}{\mathbf{g}}
\newcommand{\bv}{\mathbf{v}}
\newcommand{\bgamma}{\boldsymbol{\gamma}}
\newcommand{\1}{\mathds{1}}
\newcommand{\I}{\mathrm{i}}

\newcommand\tbar[1]{\accentset{\rule{.8em}{.8pt}}{#1}}
\newcommand{\opluslhrim}{\mathbin{\rlap{$\Leftcircle$}{+}}}

\title{Nilpotent symmetries as a mechanism for Grand Unification}

\author[a]{Lars Andersson,}
\author[b,1]{Andr\'as L\'aszl\'o,\note{Corresponding author.}}
\author[c]{B{\l}a\.zej Ruba}

\affiliation[a]{Albert Einstein Institute,\\Am M\"uhlenberg 1, D-14471 Potsdam-Golm, Germany}
\affiliation[b]{Wigner Research Centre for Physics,\\Konkoly-Thege M u 29-33, H-1121 Budapest, Hungary}
\affiliation[c]{Institute of Theoretical Physics, Jagiellonian University,\\{\L}ojasiewicza 11, 30-348 Krak\'ow, Poland}

\emailAdd{laan@aei.mpg.de}
\emailAdd{laszlo.andras@wigner.hu}
\emailAdd{blazej.ruba@doctoral.uj.edu.pl}

\abstract{
In the classic Coleman--Mandula no-go theorem which prohibits the unification 
of internal and spacetime symmetries, the assumption of the existence of a 
positive definite invariant scalar product on the Lie algebra of the internal 
group is essential. 
If one instead allows the scalar product to be positive \emph{semi}-definite, 
this opens new possibilities for unification of gauge and spacetime symmetries. 
It follows from theorems on the structure of Lie algebras, that in the case
of unified symmetries, the degenerate directions
of the positive semi-definite invariant scalar product have to correspond to
local symmetries with nilpotent generators.
In this paper we construct 
a workable minimal toy model making use of this mechanism: it admits 
unified local symmetries having a compact ($\U(1)$) component, 
a Lorentz ($\SL(2,\C)$) component, 
and a nilpotent component gluing these together. 
The construction is such that the full unified symmetry group acts locally and 
faithfully on the matter field sector, whereas the gauge fields which would 
correspond to the nilpotent generators can be transformed out from the theory, 
leaving gauge fields only with compact charges. 
It is shown that already the ordinary Dirac equation admits an extremely 
simple prototype example for the above gauge field elimination mechanism: 
it has a local symmetry with corresponding eliminable gauge field, 
related to the dilatation group. 
The outlined symmetry unification mechanism can be used to by-pass 
the Coleman--Mandula and related no-go theorems in a way that is fundamentally 
different from supersymmetry. In particular, the mechanism avoids invocation of 
super-coordinates or extra dimensions for the underlying spacetime manifold.
}


\begin{document}

\maketitle
\flushbottom

\section{Introduction}
\label{secintroduction}

One of the most important programmes in modern physics is concerned with 
model building in particle physics. Much of this endeavor is focused 
on the search for symmetries of Lagrangian field theories, 
and their corresponding quantum field theories. 
The Lagrangian of the Standard Model (SM) 
is essentially determined, up to a number of coupling constants, by its 
local symmetry group. 
The presence of a large number of free parameters 
reduces the predictive power of a physical theory, and for this reason it 
has been a long standing question whether it is possible to find alternatives 
to the Standard Model with a reduced number of free parameters by enlarging 
the local symmetry group. The ensemble of symmetries becomes the most 
restrictive whenever they form a non-direct product (unified) group. 
This simple principle motivated the \mbox{gauge--gauge} and \mbox{gauge--spacetime} symmetry 
unification strategies, which are sometimes referred to as 
GUT (Grand Unified Theories) and ToE (Theories of Everything). 
The early no-go theorem by McGlinn \cite{McGlinn1964}, the classic QFT no-go 
theorem by Coleman and Mandula \cite{ColemanMandula1967}, as well as the 
Poincar\'e group extension classification theorem by O'Raifeartaigh 
\cite{LOR1965a,LOR1965b} strongly restrict the possibilities for 
gauge--spacetime type unification. After the invention of supersymmetry 
(SUSY) \cite{SS1974,FZW1974,Ferrara1987}, it was widely believed 
that only that concept may provide a loophole to these no-go theorems 
\cite{HLS1975}. This is, however, only true under a certain set of assumptions.

It turns out that the primary ingredients of the above restrictive no-go theorems 
come from the general structural theory of finite dimensional Lie 
algebras, and mainly not from field theory itself, as discussed in \cite{Laszlo2017}, 
Section~\ref{secliegroups}, and the Appendix~\ref{appliegroups}. 
Detailed study \cite{Weinberg2000} of the arguments of the above no-go theorems 
\cite{McGlinn1964,ColemanMandula1967} reveal that in order 
to obtain these prohibitive results, the assumption that the Lie algebra of 
the internal symmetry group admits a positive definite invariant scalar product 
is essential. That is, the above no-go theorems only follow automatically 
when the group of internal symmetries is assumed to be purely compact. 
In a previous paper \cite{Laszlo2017} it was demonstrated that whenever the 
assumption on this scalar product is somewhat weakened, by e.g.\ allowing it 
to be merely positive \emph{semi}-definite, then a loophole opens. Under the 
semi-definiteness assumption, the internal group may not only be purely 
compact, but can also contain nilpotent generators. Since the nilpotent generators 
may carry compact and Lorentz charges as well, a gauge--spacetime type 
symmetry unification becomes group-theoretically possible. 
The main point of the present paper is to construct a minimal workable 
toy model utilizing this group-theoretical loophole.

The requirement of compactness of the internal symmetry group in 
conventional gauge theories has several motivations: 
\emph{(i)} compact Lie groups are classified and their 
representation theory is well understood,
\emph{(ii)} the Standard Model gauge group ${\U(1){\times}\SU(2){\times}\SU(3)}$
is compact, and 
\emph{(iii)} Yang--Mills fields with compact gauge group 
admit a strictly positive definite energy functional. 
In the more general case when an internal Lie algebra with merely positive 
\emph{semi}-definite invariant scalar product is considered, it follows that 
besides the gauge fields with compact charges, some 
gauge fields with nilpotent charges occur, 
and these have vanishing Yang--Mills kinetic Lagrangian. 
Correspondingly, they have zero Yang--Mills kinetic energy term. 
This clearly raises the question of whether gauge fields with such 
kind of charges are acceptable from a physical point of view: 
how should one interpret a field theory with gauge field degrees of freedom, 
in which the gauge fields all possess non-negative energy density, as usual, 
but there are some unusual modes of the Yang--Mills fields which have zero 
kinetic energy? Surely these ``exotic'' components of the gauge 
fields cannot have an Euler--Lagrange equation similar to a conventional 
Yang--Mills equation, since they do not have a kinetic term. 

%{\color{red} On the other hand this does not happen in the matter sector, since the corresponding kinetic term (REF) is non-degenerate.}

In this paper we present a workable example of a unified local symmetry group 
of the above kind, along with a corresponding toy model, where the 
above type gauge fields with ``exotic'' (nilpotent) charges, 
necessary for a gauge--spacetime type symmetry unification, can be transformed out from the Lagrangian. This gauge field elimination mechanism is due to a shift symmetry of the Lagrangian (see Section~\ref{secLagrangian}, following Eq.(\ref{eq:20})). 
Therefore, in the resulting field theory, the full unified symmetry group 
acts locally and faithfully on the matter fields, but only the compact part 
of the internal symmetries has corresponding physical gauge fields. The matter field sector, on the other hand, behaves as usual in gauge theory, since it has a non-degenerate kinetic term.

%{\color{green} This makes it possible to gauge away a part of matter degrees of freedom.}
% , in contrast to the Higgs mechanism treated in the unitary gauge
% without transferring them to the gauge field
% \color{red} Furthermore the matter sector is given a non-degenerate kinetic term (REF).

The above mentioned fact, that there exists a Lagrangian with a local 
symmetry without corresponding gauge field, is quite striking, and at a 
first glance it might seem that such a theory must be very artificial. 
In Section~\ref{secdirac}, however, we show that already the ordinary 
Dirac kinetic Lagrangian, when viewed in appropriate field variables, 
does admit an extremely simplified version of the above gauge field 
elimination mechanism, related to the dilatation group.

In Section~\ref{secSchC} we construct the above mentioned unified structure 
group of our toy model, involving compact ($\U(1)$), Lorentz ($\SL(2,\C)$), 
and nilpotent generators, and then 
in Section~\ref{secLagrangian} we constuct a corresponding invariant 
Lagrangian, with eliminable nilpotent gauge fields. 
It is seen that the proposed symmetry unification mechanism allows for nilpotent 
generators, and therefore may seem distantly analogous to SUSY. The main difference 
is, however, that the base manifold of the constructed model is the 
ordinary 4-dimensional Lorentzian spacetime, without super-coordinates or 
other extra dimensional objects.

In Section~\ref{secNFreeParams} we show that at the classical level the 
constructed Lagrangian has a single independent coupling constant. 
Finally, in Section~\ref{secConclusion} we present our conclusions.

The paper is closed by Appendix~\ref{appliegroups}, reviewing the structural theory of generic 
Lie algebras (not necessary semisimple), and some recent results concerning 
that in more details. These are relevant for applications of Lie algebra 
theory in model building.

\section{Structural theorems for Lie groups and Lie algebras}
\label{secliegroups}

Whenever some particle field theory has a classical field theory limit, 
one has a firm mathematical handle on the notion of its symmetry generators: 
the generators of the continuous symmetries of the theory 
are smooth vector fields on some kind of a total space 
of fields of the theory, which respect certain mathematical structures 
associated to the model. 
The spacetime manifold can be thought of, at least locally, as an immersed 
submanifold in the total space. 
Important information on the Lie algebra of 
these symmetry generating vector fields of the total space is present in 
the first order factor Lie algebra, carrying information about their formal Taylor 
expansion around a point of the spacetime manifold. 
In a classical field theory, by construction, this first order Lie algebra 
is always a finite dimensional real Lie algebra. 
Therefore, in this section we recall some facts about the structure of finite dimensional 
real Lie algebras \cite{Onishchik1990,IseTakeuchi1991,Jacobson1962,Snobl2014} 
that we shall need to discuss for model building in physics 
(see Appendix~\ref{appliegroups} for more details).

For a relativistic physical theory based on fields without internal structure, 
one can argue that the generators of first order local symmetries $\mathfrak{e}$ 
must be the Poincar\'e Lie algebra $\mathfrak{p}$. 
For fields with internal structure it is of interest to consider extensions 
of the Poincar\'e Lie algebra, i.e.\ Lie algebras $\mathfrak{e}$ with an 
injective homomorphism ${i:\,\mathfrak{p}\rightarrow\mathfrak{e}}$, 
and the investigation of such extensions has been an important strategy of 
modern particle physics. For example, the local symmetry algebra of the 
Standard Model is of the form 
${\mathfrak{e}=\mathfrak{p}\oplus\mathfrak{u}(1){\oplus}\mathfrak{su}(2){\oplus}\mathfrak{su}(3)}$, 
which in particular splits as a direct sum.

The strategy known as unification aims at finding a field theoretical 
description of particle physics with 
a unified local symmetry group, i.e.\ a group such that its Lie algebra 
$\mathfrak{e}$ does not admit a direct sum decomposition 
${\mathfrak{e}=\mathfrak{i}\oplus\mathfrak{c}}$ (see a detailed review on 
a large class of such models in \cite{Krasnov2018}). 
As an example of a unified extension of the Poincar\'e group, we mention the 
conformal Poincar\'e group, with Lie algebra isomorphic to $\so(2,4)$, which 
is a simple Lie algebra.

With these remarks in mind, we shall now recall the properties of 
extensions of the Poincar\'e Lie algebra, and start by recalling 
an important general result on the structure of Lie Algebras 
(see Appendix~\ref{appliegroups} for a more didactic and detailed treatment).

The Levi--Mal'cev decomposition theorem 
\cite{Snobl2014,Onishchik1990,IseTakeuchi1991,Jacobson1962}
states that any finite dimensional real Lie algebra $\mathfrak{e}$ 
admits a semi-direct sum decomposition of the form 
\begin{eqnarray}
 \mathfrak{e} & = & \rad(\mathfrak{e}) \opluslhrim \mathfrak{l}
\label{eqLevi}
\end{eqnarray}
where $\rad(\mathfrak{e})$ is the maximal solvable ideal in $\mathfrak{e}$, 
called to be the \emph{radical}, and $\mathfrak{l}$ is the maximal 
semisimple Lie sub-algebra of $\mathfrak{e}$, called to be the \emph{Levi factor}, 
which is unique up to inner automorphisms. 
The radical has a further important Lie sub-algebra, the \emph{nilradical} 
denoted by $\nil(\mathfrak{e})$, which is the maximal nilpotent ideal of 
$\mathfrak{e}$. The importance of the nilradical in gauge theory model building 
is justified by the fact that the elements of $\nil(\mathfrak{e})$ are precisely 
the nilpotent symmetry generators, and that ${\nil(\mathfrak{e})=\{0\}}$ can hold 
if and only if the Killing form on $\mathfrak{e}$ is non-degenerate. 
As an example of the Levi--Mal'cev decomposition, 
it is instructive to consider the Lie algebra of the Poincar\'e group,
\begin{eqnarray}
 \mathfrak{p} &=& \mathfrak{t} \opluslhrim \boldell,
\label{eqPoincare}
\end{eqnarray}
where the radical $\mathfrak{t}$, i.e. the translations, is in fact abelian, 
and coincides with the nilradical. 
As discussed in \cite{Laszlo2017} and the Appendix~\ref{appliegroups}, the Lie algebra of the super-Poincar\'e group can 
also be considered as an example to the Levi--Mal'cev decomposition, 
with a non-abelian, but two-step nilradical.

Based on the Levi--Mal'cev decomposition, 
the O'Raifeartaigh classification theorem \cite{LOR1965a,LOR1965b} 
states that if $\mathfrak{e}$ is a finite dimensional extension of the 
Poincar\'e Lie algebra, then one of the following three mutually exclusive 
cases must hold. 
\begin{enumerate}[(A)]
 \item Trivial extension, i.e.\ ${\mathfrak{e}=\mathfrak{p}\oplus\left\{\text{some Lie algebra}\right\}}$. 
 \item Not (A), and the translation Lie algebra $\mathfrak{t}$ is embedded 
into the radical $\rad(\mathfrak{e})$ of the enlarged Lie algebra, whereas the Lorentz Lie algebra 
$\boldell$ is embedded into one of the simple components of the 
Levi factor $\mathfrak{l}$ of the enlarged Lie algebra.
\item The entire Poincar\'e Lie algebra ${\mathfrak{p}=\mathfrak{t}{\opluslhrim}\boldell}$ is embedded 
into one of the simple components of the Levi 
factor $\mathfrak{l}$ of the enlarged Lie algebra.
\end{enumerate}

\begin{remark}
The O'Raifeartaigh theorem makes it easy to understand the principle of the 
Coleman--Mandula no-go theorem, without invoking deep field theoretical 
notions and arguments. 
The Coleman--Mandula theorem \cite{ColemanMandula1967} has a number of 
explicit and implicit assumptions, of which the following two are most 
relevant for our purpose.
\begin{enumerate}[(i)]
\item There exists 
a positive definite scalar product on the generators of the non-Poincar\'e part 
of the extended Lie algebra, which in finite dimensions implies that the 
extended part is purely compact, and therefore it is a direct sum of copies of 
$\ua(1)$ and a compact semisimple part.
\item No symmetry breaking is present.
\end{enumerate}
Assumption (i) rules out case (B) of the O'Raifeartaigh 
theorem, while assumption (ii) rules out case (C). 
Thus, the only remaining possibility is case (A).
(It is also useful to note that in the Coleman--Mandula theorem there is 
another important implicit assumption as well: it is assumed that symmetry generators 
preserve the one-particle Fock subspace. This prohibits symmetry generators 
possibly stepping on the Fock space hierarchy, which can eventually also 
be an important loophole.)
\end{remark}

As noted in \cite{Laszlo2017} and the Appendix~\ref{appliegroups}, the 
case (B) of the O'Raifeartaigh theorem 
opens the Lie algebra theoretical backdoor for the 
existence of the super-Poincar\'e group (SUSY). Namely, when presented in 
appropriate variables, the SUSY algebra can be cast into the form of a finite 
dimensional real Lie algebra extension of the Poincar\'e Lie algebra, 
with nontrivial, two-step nilradical. 
It is also instructive to note that an example for case (C) is 
the conformal Poincar\'e Lie algebra, isomorphic to the simple Lie algebra 
$\so(2,4)$.

If we restrict to relativistic field theories based on 
fields taking values in a vector bundle over a 4-dimensional spacetime, as is 
the case for the Standard Model, then there must be Lie algebra homomorphisms
\begin{eqnarray}
 \mathfrak{p}\overset{i}{\longrightarrow}\mathfrak{e}\overset{o}{\longrightarrow}\mathfrak{p}
\label{eqcons}
\end{eqnarray}
such that ${o\circ i:\,\mathfrak{p}\rightarrow\mathfrak{p}}$ is the identity map, 
see \cite{Laszlo2017,Laszlo2018}. 
We shall call such extensions \emph{conservative}. 
Conservative extensions can always be cast in the form 
${\mathfrak{e}=\mathfrak{t}\opluslhrim\mathfrak{g}}$, where $\mathfrak{g}$ 
is the Lie algebra of the structure group. 
In this paper, we construct a unified conservative extension of the Poincar\'e 
Lie algebra, along with a corresponding minimal toy model Lagrangian. 
We remark that for instance, the Lie algebra of the super-Poincar\'e group 
is not a conservative extension of the Poincar\'e Lie algebra 
\cite{Laszlo2017,Laszlo2018}: it does not admit a surjective 
homomorphism ${o:\,\mathfrak{e}\rightarrow\mathfrak{p}}$ as in Eq.(\ref{eqcons}), 
since it contains non-Poincar\'e generators whose commutator is a Poincar\'e generator. 
Neither are the symmetries of extra dimensional, Kaluza--Klein-like theories 
conservative, for the same reason.

\begin{remark}
In model building one often invokes a Yang--Mills-like kinetic Lagrangian 
term, with the requirement that all gauge fields propagate. This requirement 
is satisfied if and only if the Lie algebra of the internal group 
has an invariant, non-degenerate scalar product. Such Lie algebras are called 
\emph{quadratic}. Not all quadratic Lie algebras are classified as of now. 
An important sub-class of quadratic Lie algebras are the \emph{reductive} ones, 
admitting faithful finite dimensional completely reducible representations, 
which are most commonly used in model building, and are always direct sums 
of copies of $\ua(1)$ and of simple Lie algebras. For instance, 
the Lie algebra of the Standard Model structure group, 
$\sla(2,\C)\oplus\mathfrak{u}(1){\oplus}\mathfrak{su}(2){\oplus}\mathfrak{su}(3)$, 
is reductive. A quadratic Lie algebra is 
\emph{compact} if its invariant scalar product is positive definite. These 
are always reductive, and the Standard Model internal Lie algebra 
$\mathfrak{u}(1){\oplus}\mathfrak{su}(2){\oplus}\mathfrak{su}(3)$ 
provides an example. 
Thus, in traditional model building, which involves only reductive 
Lie algebras, the radical must vanish or be central (and hence abelian). 
Therefore due to the Levi--Mal'cev decomposition Eq.(\ref{eqLevi}), 
nilpotent generators cannot play an important role in symmetry unification 
if only reductive Lie algebras are considered. 
The mechanism outlined in the present paper hinges on the idea of 
considering conservative Poincar\'e extensions. Due to the O'Raifeartaigh theorem, 
these have to carry a nontrivial nilradical, if they are indecomposable (unified).
\end{remark}

\section{A hidden symmetry of the general relativistic Dirac kinetic Lagrangian}
\label{secdirac}

In this section we recall a result from \cite{Laszlo2021}, namely a hidden 
symmetry of the general relativistic Dirac kinetic Lagrangian. It is shown 
that the Dirac kinetic Lagrangian is insensitive to the $\D(1)$ part\footnote{The group 
$\D(1)$ is defined to be $\R^{+}$ with the real multiplication.} 
of the spinor connection. That example serves as a prototype for the gauge 
field elimination mechanism, which will be crucial in 
the toy model presented in Section~\ref{secSchC} and after.

In order to show the hidden symmetry, let us formally define the 
general relativistic Dirac kinetic Lagrangian \cite{Trautman2008}. 
We use Penrose abstract indices for the tangent bundle. 
Let $\M$ be a four dimensional real smooth manifold. 
Assume it to be non-compact, and to admit a Lorentz signature spin structure.\footnote{Geroch's 
theorem states that such manifolds are precisely the parallelizable ones.} 
Let ${(D(\M),\,\gamma_{a})}$ be a Lorentzian Dirac bispinor bundle over it, 
i.e.\ $D(\M)$ is a complex vector bundle with four dimensional fibers, and 
${\gamma_{a}:\,T(\M)\rightarrow D(\M){\otimes}D^{*}(\M)}$ is a pointwise 
real-linear vector bundle homomorphism, with the Clifford property 
against some Lorentz metric. That is, the existence of a 
Lorentz signature metric tensor field $g(\gamma)_{ab}$ on $\M$ is required, such that 
${\gamma_{a}\gamma_{b}+\gamma_{b}\gamma_{a}=2\,g(\gamma)_{ab}}\,I$ holds. 
In this presentation the fundamental field is $\gamma_{a}$ and not $g(\gamma)_{ab}$. 
It is well known \cite{Trautman2008}, that covariant derivations $\nabla_{a}$ on the 
vector bundle $D(\M)$ exist which are lifts of the unique Levi-Civita covariant 
derivation on $T(\M)$ associated to $g(\gamma)_{ab}$. More concretely, these 
covariant derivations are defined by the property that they are compatible 
with the Clifford map $\gamma_{a}$, with the metric $g(\gamma)_{ab}$, and are torsion-free on $T(\M)$. 
Such lifts of the Levi-Civita covariant derivations are uniquely determined, 
up to adding a complex valued covector field, 
which can be though of as a ${\D(1){\times}\U(1)}$ gauge potential. 
Given a Dirac bispinor bundle ${(D(\M),\,\gamma_{a})}$, there exists 
a compatible pointwise antilinear injective vector bundle homomorphism 
${\tbar{(\cdot)}:D(\M)\rightarrow D^{*}(\M)}$, called the Dirac adjoint, which 
is uniquely determined up to a pointwise real smooth nonzero scaling field.

Let us fix a Dirac adjoint together with the Dirac bispinor bundle, so that we 
have ${(D(\M),\,\gamma_{a},\,\tbar{(\cdot)})}$ given. Then, the covariant derivations 
$\nabla_{a}$ compatible with these structures are unique, up to adding an imaginary 
valued covector field. That is, they form an affine space over the gauge 
potentials with $\U(1)$ charge. That ambiquity can be used to encode a $\U(1)$ 
internal charge of the Dirac fields.\footnote{Alternatively, 
as rather done in the particle physics literature, one may fix such a 
reference covariant derivation $\nabla_{a}$ on $D(\M)$, and add by hand an 
imaginary covector field $A_{a}$, in order to encode the $\U(1)$ gauge fields. 
We choose here, however, the notation not splitting $\nabla_{a}$. These two 
choices are mathematically equivalent.}
As such, $\nabla_{a}$ encodes a combined gravitational and $\U(1)$ gauge connection, acting on the Dirac 
fields, being smooth sections $\Psi$ of $D(\M)$. 
Then, one can define the Dirac kinetic Lagrangian
\begin{eqnarray}
 \mathrm{L}_{\text{Dirac}}(\gamma,\Psi,\nabla\Psi) & \quad:=\quad & \mathrm{v}_{\gamma}\,\Real\left(\,\tbar{\Psi}\,\gamma^{a}\,\I\nabla_{a}\Psi\right)
\label{eqDirac}
\end{eqnarray}
being a spacetime pointwise bundle morphism into the real volume forms. 
Here, $\mathrm{v}_{\gamma}$ denotes the volume form field uniquely associated 
to the spacetime metric subordinate to the Clifford map $\gamma_{a}$ 
and to a chosen fixed spacetime orientation. The action functional is then 
local integrals of the volume form Eq.(\ref{eqDirac}) over the compact regions 
of the spacetime $\M$.

Consider now the Lagrangian Eq.(\ref{eqDirac}) as part of a larger theory in 
which case the Clifford map $\gamma_{a}$ is also dynamical. 
Then, besides the $\U(1)$ internal charges of $\Psi$, one may assign an action of the $\D(1)$ 
group on the fields in the following way:
\begin{eqnarray}
\left(\begin{array}{l} \Psi \cr \gamma_{a} \cr \nabla_{b} \end{array}\right) & \;\overset{\text{\tiny$\Omega{\in}\R^{{}^{+}}$}}{\text{\large$\longmapsto$}}\; & \left(\begin{array}{l} \Omega\!\!^{{}^{-\frac{3}{2}}}\;\Psi \cr \Omega\,\gamma_{a} \cr \nabla_{b} \end{array}\right),
\end{eqnarray}
which can be considered as a $\D(1)$ gauge transformation with a constant 
${\Omega\in\R^{+}}$, and the Dirac Lagrangian Eq.(\ref{eqDirac}) is evidently invariant to it. 
As it is well known, even more is true: the Dirac Lagrangian 
Eq.(\ref{eqDirac}) is conformally invariant. This means that
the positive scaling field ${\Omega{>}0}$ may be taken to be
not necessarily constant, at the price of making the transformation
rule only slightly more complicated:
\begin{eqnarray}
\left(\begin{array}{l} \Psi \cr \gamma_{a} \cr \nabla_{b} \end{array}\right) & \;\overset{\text{\tiny$\Omega{>}0\;$}}{\text{\large$\longmapsto$}}\; & \left(\begin{array}{l} \Omega\!\!^{{}^{-\frac{3}{2}}}\;\Psi \cr \Omega\,\gamma_{a} \cr \nabla_{b} - \frac{1}{2}\left(\I\Sigma{}_{b}{}^{c}-\delta{}_{b}{}^{c}I\right)(\Omega\!\!^{{}^{-1}}\!\!\mathrm{d}_{c}\Omega) \end{array}\right),
\label{eqconform}
\end{eqnarray}
where ${\Sigma_{ab}:=\frac{\I}{2}\left(\gamma_{a}\gamma_{b}-\gamma_{b}\gamma_{a}\right)}$ is 
the spin tensor. The transformation rule of the covariant derivation 
$\nabla$ comes from the requirement that its 
metricity, torsion and compatibility with the Clifford map 
be unaffected by the rescaling, which unambiguously 
determines the pertinent term. In the following we show that one can also 
endow the fields ${(\Psi,\,\gamma_{a},\,\nabla_{b})}$ with local $\D(1)$ 
charges in a different way, in which scenario the Dirac kinetic Lagrangian 
Eq.(\ref{eqDirac}) manifests a hidden symmetry concerning the local $\D(1)$ 
rescaling, which is related to spacetime pointwise rescaling of the physical 
measurement units.

\subsection{The measure line bundle}
\label{secmeasureline} 

In the works of Matolcsi \cite{Matolcsi1993} and of 
Jany\u{s}ka, Modugno, Vitolo \cite{Modugno2010}, 
a simple mathematical framework was proposed which formalizes 
the notion of physical dimensional analysis. 
In their formulation, the mathematical model of special relativistic 
spacetime is considered to be a triplet $(\M,L,\eta)$, where $\M$ is a four 
dimensional real affine space (modeling the flat spacetime), $L$ is a one 
dimensional oriented vector space (modeling the one dimensional vector space of 
length values), 
and $\eta:\mathop{\vee}^{2}T\rightarrow \mathop{\otimes}^{2}L$ 
is the flat Lorentz signature metric (constant throughout the spacetime), where 
$T$ is the underlying vector space of $\M$ (``tangent space''). 
The key idea in that construction is that the 
field quantities, such as the metric tensor $\eta$, are not simply 
real valued, but they take their values in the tensor powers of the 
\emph{measure line} $L$.\footnote{The term 
\emph{measure line} was introduced by \cite{Matolcsi1993}, whereas the same 
concept is called \emph{scale space} by \cite{Modugno2010}. Apparently, 
these two group of authors discovered the pertinent rather useful notion independently.} 
Due to the one-dimensionality of $L$, it can be shown that all rational tensor 
powers of it makes sense as distinct vector spaces.\footnote{Indeed, 
$L^{*}$ denoting the dual vector space of $L$, for any non-negative integer 
$n$ one can set $L^{n}:=\mathop{\otimes}^{n}L$ 
and $L^{-n}:=\mathop{\otimes}^{n}L^{*}$ in order to make sense of 
any signed integer tensor powers of $L$. Moreover, due to the one-dimensionality 
of $L$, the $n$-th tensorial root $\sqrt[m]{L}$ of $L$ also can be shown to make 
sense uniquely \cite{Matolcsi1993,Modugno2010}, via requiring 
the defining property $\mathop{\otimes}^{m}\big(\sqrt[m]{L}\big)=L$. 
As such, all rational tensor powers $L^{n/m}$ of a one dimensional oriented vector space $L$ 
makes sense, and they define distinct (not naturally isomorphic) vector spaces 
with respect to the canonical action of $\GL(L)$.}
Such a setting formalizes the physical expectation 
that quantities actually have physical dimensions (the metric carries 
length-square dimension in this case), and that quantities with different physical 
dimensions cannot be added since they reside in different vector spaces. 
It is seen that the technique of measure lines is nothing but 
the precise mathematical formulation of ordinary dimensional analysis in physics.

This formulation of dimensional analysis, although it
may seem relatively obvious, nearly tautological idea at a first 
glance, becomes a powerful tool when applied in a general relativistic setting. 
Namely, let our spacetime manifold $\M$ be some four dimensional real manifold, and let $L(\M)$ be a real 
oriented vector bundle over $\M$, with one dimensional fiber. The fiber of $L(\M)$ over each point of $\M$ shall model 
the oriented vector space of length values, and the pertinent line bundle shall be called 
the \emph{measure line bundle}, or \emph{line bundle of lengths}. We do not assume 
anything more about the line bundle $L(\M)$, and in particular, we do not 
assume that a preferred trivialization is given. 
Just as in \cite{Matolcsi1993,Modugno2010}, the 
field quantities shall carry certain tensor powers of $L(\M)$. 

For instance, considering the Dirac action discussed above, we assume that a Dirac 
field $\Psi$ is a section of the vector bundle 
\begin{eqnarray}
L^{{}^{-\frac{3}{2}}}(\M)\mathbin{\otimes}D(\M),
\label{eqDiracDim}
\end{eqnarray}
where $D(\M)$ is an ordinary (dimension-free) Dirac bispinor 
vector bundle. Similarly, one can assume that the spacetime metric $g_{ab}$ is a 
section of the vector bundle ${L^{2}(\M)\mathbin{\otimes}\mathop{\vee}^{2}T^{*}(\M)}$, 
and that the Clifford map $\gamma_{a}$ is a section of the vector bundle 
${L(\M)\mathbin{\otimes}T^{*}(\M){\otimes}D(\M){\otimes}D^{*}(\M)}$. 
This differential geometrical formulation encodes the physical idea that 
quantities occurring in the field theory have physical dimensions, and that the units of 
measurements can only be a priori defined spacetime pointwise. 
In order to transport the unit length to different spacetime points, a connection on 
$L(\M)$ must be specified. Therefore, to make sense of the covariant 
derivative ${\nabla_{a}\Psi}$ of a section $\Psi$ of Eq.(\ref{eqDiracDim}), 
$\nabla_{a}$ must be understood as the joint covariant derivation of 
the usual Clifford connection on $D(\M)$, and some connection on the line bundle 
of lengths $L(\M)$, the two being naturally joined via the Leibniz rule. Since the 
natural structure group of the vector bundle $L(\M)$ is $\D(1)$, one can think 
of this as assigning local $\D(1)$ gauge charges to $\Psi$ and $\gamma_{a}$ and 
also including a corresponding $\D(1)$ gauge field within $\nabla_{a}$.

When constructing the Lagrangian as a volume form valued bundle morphism, 
one should keep in mind that it must be dimension-free 
(carrying zero tensor powers of $L(\M)$), 
since only pure volume forms may be integrated over a manifold without 
any further assumptions, so that the action functional can be defined. 
As such, with the above assignment of dimensions, 
our example Lagrangian for the Dirac kinetic term Eq.(\ref{eqDirac}) 
indeed takes its values purely as section of ${\mathop{\wedge}^{{}^{{}_{\dim(\M)}}}\!T^{*}(\M)}$, 
i.e.\ as a pure volume form.

On the above fields $(\Psi,\,\gamma_{a},\,\nabla_{b})$, one finds 
that an ${L(\M)\rightarrow L(\M)}$ pointwise vector bundle automorphism 
acts by a smooth positive real valued field $\Omega$ over the spacetime manifold 
$\M$, i.e.\ via a local $\D(1)$ gauge transformation
\begin{eqnarray}
\left(\begin{array}{l} \Psi \cr \gamma_{a} \cr \nabla_{b} \end{array}\right) & \;\overset{\text{\tiny$\Omega{>}0\;$}}{\text{\large$\longmapsto$}}\; & \left(\begin{array}{l} \Omega\!\!^{{}^{-\frac{3}{2}}}\;\Psi \cr \Omega\,\gamma_{a} \cr \Omega\!\!^{{}^{-\frac{3}{2}}}\,\nabla_{b}\,\Omega\!^{{}^{\frac{3}{2}}} = \nabla_{b} + \Omega\!\!^{{}^{-\frac{3}{2}}}\mathrm{d}_{b}\Omega\!^{{}^{\frac{3}{2}}} \end{array}\right).
\label{eqscale}
\end{eqnarray}
As trivially seen, Eq.(\ref{eqDirac}) is invariant to these, which means that 
the Lagrangian is invariant to the pointwise rescaling of the measurement unit 
of lengths, and not only to Eq.(\ref{eqconform}).

\subsection{Connection shift invariance of the Dirac Lagrangian}
\label{secshift}

An interesting observation, not yet emphasized 
in the literature, is that the Dirac Lagrangian Eq.(\ref{eqDirac})
understood in such variables, has a further hidden symmetry: it is invariant to the choice of 
the measure line bundle connection. Quite naturally, a change in the 
$L(\M)$ connection is uniquely described by an affine shift transformation 
${\nabla_{a}\mapsto\nabla_{a}+C_{a}}$, where $C_{a}$ is a smooth 
real-valued covector field over the spacetime. Direct evaluation shows that 
the Dirac Lagrangian Eq.(\ref{eqDirac}) is invariant with respect to such a shift transformation
\begin{eqnarray}
\left(\begin{array}{l} \Psi \cr \gamma_{a} \cr \nabla_{b} \end{array}\right) & \;\overset{\text{\tiny$C_{d}$}}{\text{\large$\longmapsto$}}\; & \left(\begin{array}{l} \Psi \cr \gamma_{a} \cr \nabla_{b} \;+\; C_{b} \end{array}\right).
\label{eqshift}
\end{eqnarray}
In other terms, one could say that the Dirac Lagrangian Eq.(\ref{eqDirac}) 
is invariant with respect to the choice of a $\D(1)$ gauge connection. 
The physical meaning of this fact is that the Lagrangian is invariant to 
the choice of any parallel transport rule of measurement units throughout 
spacetime, which is an additional symmetry on top of the usual conformal 
invariance Eq.(\ref{eqconform}) or pointwise measurement unit rescaling 
invariance Eq.(\ref{eqscale}). It can be shown \cite{Laszlo2021}, that all 
the Standard Model kinetic terms, when viewed in such variables, admit this symmetry.

It is seen that due to the ${\nabla_{a}\mapsto\nabla_{a}+C_{a}}$ 
shift symmetry of the Lagrangian, 
$C_{a}$ being $\D(1)$ valued, the original 
${\D(1){\times}\U(1)}$ internal symmetry group, acting locally 
and faithfully on the matter fields, gives rise to a gauge field 
only for the compact direction, i.e.\ with $\U(1)$ degrees of freedom only. 
In our more complex toy model in this paper, we will show that such a 
forgetting mechanism can also be invoked for larger internal groups, 
and even with non-direct product (unified) group structure. 
By construction, however, it follows that the generators of the local 
symmetries whose gauge fields can be eliminated in such a manner, 
must sit in an $\ad$-invariant sub-Lie algebra. Because of that, 
the Levi--Mal'cev decomposition theorem leads to strong constraints on how 
local internal symmetry generators deprived of corresponding 
gauge bosons can accompany the usual ones.

\section{The structure group of the proposed toy model}
\label{secSchC}

The toy model presented here will be a general relativistic spinorial (Dirac-like) 
classical field theory of a fermion particle, invariant to some local 
nilpotent symmetry generators in addition to the usual local symmetries. 
The mathematically simplest, i.e.\ lowest dimensional nonabelian 
nilpotent Lie algebra is the so-called \emph{Heisenberg Lie algebra} 
with $3$ generators, denoted by $\h_{3}$. 
The name Heisenberg Lie algebra of $\h_{3}$ comes from the formal resemblance 
of its Lie algebra relations to the Heisenberg exchange relations: 
$\h_{3}$ is spanned by three elements $q$, $p$ and $e$, 
the only nonvanishing bracket relation being ${[p,q]=K\,e}$ 
where $K$ is some nonzero real number. 
For different values of $K$ the instances of $\h_{3}$ are naturally 
isomorphic to each-other, therefore 
one can fix the value of the constant $K$ to an arbitrary preferred nonzero 
real number. The complexified $3$-generator Heisenberg Lie algebra 
is denoted by $\h_{3}(\C)$, and that shall be the nilradical of our 
example group. The Lie group corresponding to $\h_{3}(\C)$ is denoted 
by $\mathrm{H}_{3}(\C)$.

It is straightforward to check, that the Lie algebra 
${\gl(2,\C)\equiv\ua(1){\oplus}\,\da(1){\oplus}\,\sla(2,\C)}$ can act as 
outer derivations on $\h_{3}(\C)$, 
via linearly mixing the first two generators 
$q$ and $p$, while merely scaling the third generator $e$ with the trace.\footnote{The symbol 
$\da(1)$ denotes the Lie algebra of $\D(1)$. In a purely Lie algebraic sense 
it is isomorphic to $\ua(1)$, but for clarity we distinguish the two, 
understood as the concrete Lie algebras of the distinct Lie groups 
$\D(1)$ and $\U(1)$, respectively.} 
In fact, e.g.\ via the \texttt{LieAlgebras} Maple package \cite{Anderson2016}, 
one may verify that the Lie algebra of outer derivations of $\h_{3}(\C)$ is $\gl(2,\C)$. 
Thus, the largest indecomposable semi-direct sum Lie algebra with nilradical 
$\h_{3}(\C)$ is nothing but ${\h_{3}(\C)\opluslhrim\gl(2,\C)}$. This Lie algebra 
is an indecomposable conservative unification of the compact $\ua(1)$ and of 
the Weyl Lie algebra ${\da(1){\oplus}\,\sla(2,\C)}$, since one has
\begin{eqnarray}
 \h_{3}(\C)\opluslhrim\gl(2,\C) & \equiv & \h_{3}(\C)\opluslhrim\big(\ua(1){\oplus}\,\da(1){\oplus}\,\sla(2,\C)\big).\qquad
\end{eqnarray}
The Lie group corresponding to the Lie algebra ${\h_{3}(\C)\opluslhrim\gl(2,\C)}$ 
is the indecomposable, semi-direct product group ${\mathrm{H}_{3}(\C)\rtimes\GL(2,\C)}$. 
The key ingredient for the structure group of our toy model shall be that 
group. In order to construct the model, we first show that the above is a 
matrix group, i.e.\ has a faithful linear representation. 
Then, we will demonstrate that its lowest dimensional faithful linear 
representation, i.e.\ its defining representation, 
carries a quite natural field theoretical meaning.

In the following, we shall use the ordinary two-spinor calculus \cite{PR1984,Wald1984}, and 
in particular its variant which is most wide spread in general relativity 
(GR) literature. 
Fix an abstract two dimensional complex vector space $S$, i.e.\ $S\cong\C^{2}$. 
The space $S$ is called the \emph{two-spinor space} or simply \emph{spinor space}, and 
its dual space $S^{*}$ is called the \emph{co-spinor space}. Their complex conjugate 
vector spaces are denoted by $\bar{S}$ and $\bar{S}^{*}$, respectively. 
In the Penrose abstract index notation \cite{PR1984,Wald1984}, elements of 
${S,\,S^{*},\,\bar{S},\,\bar{S}^{*}}$ are denoted with upper index ($\xi^{A}$), 
lower index ($\xi_{A}$), primed upper index ($\bar{\xi}^{A'}$), and 
primed lower index ($\bar{\xi}_{A'}$) spinors, respectively, with the spinor 
indices being based on upper case latin letters. 
The symbol $T$ will denote a four dimensional real vector space 
(``tangent space''), with $T^{*}$ being its dual. As is common in the GR 
literature, Penrose abstract indices of elements of $T$ and $T^{*}$ are 
denoted with lower case latin letter upper ($t^{a}$) and lower ($t_{a}$) indices. 
As usual, the index symmetrization and antisymmetrization are be denoted by 
enclosing the indices in round ${}_{{}^{(\,)}}$ or square ${}_{{}^{[\,]}}$ brackets, 
respectively.

Let $\A$ be a complex Grassmann algebra with 2 generators ($\A\cong\Lambda(\C^{2})$), 
i.e.\ $\A$ an exterior algebra of a two-dimensional complex vector space 
without a fixed preferred $\Z$-grading. Whenever a preferred 
$\Z$-grading is chosen, then $\A$ may be identified as ${\A\equiv\Lambda(S^{*})}$, 
i.e.\ a spinorial representation of it can be given. 
Motivated by this, we shall call $\A$ the space of \emph{generalized co-spinors}. 
(The convention that we are representing $\A$ as $\Lambda(S^{*})$ and not as e.g.\ $\Lambda(S)$ 
is merely a matter of convenience for the Penrose abstract index formalism.) 
Given an element ${a\in\A}$, denote by ${L_{a}\in\Lin(\A)}$ the linear operator of left multiplication by $a$ on $\A$. 
Since $\A$ is a four dimensional complex unital associative algebra, the group 
of its invertible elements can act on the space $\A$ via $L$.\footnote{It is well known, and easy to check, that the invertible 
elements of a Grassmann algebra are those which have nonvanishing scalar 
(zero-form) component. To put it differently: invertible elements are those which are 
exponentials of any elements.}
Denote by ${M(\A)\subset\A}$ the so-called maximal ideal of $\A$, which happens 
to be the subspace of order at least one forms within $\A$. 
Then, all the invertible elements of $\A$ can be uniquely written as a nonzero 
complex number times $\exp(m)$, with some element ${m\in M(\A)}$. 
Thus, the group action of the left multiplication by an invertible element of $\A$ 
on $\A$ can be uniquely written as a nonzero complex scaling times 
${\exp(L_{m})\in\GL(\A)}$ with ${m\in M(\A)}$.

The group ${\{\exp(L_{m})\,\vert\,m\in M(\A)\}}$ can be easily seen to be isomorphic 
to $\mathrm{H}_{3}(\C)$. In order to show this fact, it is enough to see that 
the Lie algebra defined by ${\{L_{m}\,\vert\,m\in M(\A)\}}$ is isomorphic to 
$\h_{3}(\C)$. That is most easily demonstrated by fixing some $\Z$-grading 
${\A\equiv\mathop{\bigoplus}_{p=0}^{2}\Lambda_{p}}$, and then taking 
the unit element ${\1\in\Lambda_{0}}$, and canonical generators 
${a_{1},a_{2}\in\Lambda_{1}}$, with which the basis ${\{\1,a_{1},a_{2},a_{1}a_{2}\}}$ 
spans the algebra $\A$, whereas the basis ${\{a_{1},a_{2},a_{1}a_{2}\}}$ spans 
its maximal ideal $M(\A)$. Since $\A$ was defined to be a Grassmann algebra 
with two generators, it directly follows from the Grassmann relations that the Lie 
algebra spanned by ${\{L_{a_{1}},L_{a_{2}},L_{a_{1}a_{2}}\}}$ has the same commutation relations as $\h_{3}(\C)$, 
and therefore ${L_{M(\A)}\equiv\h_{3}(\C)}$, and correspondingly one has ${\exp(L_{M(\A)})\equiv\mathrm{H}_{3}(\C)}$. 
As a consequence, one has natural faithful linear representations of 
$\h_{3}(\C)$ and $\mathrm{H}_{3}(\C)$ on the space $\A$.

\begin{figure*}
\begin{center}
\begin{minipage}{0.9cm}(a)\end{minipage}\begin{minipage}{2cm}\includegraphics[width=2cm]{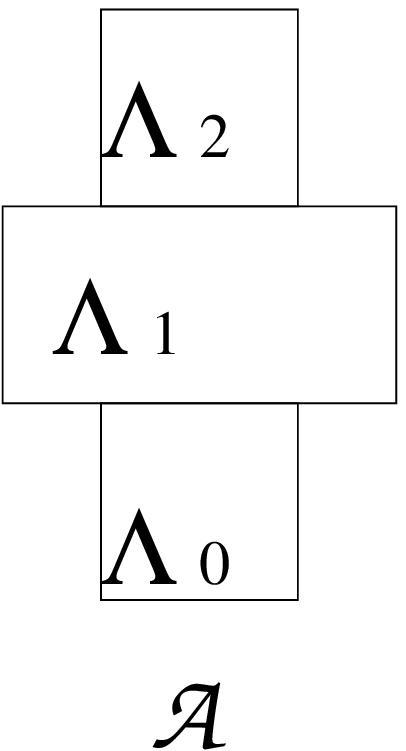}\end{minipage}
\begin{minipage}{0.6cm}$\;$\end{minipage}
\begin{minipage}{0.9cm}(b)\end{minipage}\begin{minipage}{2cm}{\includegraphics[width=2cm]{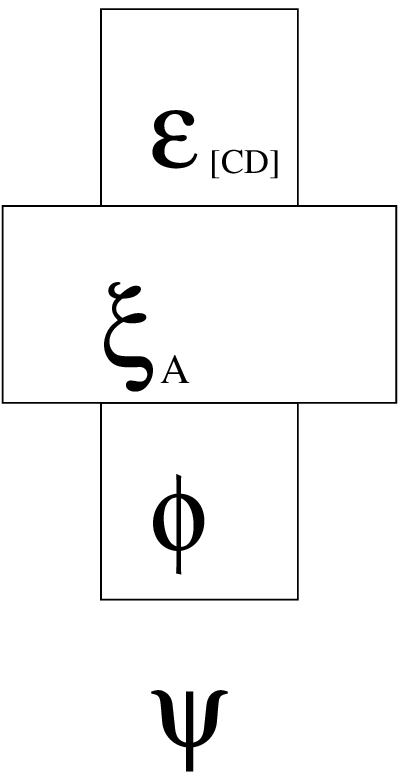}}\end{minipage}
\begin{minipage}{0.6cm}$\;$\end{minipage}
\begin{minipage}{0.9cm}(c)\end{minipage}\begin{minipage}{2cm}{\includegraphics[width=2cm]{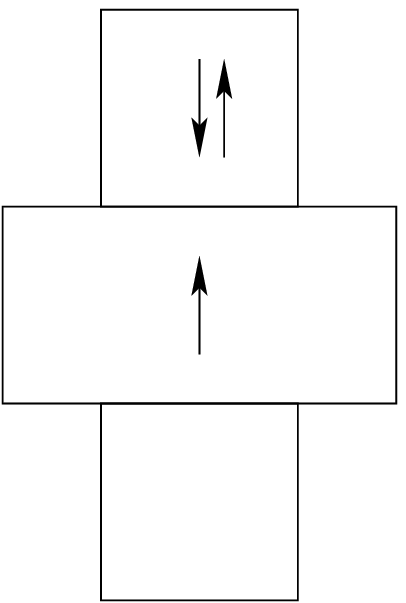}\vspace{7mm}}\end{minipage}\newline
$\;$\newline$\;$\newline
\begin{minipage}{0.9cm}(d)\end{minipage}\begin{minipage}{2cm}\includegraphics[width=2cm]{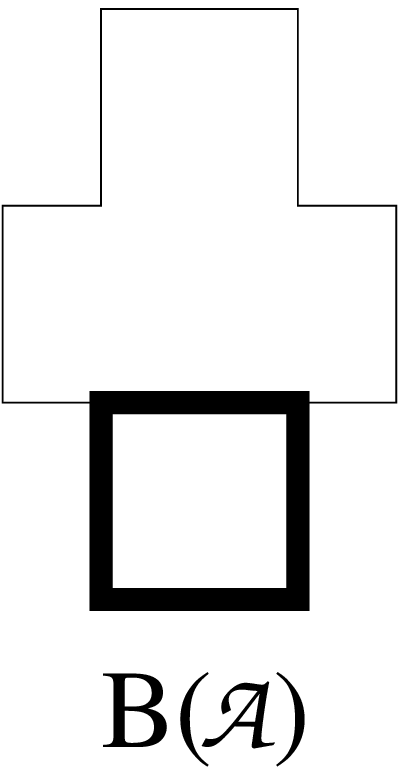}\end{minipage}
\begin{minipage}{0.6cm}$\;$\end{minipage}
\begin{minipage}{0.9cm}(e)\end{minipage}\begin{minipage}{2cm}\includegraphics[width=2cm]{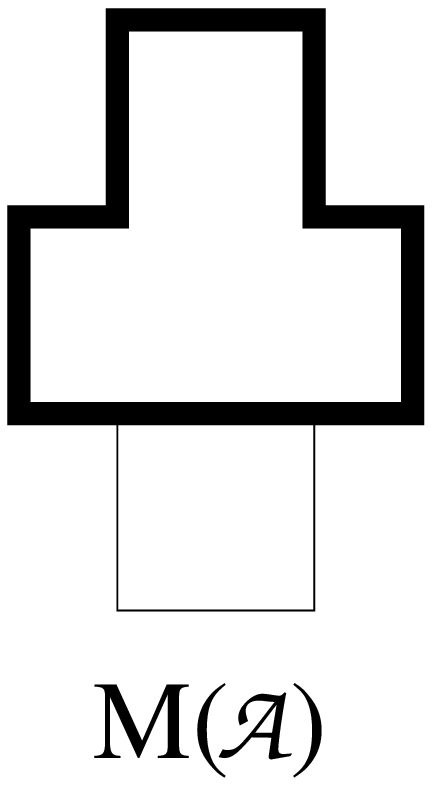}\end{minipage}
\begin{minipage}{0.6cm}$\;$\end{minipage}
\begin{minipage}{0.9cm}(f)\end{minipage}\begin{minipage}{2cm}\includegraphics[width=2cm]{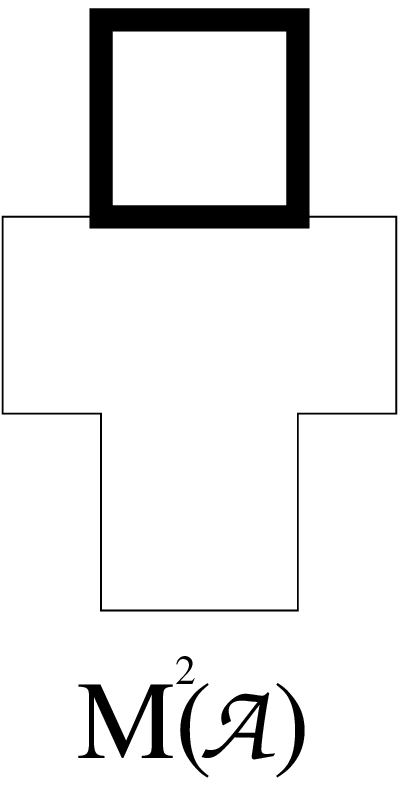}\end{minipage}
\begin{minipage}{0.6cm}$\;$\end{minipage}
\begin{minipage}{0.9cm}(g)\end{minipage}\begin{minipage}{2cm}\includegraphics[width=2cm]{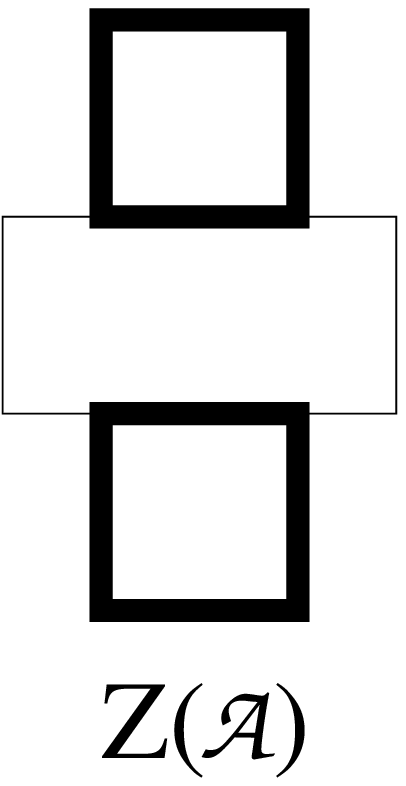}\end{minipage}\newline
$\;$\newline$\;$\newline
\begin{minipage}{0.9cm}(h)\end{minipage}\begin{minipage}{2cm}{\includegraphics[width=2cm]{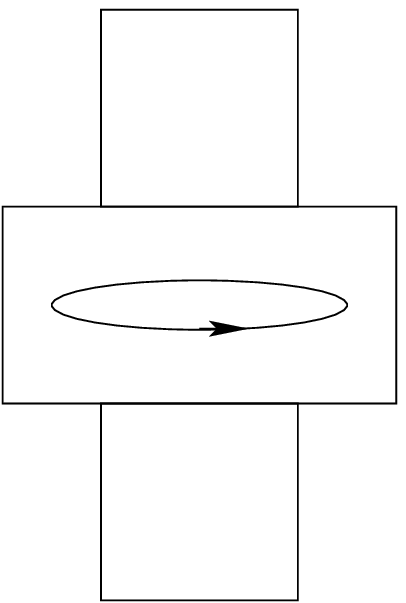}\vspace{7mm}}\end{minipage}
\begin{minipage}{0.3cm}$\;$\end{minipage}
\begin{minipage}{0.9cm}(i)\end{minipage}\begin{minipage}{3.5cm}{\includegraphics[width=3.5cm]{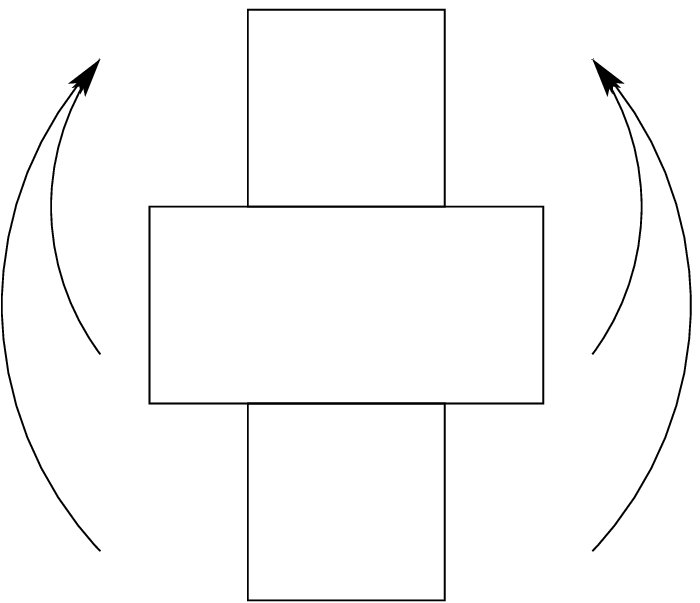}\vspace{7mm}}\end{minipage}
\begin{minipage}{0.3cm}$\;$\end{minipage}
\begin{minipage}{0.9cm}(j)\end{minipage}\begin{minipage}{2cm}\includegraphics[width=2cm]{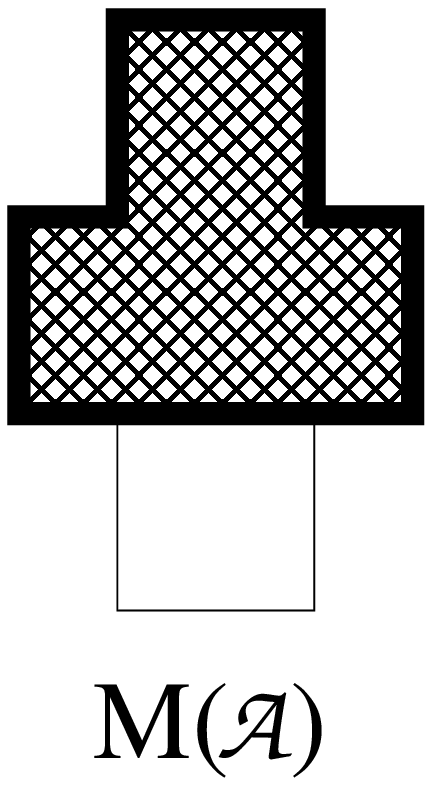}\end{minipage}
\begin{minipage}{0.3cm}$\;$\end{minipage}
\begin{minipage}{0.9cm}(k)\end{minipage}\begin{minipage}{2cm}\includegraphics[width=2cm]{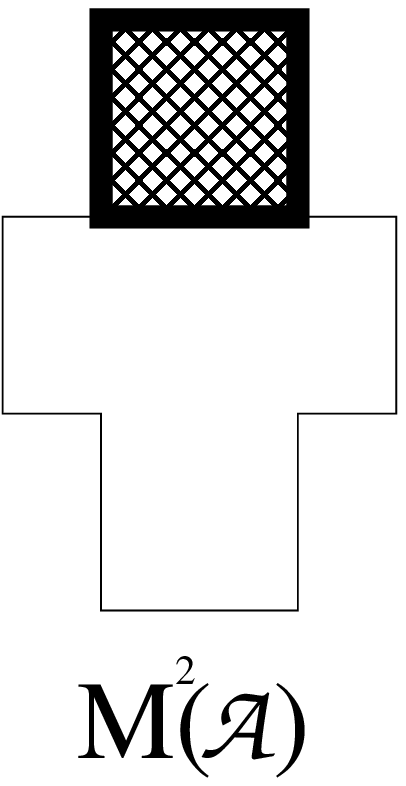}\end{minipage}
\end{center}
\caption{Illustration of the structure of the complex unital associative algebra 
${\A\equiv\Lambda(S^{*})}$ and the natural group action of the conservative Lorentz group extension 
${\exp(L_{M(\A)})\rtimes\GL(\Lambda_{1})}$ over it.
Panel~(a): the algebra $\A$ with a 
fixed $\Z$-grading (${\Lambda_{p}\equiv\mathop{\wedge}^{p}S^{*}}$). 
Panel~(b): whenever a fixed $\Z$-grading is taken, an element $\psi$ of $\A$ 
can be represented by a tuple of spinors. 
Panel~(c): heuristically speaking, the algebra $\A$ can be considered as a 
creation operator algebra of fermions with 2 fundamental degrees of freedom.
Panels~(d)--(e)--(f)--(g): important subspaces of the algebra $\A$, 
namely the scalar sector $B(\A)$, the maximal ideal $M(\A)$, and its second power $M^{2}(\A)$, 
moreover the center $Z(\A)$.
Panels~(h)--(i): illustration of the group action of the grading preserving part 
($\GL(\Lambda_{1})$) and of the grading non-preserving part ($\exp L_{M(\A)}$) of the 
full symmetry group ${\exp(L_{M(\A)})\rtimes\GL(\Lambda_{1})}$. The grading preserving part, 
by definition conserves the $p$-form subspaces, whereas the grading 
non-preserving part mixes higher forms to lower forms.
Panels~(j)--(k): list of all the invariant subspaces, which are invariant 
to the group action of the full 
symmetry group ${\exp(L_{M(\A)})\rtimes\GL(\Lambda_{1})}$. It is seen that none of the 
invariant subspaces possess an invariant complementing subspace, and thus 
the defining representation on $\A$ is indecomposable. In other words: 
the pertinent group action puts $\A$ into a single multiplet. Note that in the 
representation space of a non-semisimple Lie group an invariant subspace might 
not have invariant complement, i.e.\ a reducible representation might still 
be an indecomposable (non-direct sum) multiplet.}
\label{figAcal}
\end{figure*}

On the algebra $\A$, the group ${\GL(2,\C)\equiv\GL(S^{*})}$ also has a natural representation. 
That is because ${\GL(S^{*})}\equiv{\GL(\Lambda_{1})}\equiv{\GL(M(\A)/M^{2}(\A))}$ describes the 
$\Z$-grading preserving algebra automorphisms of $\A$ \cite{Djokovic1978}. 
Therefore, one can construct the semi-direct product group 
${\exp(L_{M(\A)})\rtimes\GL(\Lambda_{1})}\equiv{\mathrm{H}_{3}(\C)\rtimes\GL(2,\C)}$, 
which then by construction has a natural faithful complex-linear representation on $\A$, 
which happens to be the defining representation, i.e.\ the smallest dimensional 
faithful linear representation of ${\mathrm{H}_{3}(\C)\rtimes\GL(2,\C)}$. 
The structure of the algebra $\A$ along with the natural action of 
the group ${\exp(L_{M(\A)})\rtimes\GL(\Lambda_{1})}$ on $\A$ is illustrated in Figure~\ref{figAcal}.

Since the group ${\exp(L_{M(\A)})\rtimes\GL(\Lambda_{1})}$ has a linear 
action on $\A$, there is a canonical faithful representation also on its complex conjugate 
space $\bar{\A}$, via the requirement of being compatible with the natural 
${\A\rightarrow\bar{\A}}$ complex conjugation map. This is in analogy of 
$\GL(S^{*})$ having its canonical representation on $S^{*}$, and consequently 
having its canonical representation on $\bar{S}^{*}$, via requiring the 
invariance of the ${S^{*}\rightarrow\bar{S}^{*}}$ complex conjugation map.

The actual representation space in our toy model shall be ${A:=\bar{\A}\otimes\A}$, 
where $\otimes$ denotes ordinary, i.e.\ vector space sense tensor product 
(not a graded tensor product).\footnote{If 
$\otimes$ were a graded tensor product, then ${\bar{\A}{\otimes}\A}$ could be viewed as superfields. Here, we are not considering that situation, since we would like $\A$ and $\bar{\A}$ to describe fermionic degrees of freedom, and their charge conjugates, respectively, and we would like to impose Pauli principle for these fields separately.} The algebra $A$ is a kind of doubled exterior algebra, 
which we shall call \emph{spin algebra}, being a 16 dimensional complex unital associative algebra. 
Since its components $\A$ and $\bar{\A}$ play the role of generalizations of the co-spinor space 
$S^{*}$ and the complex conjugate co-spinor space $\bar{S}^{*}$, the spin algebra 
${A=\bar{\A}\otimes\A}$ can be considered as a generalization of the mixed 
co-spinor space $\bar{S}^{*}\otimes S^{*}$. In fact, whenever a preferred 
$\Z$-grading of $\A$ is fixed, the spin algebra may be identified as 
${A\equiv\mathop{\bigoplus}_{p,q=0}^{2}\mathop{\wedge}^{p}\bar{S}^{*}\mathop{\otimes}\mathop{\wedge}^{q}S^{*}}$, 
i.e.\ its representation can be given in terms of ordinary two-spinors. 
By construction, the spin algebra $A$ also carries a natural antilinear 
involution ${\overline{(\cdot)}:A\rightarrow A}$, which we call \emph{charge conjugation}, 
and which has the property ${\overline{x\,y}=\overline{x}\,\overline{y}}$ for all ${x,y\in A}$. 
The pertinent charge conjugation map is simply defined by the composition of the natural complex conjugation as a 
${\bar{\A}\otimes\A\rightarrow \A\otimes\bar{\A}}$ map and of the natural tensor 
product swapping as a ${\A\otimes\bar{\A}\rightarrow\bar{\A}\otimes\A}$ map, hence 
giving rise to a natural ${\bar{\A}\otimes\A\rightarrow\bar{\A}\otimes\A}$ antilinear 
involution on $A$. It can be considered as a generalization of the hermitian 
conjugation ${\bar{S}^{*}\otimes S^{*}\rightarrow\bar{S}^{*}\otimes S^{*}}$ on the 
space of mixed co-spinors $\bar{S}^{*}\otimes S^{*}$, as usual in the ordinary two-spinor calculus. 
Since the group ${\exp(L_{M(\A)})\rtimes\GL(\Lambda_{1})}\equiv{\mathrm{H}_{3}(\C)\rtimes\GL(2,\C)}$ 
has a natural linear representation both on $\A$ and $\bar{\A}$, it also has a corresponding 
linear representation on the spin algebra ${A=\bar{\A}\otimes\A}$.

\begin{figure*}
\begin{center}
\begin{minipage}{0.7cm}(a)\end{minipage}\begin{minipage}{4cm}\includegraphics[width=4cm]{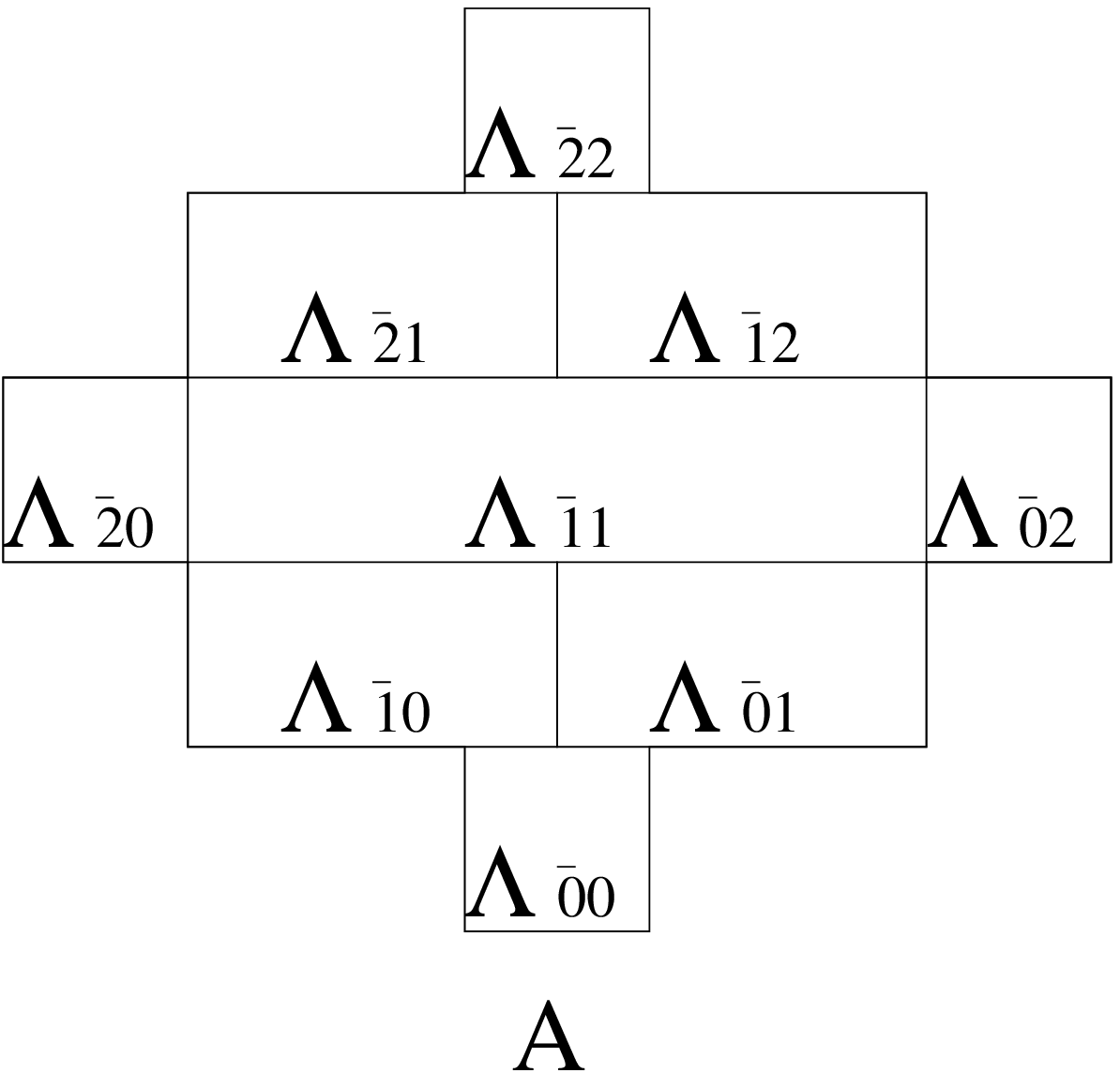}\end{minipage}
\begin{minipage}{0.2cm}$\;$\end{minipage}
\begin{minipage}{0.7cm}(b)\end{minipage}\begin{minipage}{4cm}{\includegraphics[width=4cm]{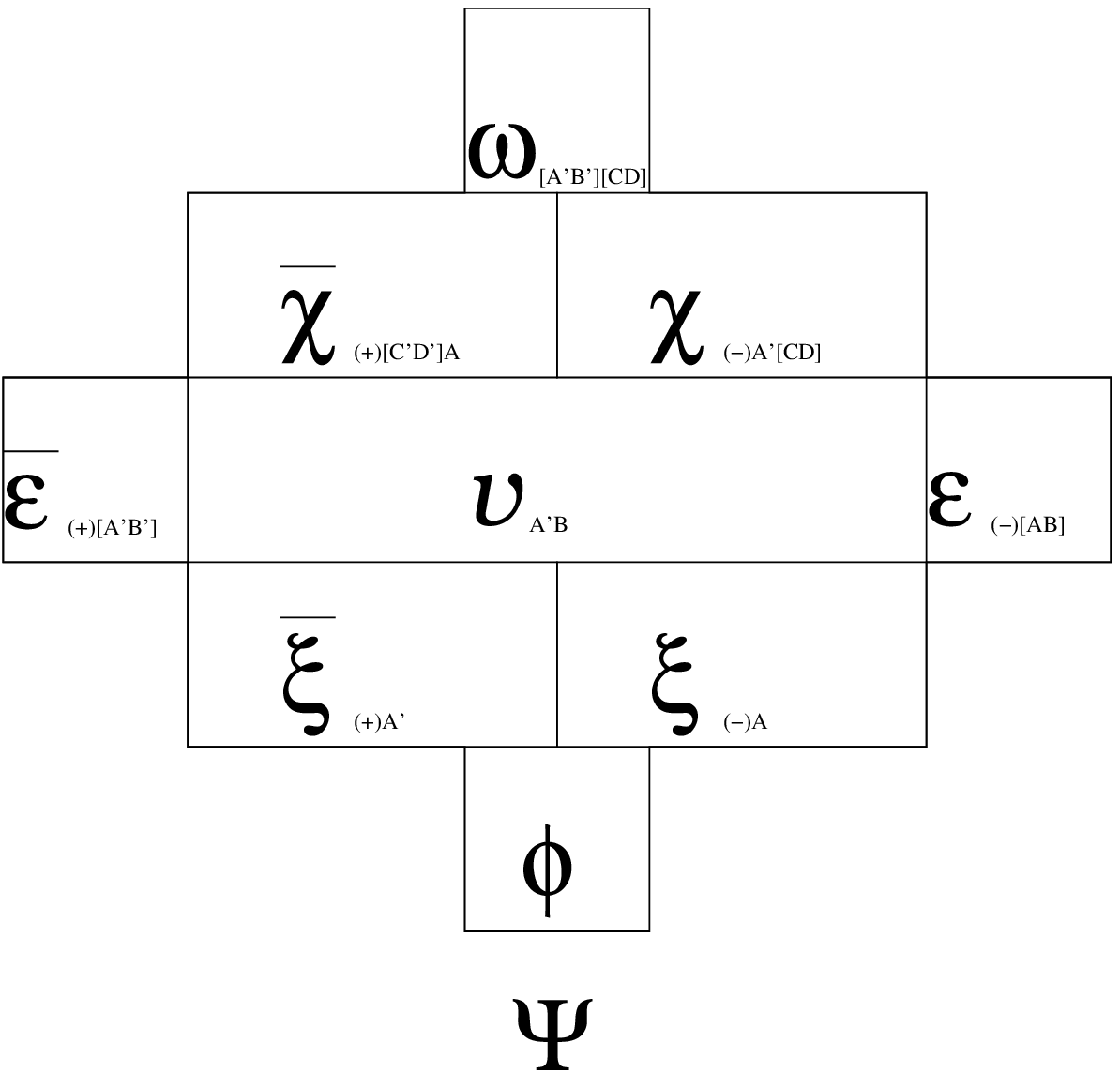}}\end{minipage}
\begin{minipage}{0.2cm}$\;$\end{minipage}
\begin{minipage}{0.7cm}(c)\end{minipage}\begin{minipage}{4cm}{\includegraphics[width=4cm]{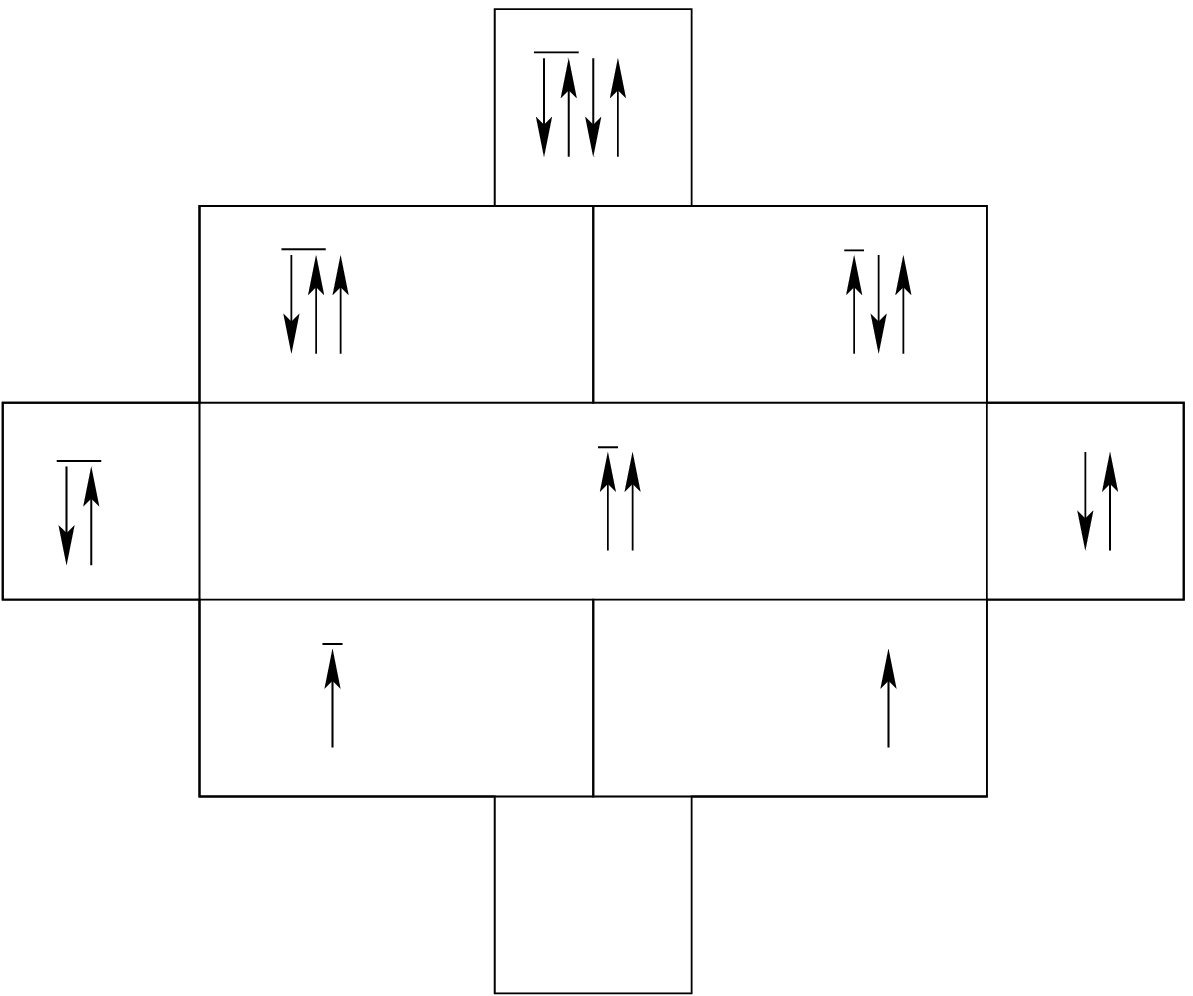}\vspace{5mm}}\end{minipage}\newline
\begin{minipage}{\textwidth}\vspace*{2mm}\end{minipage}\newline
\begin{minipage}{0.7cm}(d)\end{minipage}\begin{minipage}{4cm}\includegraphics[width=4cm]{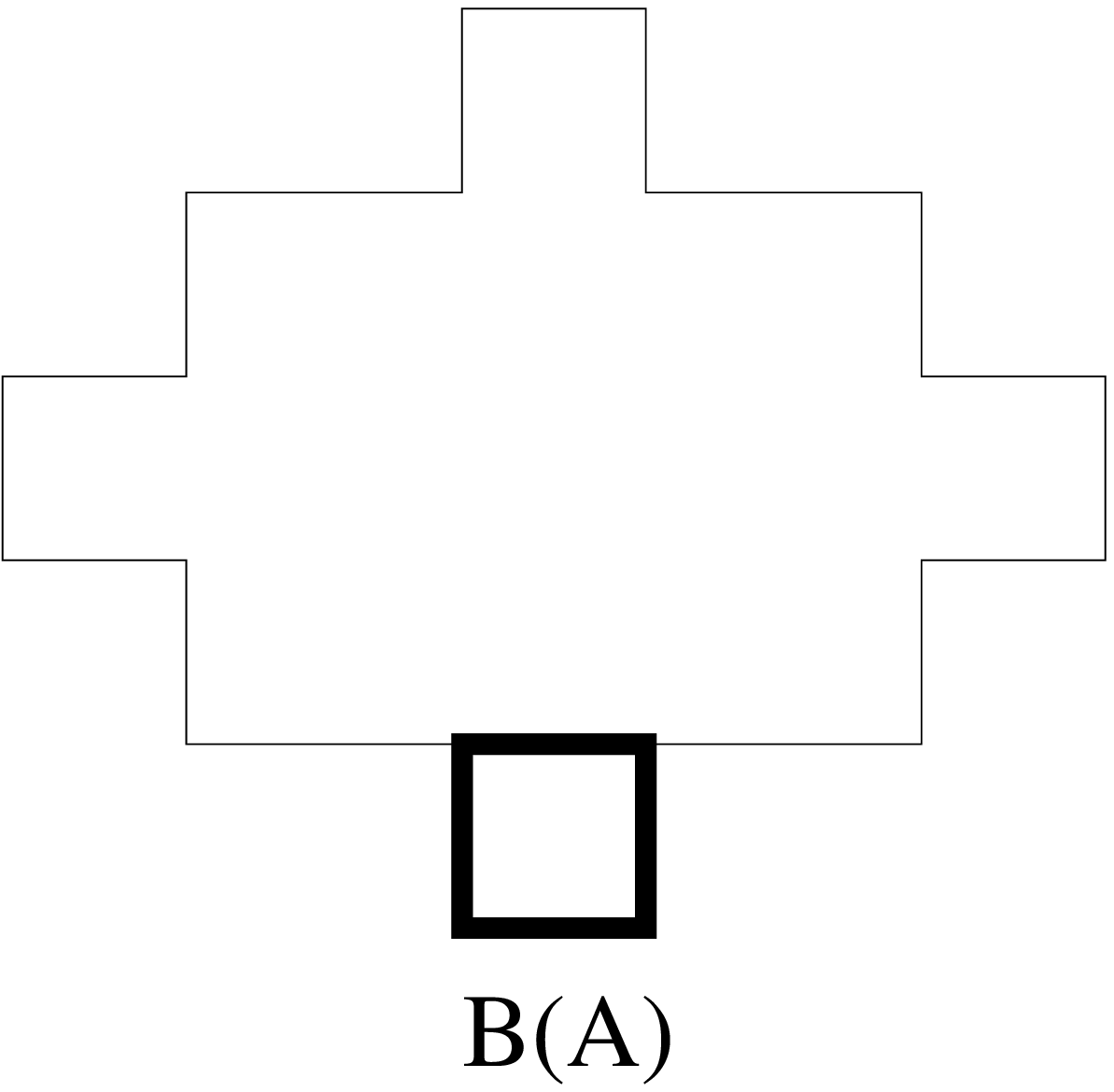}\end{minipage}
\begin{minipage}{0.2cm}$\;$\end{minipage}
\begin{minipage}{0.7cm}(e)\end{minipage}\begin{minipage}{4cm}\includegraphics[width=4cm]{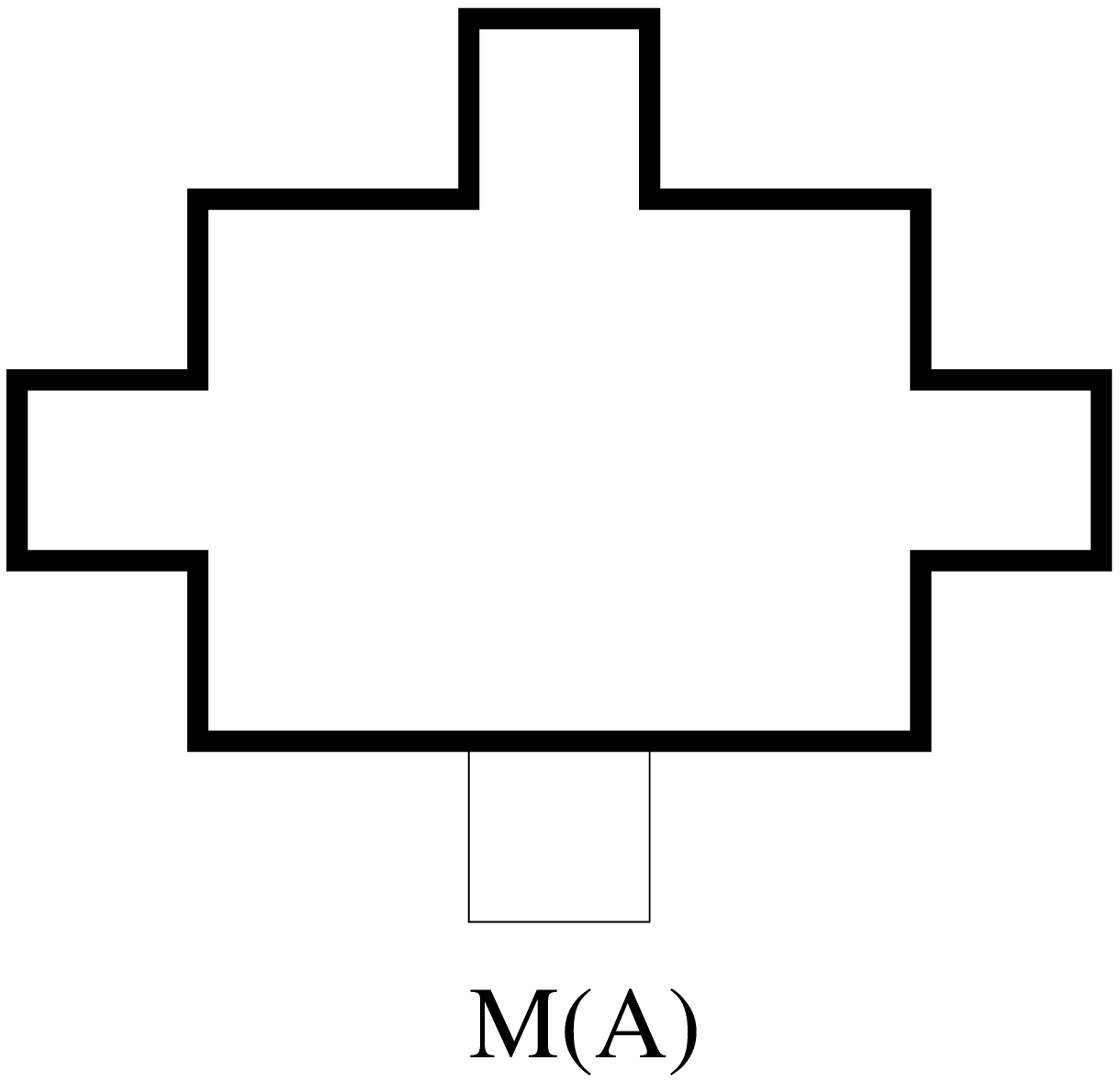}\end{minipage}
\begin{minipage}{0.2cm}$\;$\end{minipage}
\begin{minipage}{0.7cm}(f)\end{minipage}\begin{minipage}{4cm}\includegraphics[width=4cm]{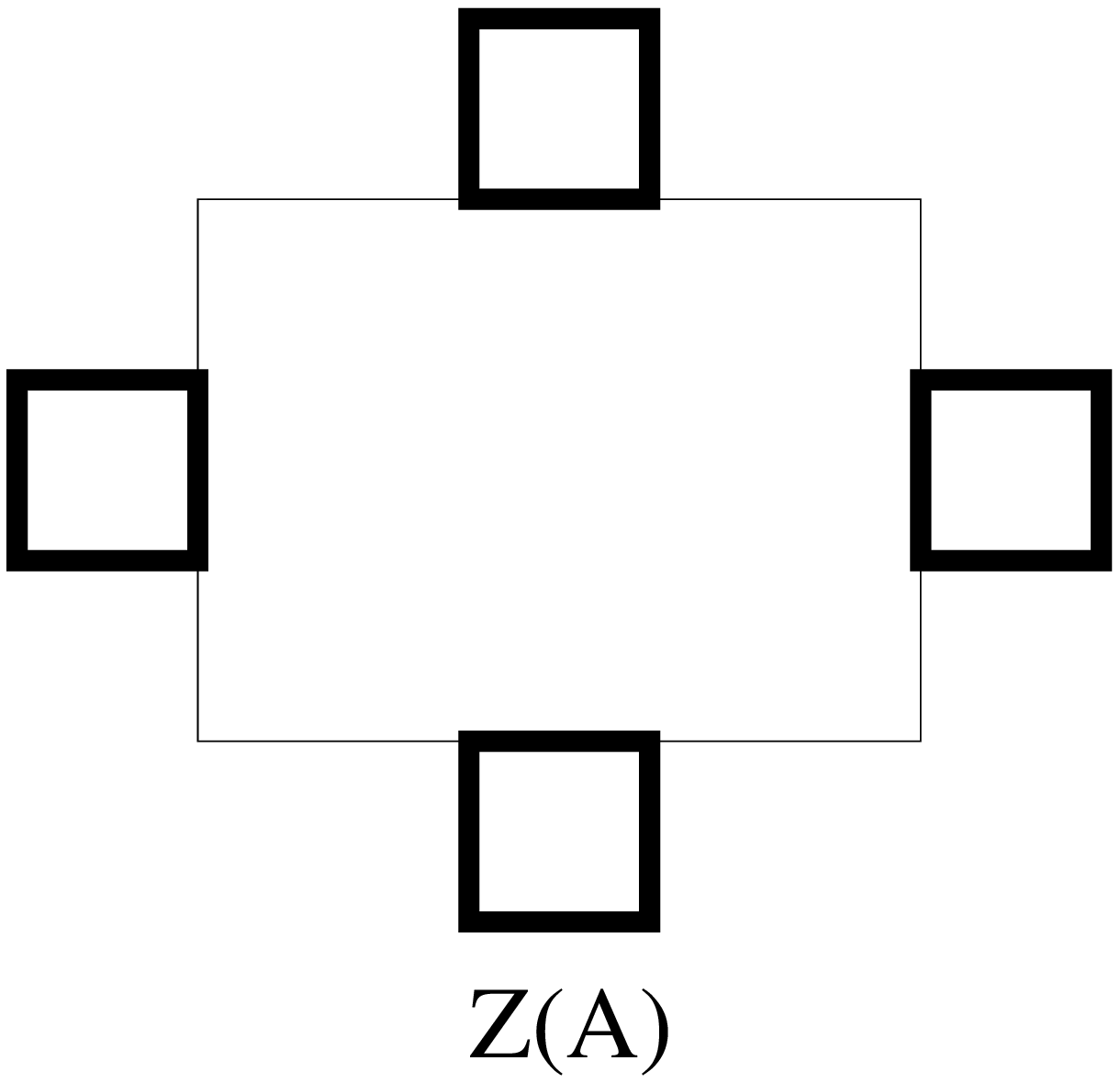}\end{minipage}\newline
\begin{minipage}{\textwidth}\vspace*{2mm}\end{minipage}\newline
\begin{minipage}{0.7cm}(g)\end{minipage}\begin{minipage}{4cm}\includegraphics[width=4cm]{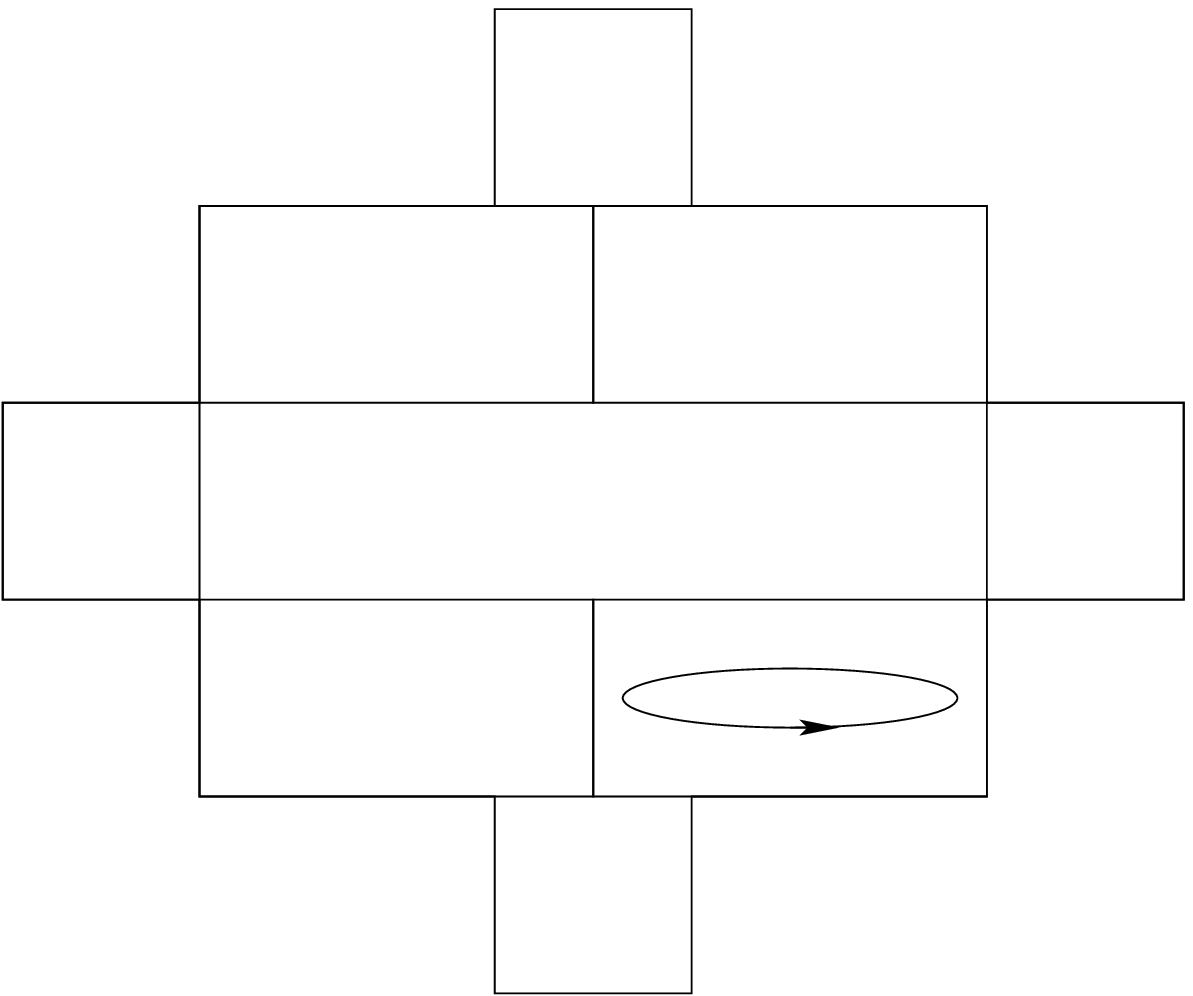}\end{minipage}
\begin{minipage}{0.2cm}$\;$\end{minipage}
\begin{minipage}{0.7cm}(h)\end{minipage}\begin{minipage}{5.2cm}\includegraphics[width=5.2cm]{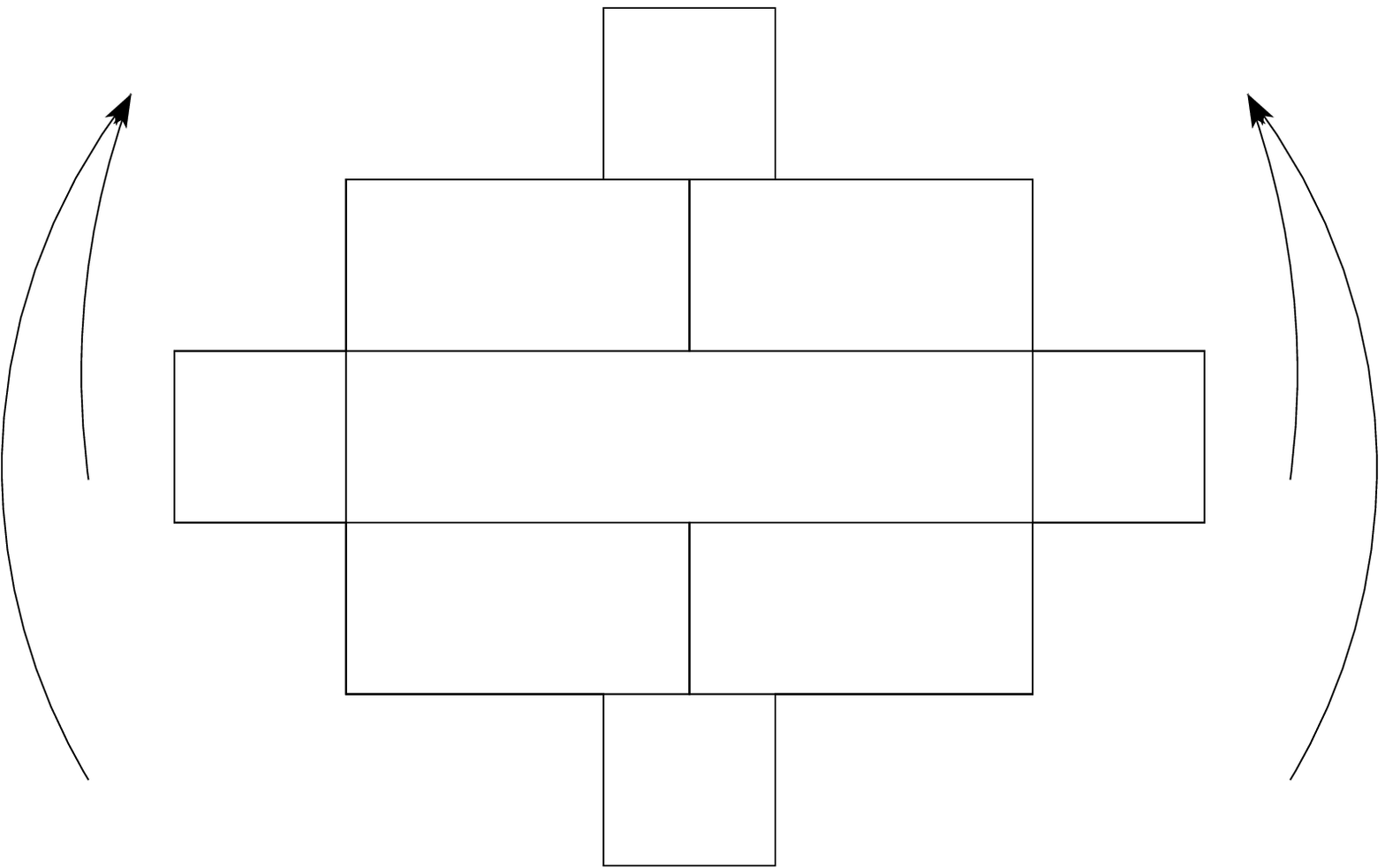}\end{minipage}
\begin{minipage}{0.2cm}$\;$\end{minipage}
\begin{minipage}{\textwidth}$\;$\end{minipage}\newline
\begin{minipage}{0.7cm}(i)\end{minipage}\begin{minipage}{2.7cm}\includegraphics[width=2.7cm]{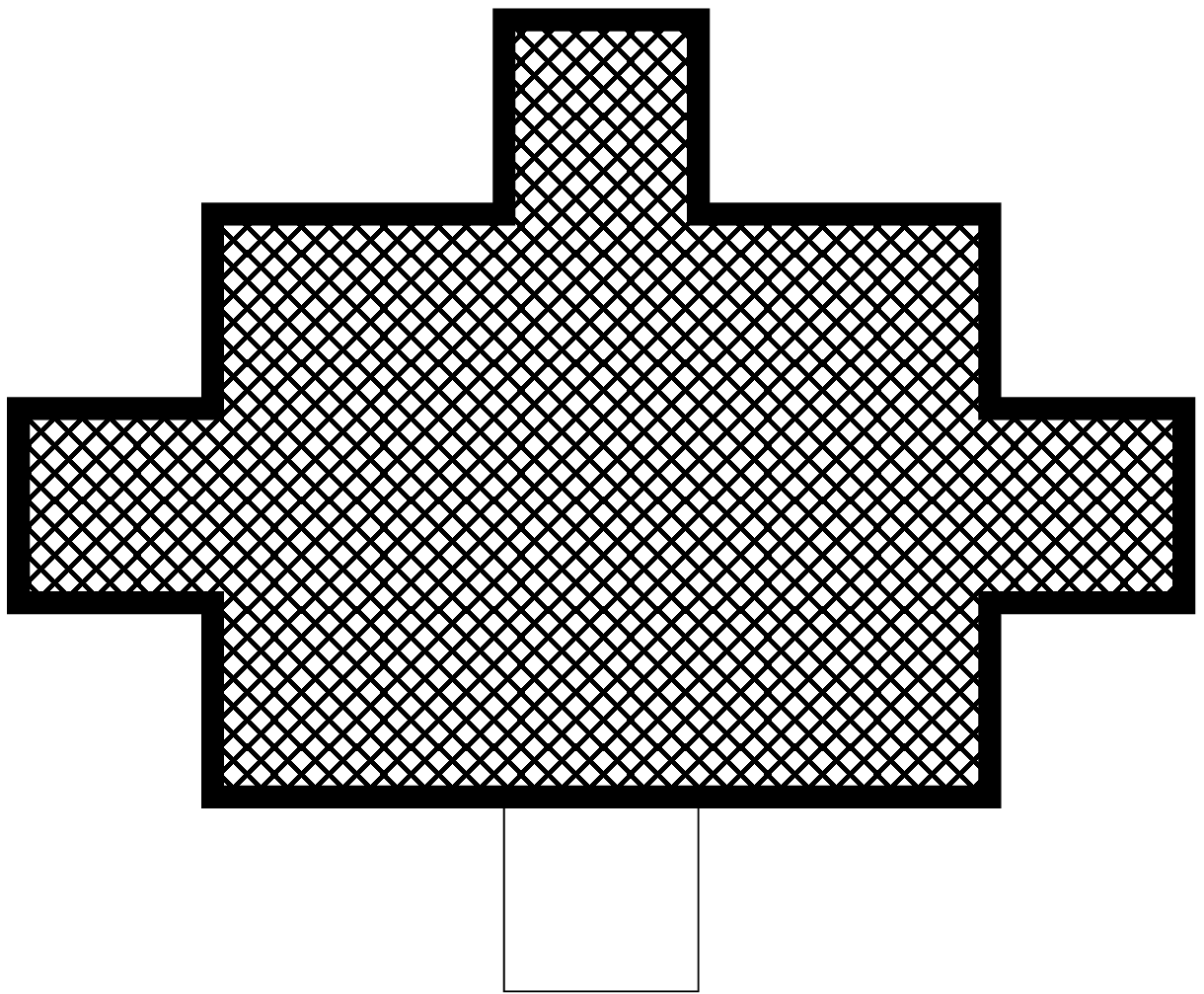}\end{minipage}
\begin{minipage}{0.2cm}$\;$\end{minipage}
\begin{minipage}{0.7cm}(j)\end{minipage}\begin{minipage}{2.7cm}\includegraphics[width=2.7cm]{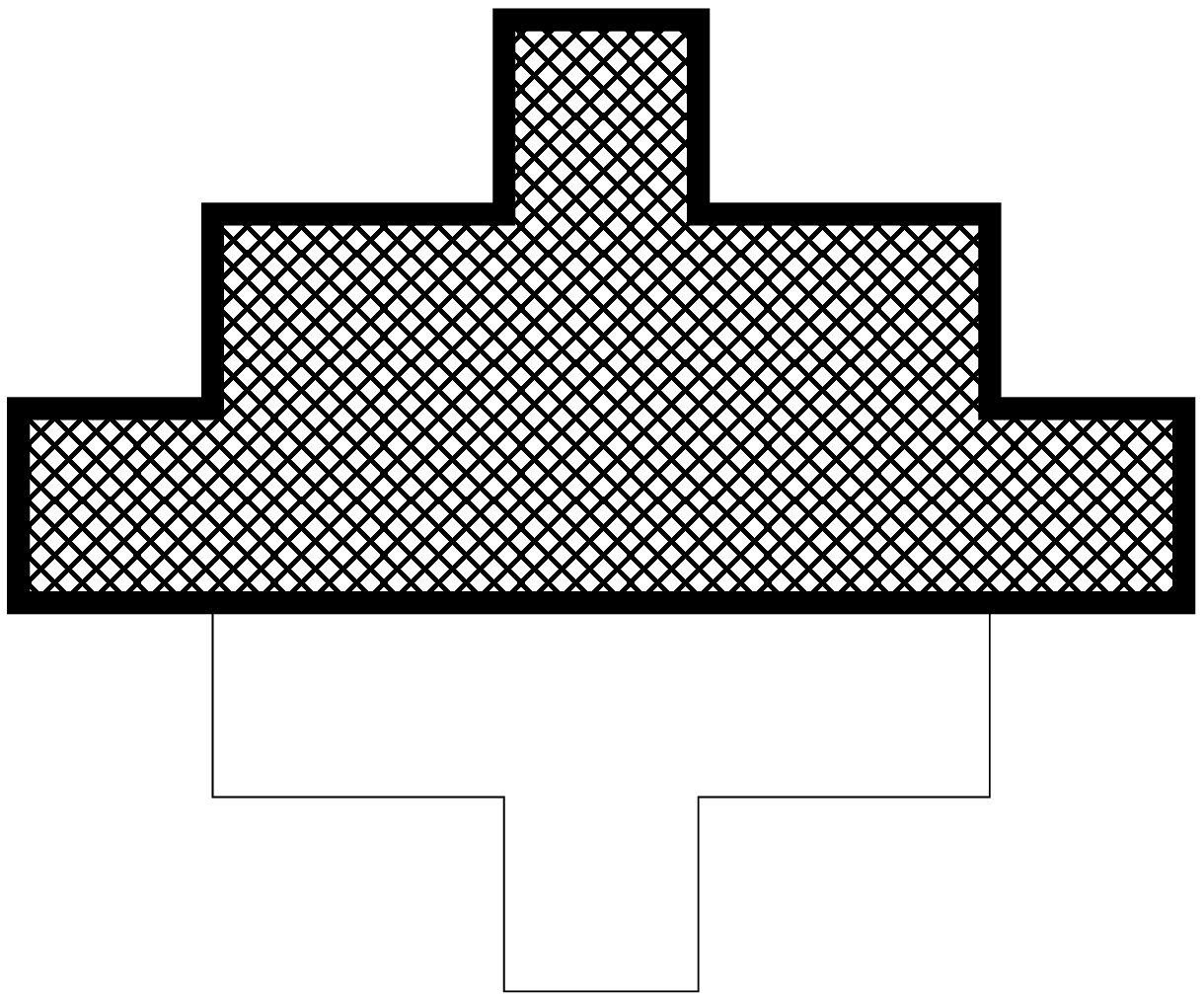}\end{minipage}
\begin{minipage}{0.2cm}$\;$\end{minipage}
\begin{minipage}{0.7cm}(k)\end{minipage}\begin{minipage}{2.7cm}\includegraphics[width=2.7cm]{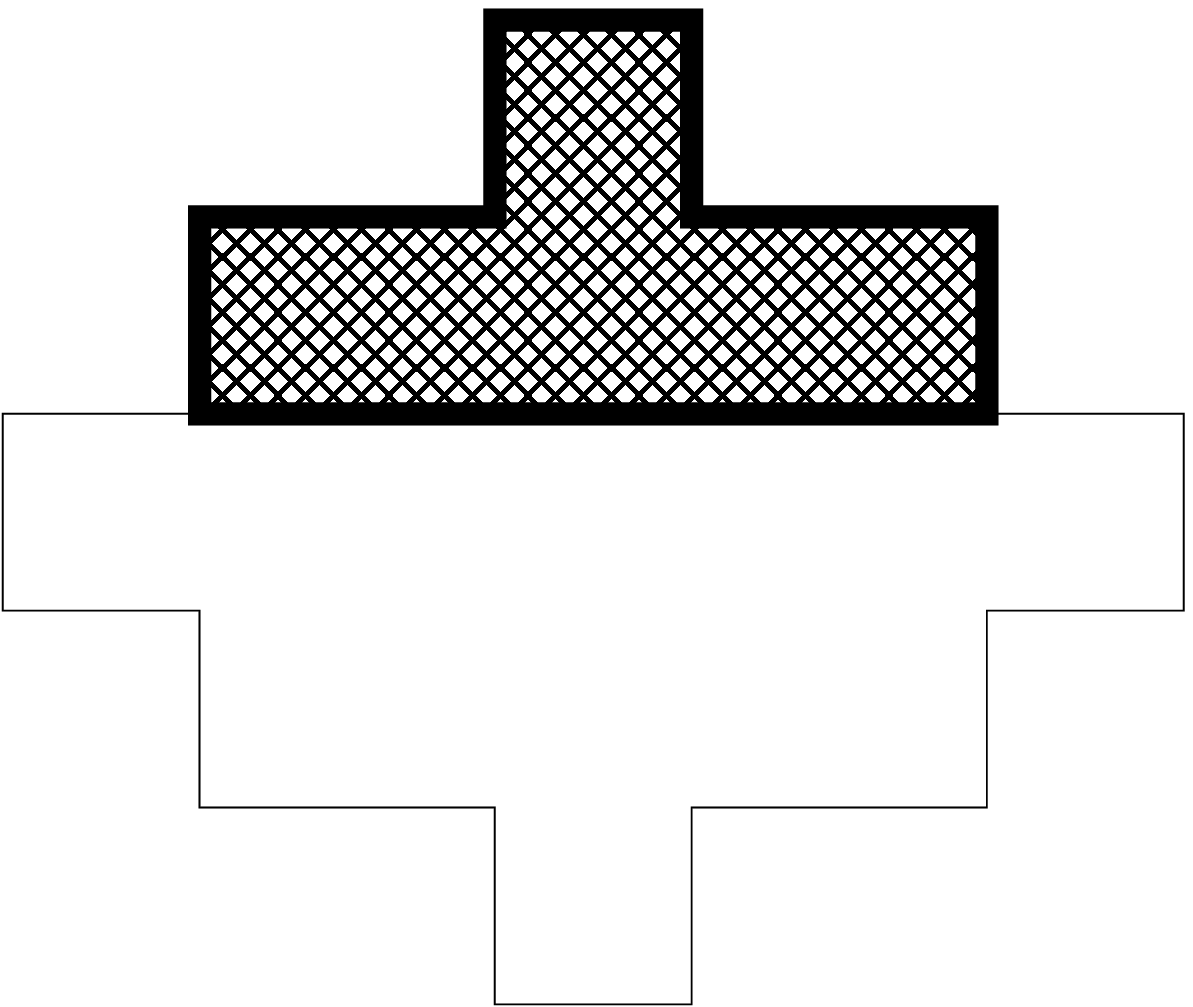}\end{minipage}
\begin{minipage}{0.2cm}$\;$\end{minipage}
\begin{minipage}{0.7cm}(l)\end{minipage}\begin{minipage}{2.7cm}\includegraphics[width=2.7cm]{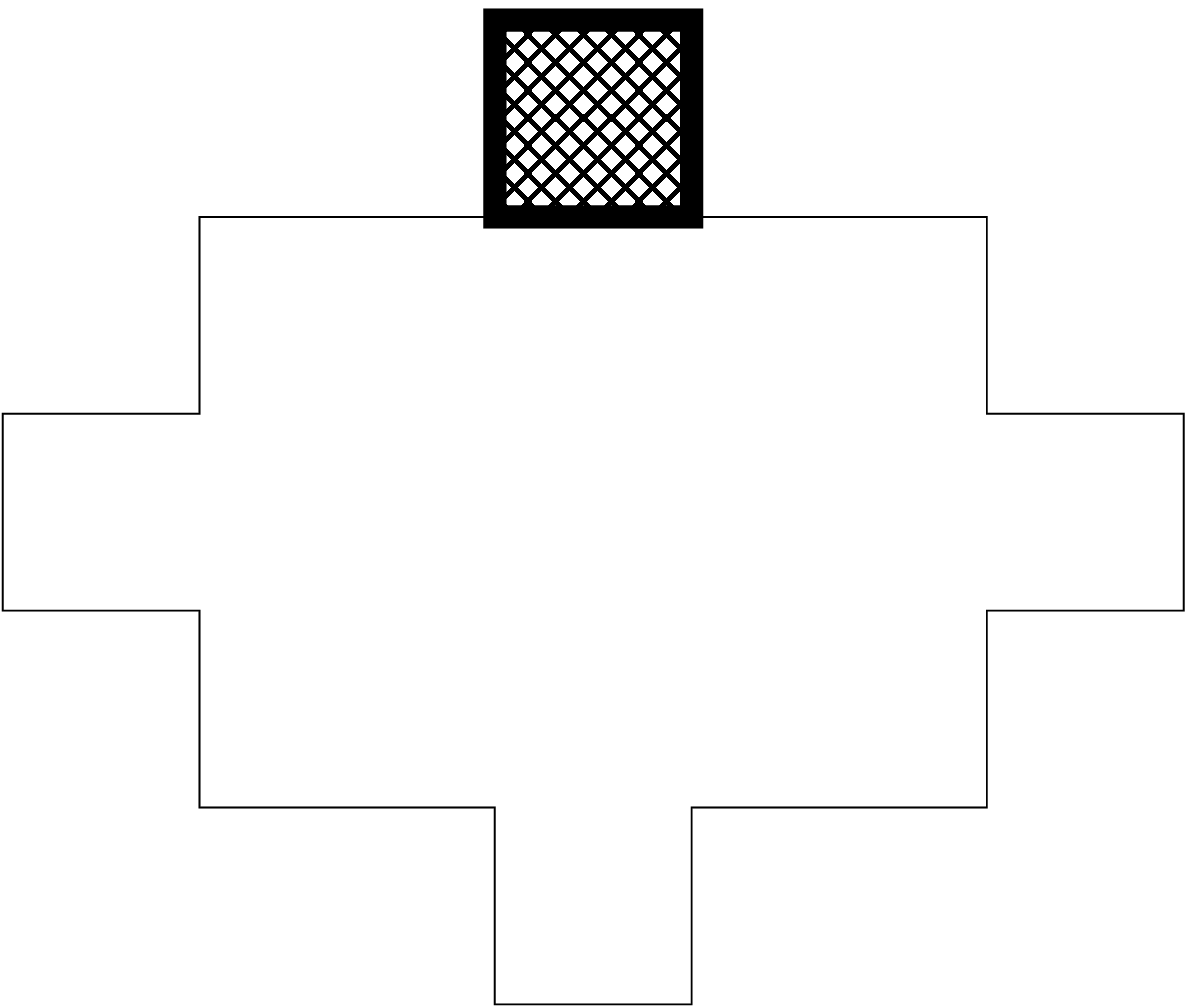}\end{minipage}\newline
\newline
\begin{minipage}{0.7cm}(m)\end{minipage}\begin{minipage}{2.7cm}\includegraphics[width=2.7cm]{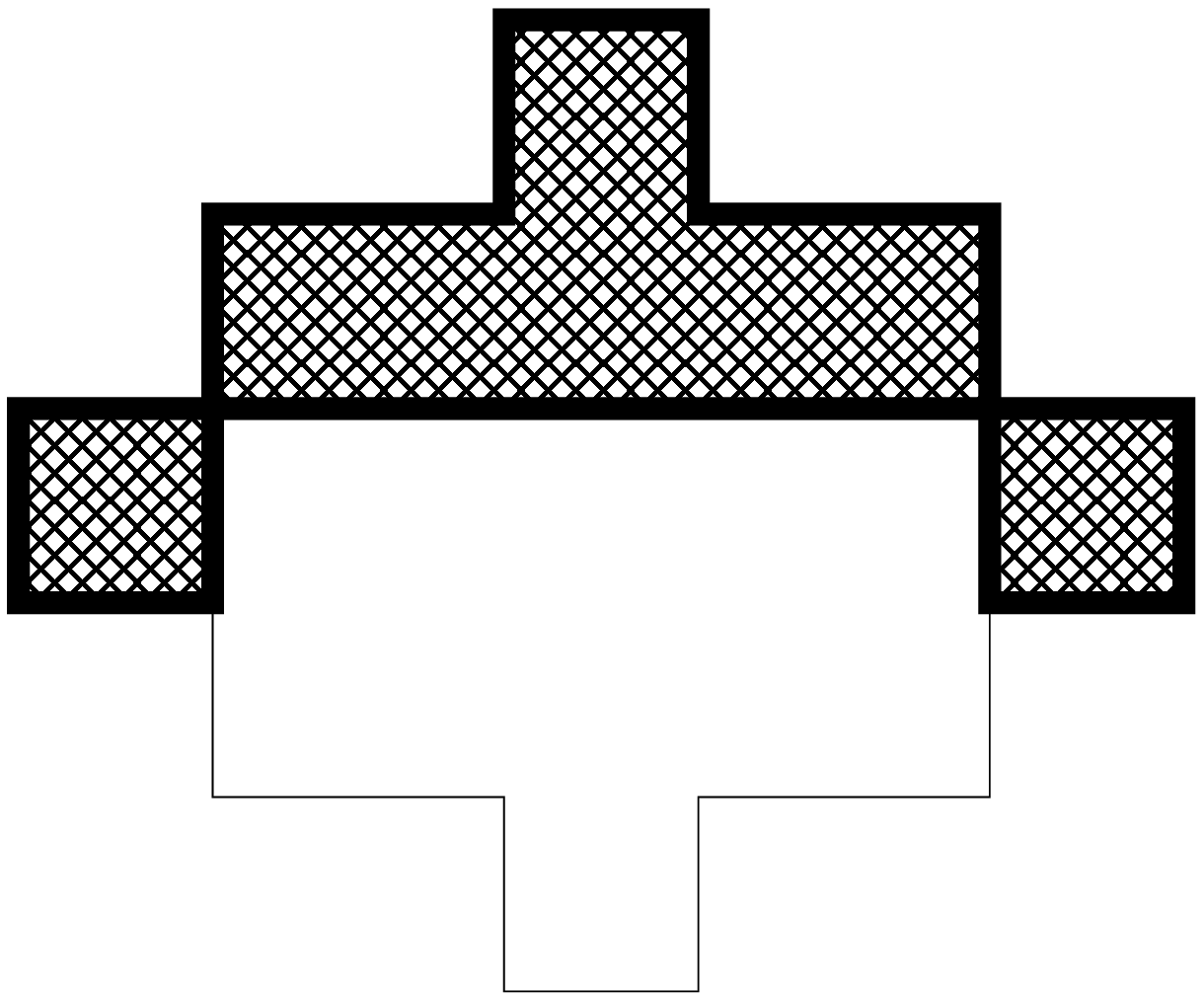}\end{minipage}
\begin{minipage}{2cm}$\;$\end{minipage}
\begin{minipage}{0.7cm}(n)\end{minipage}\begin{minipage}{2.7cm}\includegraphics[width=2.7cm]{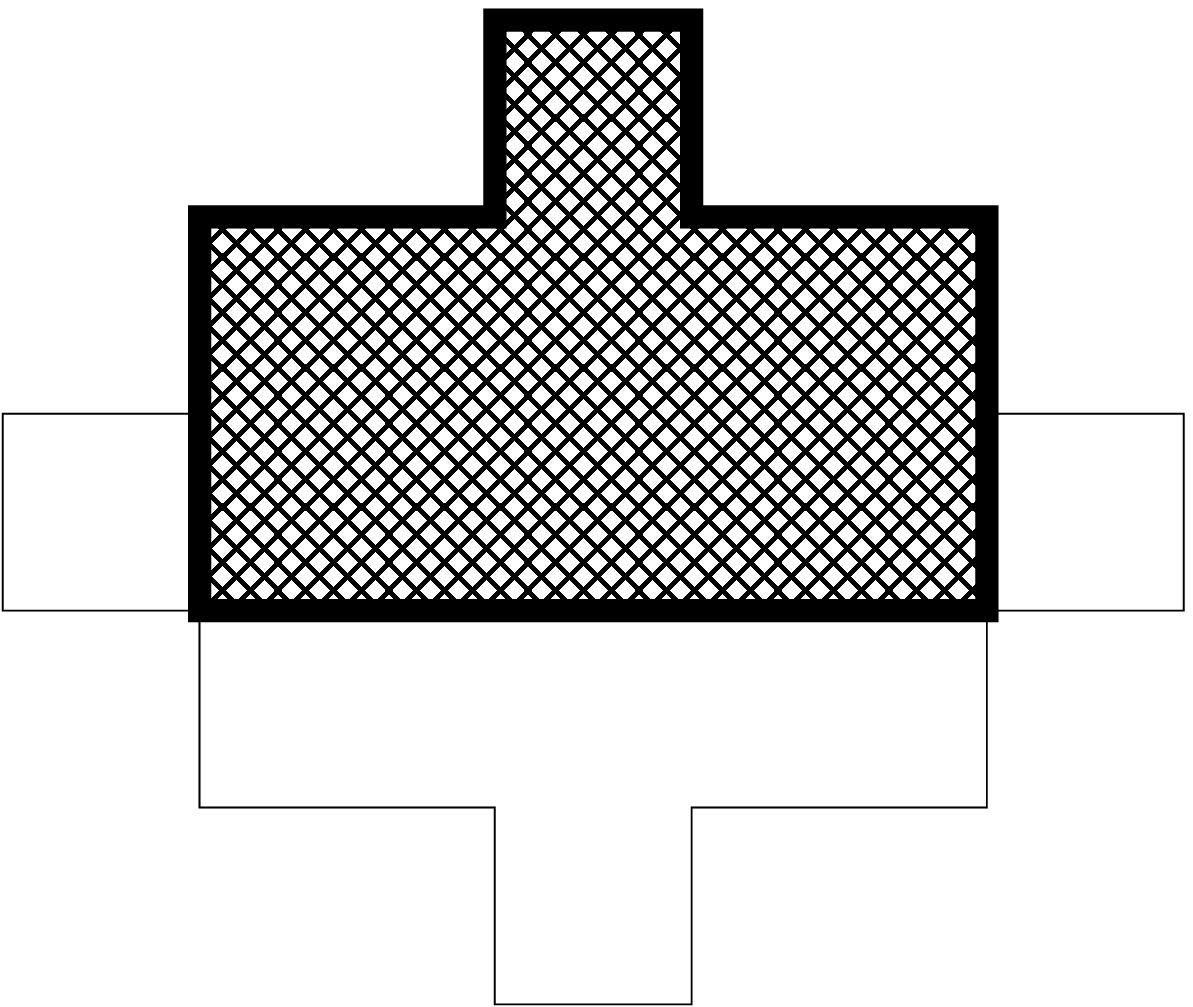}\end{minipage}
\end{center}
\caption{Illustration of the structure of the \emph{spin algebra} 
${A\equiv\Lambda(\bar{S}^{*})\otimes\Lambda(S^{*})}$ and the natural group action of ${\G\rtimes\{I,\,\overline{(\cdot)}\}}$ over it. 
Panel~(a): the algebra $A$ with a 
fixed $\Zs$-grading (${\Lambda_{\bar{p}q}\equiv\mathop{\wedge}^{p}\bar{S}^{*}\otimes\mathop{\wedge}^{q}S^{*}}$). 
Panel~(b): whenever a fixed $\Zs$-grading is taken, an element $\Psi$ of $A$ 
can be represented by a tuple of spinors. 
Panel~(c): heuristically speaking, the algebra $A$ can be considered as a 
creation operator algebra of two distinct kind of fermions with 2 fundamental degrees of 
freedom each, and the two kinds being charge conjugate to each-other. 
Panels~(d)--(e)--(f): important subspaces of the algebra $A$, 
namely the scalar sector $B(A)$, the maximal ideal $M(A)$, moreover the center $Z(A)$.
Panels~(g)--(h): illustration of the group action of the grading preserving part 
and of the grading non-preserving part of the symmetry group ${\G\rtimes\{I,\,\overline{(\cdot)}\}}$. 
Panels~(i)--(n): list of all the subspaces of $A$, which are invariant 
under the group action of the symmetry group ${\G\rtimes\{I,\,\overline{(\cdot)}\}}$. It is seen that 
no invariant complementing subspaces exist, i.e.\ $A$ is an indecomposable multiplet.}
\label{figA}
\end{figure*}

The structure group of our toy model will be specified via its faithful linear 
representation on the spin algebra ${A=\bar{\A}\otimes\A}$. It is defined to be the group
\begin{eqnarray}
 \G & \;\;:=\;\;             &       \C^{{}^{\times}} \;\times\; \big(\, \exp(L_{M(\A)}) \;\rtimes\; \GL(\Lambda_{1}) \,\big) \cr
        & \;\;\;\;\equiv\;\;     & \Big. \C^{{}^{\times}} \;\times\; \big(\, \mathrm{H}_{3}(\C) \;\rtimes\; \GL(2,\C) \,\big) \cr
        & \;\;\;\;\;\;\equiv\;\; &       \big(\,\C^{{}^{\times}} \;\times\; \mathrm{H}_{3}(\C) \,\big) \;\rtimes\; \GL(2,\C)
\label{eqG}
\end{eqnarray}
where $\C^{{}^{\times}}$ denotes the scaling by nonzero complex numbers on $A$. 
The factor $\C^{{}^{\times}}$ is merely present because in fact in the toy model, 
a projective representation of 
${\exp(L_{M(\A)})\rtimes\GL(\Lambda_{1})}\equiv{\mathrm{H}_{3}(\C) \;\rtimes\; \GL(2,\C)}$ 
is taken over $A$, and it is a notational convenience to view that projective 
representation instead a linear representation of $\G$ as in Eq.(\ref{eqG}). 
The Lie algebra of $\G$ is correspondingly
\begin{eqnarray}
 \mathfrak{g} & \;\;:=\;\;       &       \C \;\oplus\; \Big(\, L_{M(\A)} \;\opluslhrim\; \gl(\Lambda_{1}) \,\Big) \cr
        & \;\;\;\;\equiv\;\;     & \Big. \C \;\oplus\; \Big( \h_{3}(\C) \opluslhrim \big(\ua(1){\oplus}\,\da(1){\oplus}\,\sla(2,\C)\big) \Big) \cr
        & \;\;\;\;\;\;\equiv\;\; & \Big. \big(\C\oplus\h_{3}(\C)\big) \opluslhrim \big(\ua(1){\oplus}\,\da(1){\oplus}\,\sla(2,\C)\big) \qquad
\label{eqGa}
\end{eqnarray}
where $\C$ denotes the scaling by complex numbers on $A$. The group $\G$ is invariant under the conjugation by elements of the charge conjugation group ${\{I,\,\overline{(\cdot)}\}\equiv\Z_{2}}$, where $I$ is the identity map on $A$. Therefore, the semi-direct product ${\G\rtimes\{I,\,\overline{(\cdot)}\}}$ is meaningful. This detail will be important because we will prescribe the charge conjugation group ${\{I,\,\overline{(\cdot)}\}\equiv\Z_{2}}$ to be global symmetry of the toy model. 
The structure of the spin algebra $A$ along with the natural action of the group ${\G\rtimes\{I,\,\overline{(\cdot)}\}}$ on it is illustrated in Figure~\ref{figA}. It is seen that 
although ${\G\rtimes\{I,\,\overline{(\cdot)}\}}$-invariant subspaces within $A$ do exist, but none of them has an invariant complement, and thus the representation space $A$ is direct-indecomposable.

Before we continue, we briefly mention the heuristic meaning of the representation 
space $A$ and the group action of $\G$ on it. Since ${A=\bar{\A}\otimes\A}$, 
the algebra $A$ can be thought of as a creation operator algebra 
of two kinds of fermions, each having 2 fundamental degrees of freedom, and the two 
kinds being related to each-other via the charge conjugation operation 
$\overline{(\cdot)}$. The finite dimensional real Lie group $\G$ 
acts naturally on $A$, and the meaning of grading preserving transformations 
of $\G$ is clear: they induce $\gl(2,\C)\equiv{\ua(1){\oplus}\,\da(1){\oplus}\,\sla(2,\C)}$ transformations 
on the generating sector $\Lambda_{\bar{0}1}$ and corresponding natural action 
on all of the sectors $\Lambda_{\bar{p}q}$, and thus on the entire ${A\equiv\mathop{\bigoplus}_{p,q=0}^{2}\Lambda_{\bar{p}q}}$. 
The grading non-preserving transformations, isomorphic to $\mathrm{H}_{3}(\C)$, 
mix higher forms to lower forms, deforming the original $\Zs$-grading of $A$ to an 
other equivalent one. In the heuristic picture of creation operator algebras, 
the corresponding $\mathrm{H}_{3}(\C)$ action on an element ${\Psi\in A}$ 
would mean left insertion of equal amount of particles and corresponding 
charge conjugate particles into $\Psi$. (The spin algebra $A$ is not a 
CAR algebra, but is a related concept.)

In the following part we investigate important $\G$-invariant functions on $A$, 
which will be used to construct the invariant Lagrangian.

\subsection{Important invariant functions on representations of the example group}

In order to study the $\G$-invariant functions on $A$, it is convenient 
to first study the invariants of important ``special'' subgroup of it, 
in which the projective scaling group $\C^{{}^{\times}}$ is omitted, and 
that shall be denoted by $\G_{s}$. By construction, the special subgroup 
$\G_{s}$ may not only act on the full representation space 
${A=\bar{\A}{\otimes}\A}$, but also on its individual factors $\A$ and $\bar{\A}$ alone. 
It is a further convenience to introduce some even smaller special subgroups within 
$\G_{s}$: the subgroups $S\G_{s}$ and $S^{{}^{\times}}\!\!\G_{s}$ 
in which the $\D(1)$ and the ${\D(1){\times}\U(1)}$ component is omitted, respecively. 
These special subgroups within $\G$ are most concisely presented in terms of 
the Lie algebra structure:
\begin{eqnarray}
 \mathfrak{g} & \;\;\equiv\;\; & \underbrace{\C \;\oplus\; \Big( \underbrace{\underbrace{\underbrace{\h_{3}(\C) \opluslhrim \big(\sla(2,\C)}_{\text{Lie algebra of }S^{{}^{\times}}\!\!\G_{s}}{\oplus}\,\ua(1)}_{\text{Lie algebra of }S\G_{s}}{\oplus}\,\da(1)\big)}_{\text{Lie algebra of }\G_{s}} \Big)}_{\text{Lie algebra of }\G}. \quad
\end{eqnarray}
(The subgroup ${S\G_{s}\subset\G_{s}}$ is defined by acting trivially on $M^{4}(A)$, 
whereas ${S^{{}^{\times}}\!\!\G_{s}\subset S\G_{s}}$ is defined by acting trivially also on $M^{2}(\A)$.) 
Our strategy will be to first find invariants of the representations of the 
special subgroups ${S^{{}^{\times}}\!\!\G_{s}}$ on $\A$ and of ${S\G_{s}}$ on 
$A$. Then, we will study the action of the dilatation $\D(1)$ group and 
the projective scaling group $\C^{{}^{\times}}$ on the ensemble of the found 
invariants, in order to construct invariants of the full group $\G$.

Using the \texttt{LieAlgebras} Maple package \cite{Anderson2016}, one 
can search for invariant functions of the pertinent special groups. 
For instance, one can show that there is a single functionally independent 
${\A\rightarrow\C}$ map, which is invariant to the group action 
of $S^{{}^{\times}}\!\!\G_{s}$, and is nothing but the scalar component function 
${b:\A\rightarrow\C,\psi\mapsto b\psi}$, where $b$ picks out the scalar component 
(bottom-form or zero-form) of an element of $\A$. In a two-spinor representation 
${\psi\equiv(\phi,\,\xi_{{}_{A}},\,\varepsilon_{{}_{BC}})}$ of an element ${\psi\in\A}$, one has 
that ${b\psi=\phi}$. Similarly, one can search for 
${\A\times\A\rightarrow\C}$ functions, invariant to $S^{{}^{\times}}\!\!\G_{s}$, and these 
turn out to be functional combinations of these three invariants:
\begin{eqnarray}
(\psi,\psi') & \mapsto & b\psi, \cr
(\psi,\psi') & \mapsto & b\psi', \cr
(\psi,\psi') & \mapsto & \lambda(\psi_,\psi') := 
(b\partial_{1}\psi)(b\partial_{2}\psi') - (b\partial_{2}\psi)(b\partial_{1}\psi') 
 - (b\psi) (b\partial_{2}\partial_{1}\psi') + (b\partial_{2}\partial_{1}\psi) (b\psi') \cr
 & & 
\end{eqnarray}
where ${\partial_{1},\partial_{2}}$ denote stepping down operators associated to 
some arbitrarily chosen generators ${a_{1},a_{2}\in\Lambda_{1}}$.\footnote{The 
$S^{{}^{\times}}\!\!\G_{s}$ invariance of the bilinear form ${\lambda:\,\A\times\A\rightarrow\C}$ 
can be easily understood via first verifying the identity
${\lambda(\psi,\psi')}={(b\psi)^{2}\;\big(b\partial_{2}\partial_{1}(\psi^{-1}\psi')\big)}$
for any element ${\psi'\in\A}$ and any invertible element ${\psi\in\A}$, where 
$(\cdot)^{-1}$ denotes the algebraic inverse in $\A$. 
It is clear that the linear form ${b:\,\A\rightarrow\C}$ is invariant, moreover, by 
construction of $S^{{}^{\times}}\!\!\G_{s}$, the map 
${\A\times\A\rightarrow\A,\,(\psi,\psi')\rightarrow\psi^{-1}\psi'}$ 
is invariant, thus the map $\lambda$ indeed has to be invariant when its first 
argument is restricted to the invertible elements. 
Then, one may drop the assumption of the invertibility of the first argument, 
because any non-invertible element of $\A$ may be written 
as difference of two invertible elements and because $\lambda$ is linear in 
its arguments, in particular, in its first argument.} 
In two-spinor representation by setting 
${\psi\equiv(\phi,\,\xi_{{}_{A}},\,\varepsilon_{{}_{BC}})}$ and 
${\psi'\equiv(\phi',\,\xi'_{{}_{A}},\,\varepsilon'_{{}_{BC}})}$ one has that 
\begin{eqnarray}
\lambda(\psi,\psi') = \tfrac{1}{2}\bepsilon^{{}_{AB}}\left(\xi{}_{{}_{A}}\xi'{}_{{}_{B}}-\xi'{}_{{}_{A}}\xi{}_{{}_{B}} + \phi\,\varepsilon'{}_{{}_{AB}} - \phi'\,\varepsilon{}_{{}_{AB}}\right),
\end{eqnarray}
where ${\bepsilon_{{}_{AB}}\in\mathop{\wedge}^{2}S^{*}\equiv M^{2}(\A)}$ 
is an arbitrary but fixed nonzero maximal form in $\A$, and 
$\bepsilon^{{}_{AB}}$ is its corresponding inverse maximal form satisfying 
${\bepsilon_{{}_{AB}}\,\bepsilon^{{}_{CB}}=\delta_{{}_{A}}{}^{{}_{C}}}$. 
It is seen that $\lambda$ is a nondegenerate symplectic form, and that its 
choice is unique up to a complex multiplier, i.e.\ up to the choice of 
$\bepsilon_{{}_{AB}}$. One could say that the symplectic form $\lambda$ 
is a generalization of the symplectic form $\bepsilon^{{}_{AB}}$ from two-spinors 
to their exterior algebra. 
It is seen that $\lambda$ is uniquely determined up to 
complex normalization, where the ambiguity comes from the choice of the nonzero 
maximal form ${\bepsilon_{{}_{AB}}\in M^{2}(\A)}$. In order to fix this 
normalization ambiguity in the formalism, one could consider instead the 
``densitized version'' of $\lambda$. That can be defined to be the unique 
$\G_{s}$-invariant symplectic form 
${\blambda:\,\A\times\A\rightarrow M^{2}(\A)}$ satisfying the 
natural normalization condition ${\blambda(\1,\bepsilon)=\bepsilon}$ for all 
maximal forms ${\bepsilon\in M^{2}(\A)}$.

Using again the \texttt{LieAlgebras} Maple package \cite{Anderson2016}, one 
can search for $S\G_{s}$-invariant functions of $A$. For instance, one can show 
that there is a single functionally independent invariant ${A\rightarrow\C}$ 
function, namely ${\bar{b}{\otimes}b}$, picking out the scalar component 
(bottom-form or zero-form) of an element in $A$. In the following we shall use the 
abbreviation $b$ for ${\bar{b}{\otimes}b}$, since their distinction is not relevant. 
Similarly, one can search for 
${A\times A\rightarrow\C}$ functions, invariant to $S\G_{s}$, and these 
turn out to be functional combinations of these three invariants:
\begin{eqnarray}
(\Psi,\Psi') & \mapsto & b\Psi, \cr
(\Psi,\Psi') & \mapsto & b\Psi', \cr
(\Psi,\Psi') & \mapsto & L(\Psi_,\Psi') := \left(\bar{\lambda}{\otimes}\lambda\right)\circ\left(I_{\bar{\A}}{\otimes}J{\otimes}I_{\A}\right)(\Psi\otimes\Psi')
\label{eqLdef}
\end{eqnarray}
where $J$ denotes the ${\A\otimes\bar{\A}\rightarrow\bar{\A}\otimes\A}$ swapping map, whereas 
$I_{\bar{\A}}$ and $I_{\A}$ denote the identity map of $\bar{\A}$ and $\A$, respectively. 
If a preferred $\Zs$-grading is taken along with generators ${a_{1},a_{2}\in\Lambda_{\bar{0}1}}$, 
and corresponding stepping down operators ${\partial_{1},\partial_{2}}$, then 
the concrete expression
\begin{eqnarray}
L(\Psi,\Psi')&=&b\bar{\partial}_{2}\bar{\partial}_{1}\partial_{2}\partial_{1}\big((\Psi_{\bar{0}0}-\Psi_{\bar{1}0}-\Psi_{\bar{0}1}+\Psi_{\bar{1}1}
 -\Psi_{\bar{2}0}-\Psi_{\bar{0}2} +\Psi_{\bar{2}1}+\Psi_{\bar{1}2}+\Psi_{\bar{2}2})\,\Psi'\big)\Big. \qquad
\end{eqnarray}
holds for all ${\Psi,\Psi'\in A}$. By construction, $L$ is a nondegenerate 
symmetric bilinear form with alternating signature (${{+}1,{-}1,{+}1,{-}1,\dots}$). 
When expressed in terms of two-spinor representation 
${A\equiv\Lambda(\bar{S}^{*})\otimes\Lambda(S^{*})}$, then for two elements
\begin{equation*}
\begin{array}{l}
\Psi\equiv\Big(\phi,\,\bar{\xi}_{{}_{(+)A'}},\,\xi_{{}_{(-)A}},\,\bar{\varepsilon}_{{}_{(+)A'B'}},\,v{}_{{}_{A'B}},\,\varepsilon_{{}_{(-)AB}},
 \,\bar{\chi}_{{}_{(+)A'B'C}},\,\chi_{{}_{(-)C'AB}},\,\omega_{{}_{A'B'AB}}\Big) \cr
\end{array}
\end{equation*}
and
\begin{equation*}
\begin{array}{l}
\Psi'\equiv\Big(\phi',\,\bar{\xi}'_{{}_{(+)A'}},\,\xi'_{{}_{(-)A}},\,\bar{\varepsilon}'_{{}_{(+)A'B'}},\,v'{}_{{}_{A'B}},\,\varepsilon'_{{}_{(-)AB}},
 \,\bar{\chi}'_{{}_{(+)A'B'C}},\,\chi'_{{}_{(-)C'AB}},\,\omega'_{{}_{A'B'AB}}\Big) \cr
\end{array}
\end{equation*}
one has the identity
\begin{eqnarray}
L(\Psi,\Psi') = \tfrac{1}{4}\bomega^{{}_{A'B'CD}}\Big( \cr
 \Big. \phi\,\omega'_{{}_{A'B'CD}}
 +\omega_{{}_{A'B'CD}}\,\phi'
 +4v_{{}_{A'C}}\,v'_{{}_{B'D}}
 -\bar{\varepsilon}_{{}_{(+)A'B'}}\,\varepsilon'_{{}_{(-)CD}}
 -\varepsilon_{{}_{(-)CD}}\,\bar{\varepsilon}'_{{}_{(+)A'B'}}\cr
 -2\bar{\xi}_{{}_{(+)A'}}\,\chi'_{{}_{(-)B'CD}}
 -2\xi_{{}_{(-)C}}\,\bar{\chi}'_{{}_{(+)A'B'D}}
 +2\bar{\chi}_{{}_{(+)A'B'C}}\,\xi'_{{}_{(-)D}}
 +2\chi_{{}_{(-)A'CD}}\,\bar{\xi}'_{{}_{(+)B'}}
\Big),
\end{eqnarray}
where ${\bomega_{{}_{A'B'CD}}\in\mathop{\wedge}^{2}\bar{S}^{*}\otimes\mathop{\wedge}^{2}S^{*}\equiv M^{4}(A)}$ 
is an arbitrary but fixed nonzero positive maximal form of $A$, and 
$\bomega^{{}_{A'B'CD}}$ is its inverse maximal form 
with the normalization convention ${\bomega_{{}_{A'B'DE}}\;\bomega^{{}_{C'B'FE}}=\bar{\delta}_{{}_{A'}}{}^{{}_{C'}}\,\delta_{{}_{D}}{}^{{}_{F}}}$. 
The invariant bilinear form $L$ shall be shown to be a kind of generalization 
of the form related to the Dirac adjoint, and will be a key object in defining 
$\G$-invariant Lagrangians. 
It is seen that $L$ is uniquely determined up to 
complex normalization, where the ambiguity comes from the choice of the nonzero 
maximal form ${\bomega_{{}_{A'B'CD}}\in M^{4}(A)}$. In order to fix this 
normalization ambiguity in the formalism, one could consider instead the 
``densitized version'' of $L$. That can be defined to be the unique 
$\G_{s}$-invariant symmetric bilinear form 
${\bL:\,A\times A\rightarrow M^{4}(A)}$ with the 
natural normalization condition ${\bL(\1,\bomega)=\bomega}$ for all 
maximal forms ${\bomega\in M^{4}(A)}$.

Before we can go on to the formulation of $\G$-invariant theories, invocation 
of some further invariant objects is necessary, related to the two-spinor calculus. 
As it is well known \cite{PR1984,Wald1984}, in the ordinary two-spinor 
formalism the spinor space $S$ is considered as a representation space of 
$\GL(S)$, and by requiring the invariance of the duality pairing form and 
of the complex conjugation map, a canonical representation of $\GL(S)$ is 
defined also on $S^{*}$, $\bar{S}$, $\bar{S}^{*}$, respectively. Therefore, one 
has a canonical representation on the tensor product space 
${\bar{S}{\otimes}S}$, as well as on its real part ${\Real}\big(\bar{S}{\otimes}S\big)$. 
The formalism of two-spinor calculus is based on the fact that on the four 
dimensional real vector space ${\Real}\big(\bar{S}{\otimes}S\big)$ the canonical 
representation of $\GL(S)$ reduces to a representation of the Weyl group 
(dilatation + Lorentz group). More concretely, for any nonzero 
maximal form ${\bepsilon_{{}_{AB}}\in\mathop{\wedge}^{2}S^{*}}$ one has that 
the form ${\bomega_{{}_{A'B'AB}}:=\bar{\bepsilon}_{{}_{A'B'}}\otimes\bepsilon_{{}_{AB}}}$ 
defines a nondegenerate, symmetric, Lorentz signature 
\mbox{({+},{-},{-},{-})} real-bilinear form on 
${{\Real}\big(\bar{S}{\otimes}S\big)}$, which is preserved by the action of $\GL(S)$ 
up to positive multiplier. Therefore, if some other four dimensional real vector space 
$T$ is taken (which one may call ``tangent space''), and a linear injection 
${\sigma_{a}^{{}_{A'A}}:T\rightarrow{\Real}\big(\bar{S}{\otimes}S\big)}$ 
is fixed, then the $\GL(S)$-induced Weyl group representation is pulled back onto $T$, via 
requiring the $\sigma_{a}^{{}_{A'A}}$ to be invariant \cite{MR96276}. By construction, 
this representation of $\GL(S)$ respects the Lorentz metric 
${g(\sigma,\bomega)_{ab}:=\sigma_{a}^{{}_{A'A}}\,\sigma_{b}^{{}_{B'B}}\,\bomega_{{}_{A'B'AB}}}$ 
on $T$ up to positive multiplier. The map $\sigma_{a}^{{}_{A'A}}$ is called 
\emph{soldering form} or \emph{Pauli map} or \emph{Infeld--Van der Waerden symbol} 
in the literature. 
The pertinent philosophy naturally generalizes to the spin algebra case: 
the subspaces ${M(\A)\subset\A}$ and ${M^{2}(\A)\subset M(\A)}$ are invariant 
under the canonical representation of $\G_{s}$ on $\A$, and 
therefore one has the natural $\G_{s}$-invariant induced representation 
on the quotient space ${S^{*}\equiv M(\A)/M^{2}(\A)}$, and on its dual 
${S\equiv\big(M(\A)/M^{2}(\A)\big)^{*}}$. Because of that, one can take a real-linear injection 
${\sigma_{a}^{{}_{A'A}}:T\rightarrow{\Real}\big(\bar{S}\otimes S\big)}$ 
into the $\G_{s}$-invariant space ${\Real\big(\bar{S}\otimes S\big)}$. 
Clearly, fixing such a soldering form $\sigma_{a}^{{}_{A'A}}$ pulls back the 
natural real-linear representation of the group $\G_{s}$ onto $T$, 
via the requirement of the soldering form $\sigma_{a}^{{}_{A'A}}$ to be invariant. 
Similarly to the ordinary two-spinor case, this induced linear representation of 
$\G_{s}$ on $T$ is nothing but the Weyl group: 
the Lorentz group together with the metric rescalings.

It is sometimes useful to construct a further equivalent realization of the 
soldering form, in the spin algebra context. 
Using the \texttt{LieAlgebras} Maple package \cite{Anderson2016}, one 
can show that the subspace of elements of $\Lin(\A)$ which are invariant to the 
Heisenberg (nilpotent) group action of ${\exp L_{M(\A)}}$, is nothing but 
$R_{\A}$, i.e.\ the image of $\A$ in $\Lin(\A)$ by the right multiplication. 
Correspondingly, the Heisenberg-invariant elements in $\Lin(\bar{\A})$ span 
$R_{\bar{\A}}$. Therefore, one has the natural $\G_{s}$-invariant injections by right multiplication
\begin{eqnarray}
       S^{*}      \rightarrow R_{\A}       &\;,\quad& \xi_{{}_{B}}       \mapsto \xi_{{}_{B}}R_{\bdelta^{B}} \cr
 \Big. \bar{S}^{*}\rightarrow R_{\bar{\A}} &\;,\quad& \bar{\xi}_{{}_{B'}}\mapsto \bar{\xi}_{{}_{B'}}R_{\bar{\bdelta}^{B'}}
\label{eqR}
\end{eqnarray}
of the co-spinor spaces into the space right multiplication operators 
${R_{\A}\subset\Lin(\A)}$ and ${R_{\bar{\A}}\subset\Lin(\bar{\A})}$, respectively. We used Penrose indices on 
spinor side, and suppressed indices on the algebra side in order to introduce 
the right injection operators $R_{\bdelta^{B}}$ and $R_{\bar{\bdelta}^{B'}}$. 
Analoguously, using the \texttt{LieAlgebras} Maple package \cite{Anderson2016}, one can 
show that the subspace of elements of $\Lin(A)$ which are Heisenberg-invariant 
is $R_{A}$.
After verifying these facts, it follows that, up to a real multiplier, the only $\G_{s}$-invariant 
${T^{*}\rightarrow\Lin(A)}$ real-linear injective map is 
$\bsigma^{b}:=\sigma^{b}_{{}^{B'B}}\,R_{\bar{\bdelta}^{B'}}{\otimes}R_{\bdelta^{B}}$, 
where ${\sigma^{b}_{{}^{B'B}}:{\Real}(\bar{S}{\otimes}S)\rightarrow T}$ is the 
usual two-spinorial inverse soldering form, uniquely determined via the relation 
${\sigma^{b}_{{}^{B'B}}\,\sigma_{a}^{{}_{B'B}}=\delta^{b}{}_{a}}$. The normalization 
of $\bsigma^{b}$ and $\sigma_{a}^{{}_{A'A}}$ can be uniquely interlinked via fixing the natural 
normalization identity 
${\big((\bsigma^{b}\1)\big/ M^{3}(A)\big)_{{}^{B'B}}\,\sigma_{a}^{{}_{B'B}}=\delta^{b}{}_{a}}$.

Given a fixed soldering form $\sigma_{a}^{{}_{A'A}}$ and a fixed real maximal 
form ${\bomega\in\Real(M^{4}(A))}$, the previously introduced Lorentz metric 
${g(\sigma,\bomega)_{ab}}$ is a naturally defined $S\G_{s}$-invariant object. 
The normalization of the metric $g(\sigma,\bomega)_{ab}$ is, however, ambiguous up to the choice of $\bomega$. 
In order to fix this normalization ambiguity in the formalism, one could consider instead 
the ``densitized version'' of the metric. That can be defined to be the unique $\G_{s}$-invariant 
symmetric bilinear form ${\bg(\sigma)_{ab}:\,T\times T\rightarrow \Real(M^{4}(A))^{*}}$, 
satisfying ${(u^{a}\,v^{b}\,\bg(\sigma)_{ab}\,\vert\,\bomega)=u^{a}\,v^{b}\,g(\sigma,\bomega)_{ab}}$ for 
all ${\bomega\in\Real(M^{4}(A))}$ and ${u^{a},v^{a}\in T}$. 
The corresponding densitized inverse metric, being a symmetric bilinear 
form ${\bg(\sigma)^{ab}:\,T^{*}\times T^{*}\rightarrow \Real(M^{4}(A))}$, 
is uniquely determined by the relation ${\bg(\sigma)_{ab}\,\bg(\sigma)^{bc}=\delta^{c}{}_{a}}$. 
The densitized inverse metric can also be expressed in terms of the ordinary, real 
valued inverse metric via the identity ${\bg(\sigma)^{ab}=\bomega\,g(\sigma,\bomega)^{ab}}$, 
given any nonvanishing ${\bomega\in\Real(M^{4}(A))}$. 
Associated to the metric $g(\sigma,\bomega)_{ab}$, also a unique volume form 
in ${\mathop{\wedge}^{4}T^{*}}$ exists (up to orientation), 
and that is known to be expressable in the form
\begin{equation*}
\begin{array}{l}
\mathrm{v}(o,\sigma,\bomega)_{abcd} \;:=\; \cr
 \qquad \Big. o\;\big(\,\I\,\sigma_{a}^{{}^{{}_{E'E}}}\,\sigma_{b}^{{}^{{}_{F'F}}}\,\sigma_{c}^{{}^{{}_{B'A}}}\,\sigma_{d}^{{}^{{}_{A'B}}}\,\bomega_{{}_{E'A'EA}}\,\bomega_{{}_{F'B'FB}} 
 -\I\,\sigma_{a}^{{}^{{}_{E'E}}}\,\sigma_{b}^{{}^{{}_{F'F}}}\,\sigma_{d}^{{}^{{}_{B'A}}}\,\sigma_{c}^{{}^{{}_{A'B}}}\,\bomega_{{}_{E'A'EA}}\,\bomega_{{}_{F'B'FB}}\,\big) \cr
\end{array}
\end{equation*}
\cite{PR1984,Wald1984}, where ${o=\pm1}$ describes the chosen 
orientation sign. The normalization of the volume form also depends on the 
choice of an ${\bomega\in\Real(M^{4}(A))}$. In order to fix this normalization 
ambiguity, the corresponding densitized volume form is introduced, which is 
the unique element ${\bv(o,\sigma)\in\mathop{\wedge}^{4}T^{*}\,\otimes\,\Real(M^{4}(A))^{*}\otimes\Real(M^{4}(A))^{*}}$ 
satisfying ${(\bv(o,\sigma)\,\vert\,\bomega{\otimes}\,\bomega)=\mathrm{v}(o,\sigma,\bomega)}$ for 
all ${\bomega\in\Real(M^{4}(A))}$. By construction, the densitized volume form 
$\bv(o,\sigma)$ is also $\G_{s}$-invariant.

The \emph{spin tensor} is a further invariant function of $\sigma_{a}^{{}_{A'A}}$ 
according to the definition
\begin{equation*}
\Sigma(\sigma){}_{a}{}^{b}{}_{{}_{C}}{}^{{}_{D}} := \I\,\sigma_{a}^{{}_{A'D}}\,\sigma^{b}_{{}^{A'C}} - \I\,\bg(\sigma)^{cb}\,\bg(\sigma)_{ad}\,\sigma_{c}^{{}_{A'D}}\,\sigma^{d}_{{}^{A'C}}
\end{equation*}
which is a tensor of ${T^{*}{\otimes}T \otimes S^{*}{\otimes}}S$, using the identification 
${S^{*}\equiv M(\A)\big/M^{2}(\A)}$ as previously. The spin tensor 
$\Sigma(\sigma){}_{a}{}^{b}{}_{{}_{C}}{}^{{}_{D}}$, by construction, is also $\G_{s}$-invariant.

The introduced formalism is all as usual in the ordinary two-spinor calculus \cite{PR1984,Wald1984}, 
with the slight generalization of providing some extra representation space ${\A\equiv\Lambda(S^{*})}$ for the 
nilpotent Lie group component $\mathrm{H}_{3}(\C)$ of our symmetry group $\G$, 
where $\G$ as acting on $A$ can be considered as a generalization of $\GL(S^{*})$ as acting on $S^{*}$.

\section{The example Lagrangian}
\label{secLagrangian}

\begin{figure}[!h]
\begin{center}
\includegraphics[width=4cm]{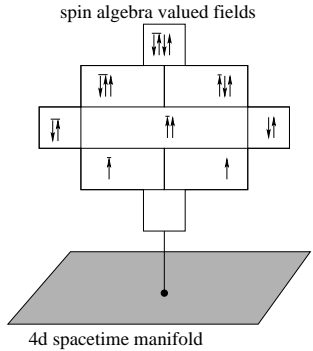}
\end{center}
\caption{Illustration of the concept of spin algebra valued fields. 
The structure group of such a theory can be set to be a conservative 
unification $\G$ of the Lorentz (or Weyl) and of the compact $\U(1)$ symmetries.}
\label{figAfield}
\end{figure}

In order to define our Lagrangian, we assume that our matter fields are sections 
of an $A$-valued vector bundle over a four dimensional spacetime, as illustrated in 
Figure~\ref{figAfield}. A distantly similar construction was considered by Anco 
and Wald \cite{Anco1989}, but the algebra they employed was too small 
in order to accommodate representation space for any symmetries larger than 
the conventional direct product symmetries, based merely on reductive Lie algebras.

In our construction, we consider a four dimensional real manifold $\M$ (which we shall call the spacetime manifold), 
and a vector bundle over $\M$ with fiber $A$ and structure group $\G$, as 
defined in Eq.(\ref{eqG}). As we shall see, such a structure exists whenever $\M$ is spin. The bundle $A(\M)$ is a spin algebra valued vector bundle of the form 
${A(\M)=\bar{\A}(\M){\otimes}\A(\M)}$ with $\A(\M)$ being a two generator complex 
Grassmann algebra bundle over $\M$. 
%The vector bundle $A(\M)$ can be considered to be associated to a principal bundle $\G(\M)$ with group $\G$.  
Analogously to 
ordinary two-spinor calculus \cite{PR1984,Wald1984}, we assume a $\sigma_{a}^{{}_{A'A}}$ pointwise injective 
${T(\M)\rightarrow\Real\big(\bar{S}(\M){\otimes}\,S(\M)\big)}$ vector bundle morphism 
(soldering form) to be present, where ${S^{*}(\M):=M(\A)(\M)\big/M^{2}(\A)(\M)}$ 
plays the role of an ordinary lower index two-spinor bundle.
We see from the above that given a spacetime $\M$ with co-spinor bundle $S^*(\M)$, and a choice of soldering form $\sigma_a^{{}_{A'A}}$, we can construct a spin algebra valued bundle $A(\M)$ over $\M$. The charge conjugation group ${\{I,\,\overline{(\cdot)}\}\equiv\Z_{2}}$, which has a canonical action on the sections of $A(\M)$, will be required to be a global symmetry of the model.

Let $\nabla_{a}$ be a covariant derivation operator on $A(\M)$. 
In the model, these will play the role of mediator fields. 
The adjoining by the discrete group of charge conjugation 
${\{I,\,\overline{(\cdot)}\}\equiv\Z_{2}}$ acts trivially on ${\G_{s}\subset\G}$, 
but acts nontrivially on the projective scaling subgroup ${\C^{{}^{\times}}\subset\G}$. 
Therefore, the charge conjugation map takes a 
covariant derivation $\nabla_{a}$ in general to a different 
covariant derivation $\overline{\nabla}_{a}$, according to the 
canonical action 
${\overline{\nabla}_{a}\Psi:=\overline{\nabla_{a}\overline{\Psi}}}$, for any 
section $\Psi$ of $A(\M)$. It is straightforward to check that the differential 
operator ${\nabla_{a}^{\R}:=\tfrac{1}{2}\big(\nabla_{a}+\overline{\nabla}_{a}\big)}$ 
also defines a covariant derivation. By construction, the charge 
conjugation $\overline{(\cdot)}$ acts trivially on $\nabla_{a}^{\R}$. 
Since on $\G_{s}$ the adjoining by charge conjugation acts trivially, and also on the 
subgroup $\vert\C^{{}^{\times}}\vert$ of the projective scaling group $\C^{{}^{\times}}$, 
one has the gauge potential ${\nabla_{a}{-}\nabla_{a}^{\R}}$ is 
simply a covector field carrying gauge charge of merely the quotient 
${\C^{{}^{\times}}/\vert\C^{{}^{\times}}\vert}$. Therefore, the mapping 
${\nabla_{a}\mapsto\nabla_{a}^{\R}}$ takes 
a covariant derivation to an other covariant derivation, with the gauge potential 
corresponding to the complex phase of the projective subgroup 
$\C^{{}^{\times}}$ zeroed. This slight complication with the distinction between 
$\nabla_{a}$ and $\nabla_{a}^{\R}$ in the formalism comes 
from the convention that we would like to handle projective representations of 
$\G_{s}$ via addressing linear representations 
of $\G$, and it does not have any particular physics meaning or relevance. 
The physically relevant part of $\nabla_{a}$ will turn out to be simply $\nabla_{a}^{\R}$ in the model.

Our action principle shall be Palatini-like, i.e.\ the metric will not be 
a distinguished field. In fact, it will be a function of an other fundamental 
field: the soldering form $\sigma_{a}^{{}_{A'A}}$. The matter field sector 
of the theory will consist of the soldering form $\sigma_{a}^{{}_{A'A}}$ and of a section 
$\Psi$ of the spin algebra bundle $A(\M)$. 
Moreover, as in the Palatini formalism, the covariant derivation $\nabla_{a}$ 
is independently varied from the matter field sector. That is, $\nabla_{a}$ physically 
describes a combined gravitational-and-gauge connection, without an \emph{a priori} splitting 
into a gravitational and an internal part. In total, the independent field 
variables form a tuple ${\big(\sigma_{a}^{{}_{A'A}},\,\Psi,\,\nabla_{b}\big)}$. 
The subgroup $\G_{s}$ of the structure group $\G$ has a canonical action on these, introduced in 
the previous sections. 
The charge conjugation group ${\{I,\,\overline{(\cdot)}\}\equiv\Z_{2}}$ as a global symmetry group also has a natural action via representing the charge conjugation map as 
${\big(\sigma_{a}^{{}_{A'A}},\,\Psi,\,\nabla_{b}\big)\,\mapsto\,\big({-}\sigma_{a}^{{}_{A'A}},\,\overline{\Psi},\,\overline{\nabla}_{b}\big)}$ 
on the fields, which can be understood as the action of a local $CPT$ transformation. 
(The field $\Psi$ is charge conjugated, and simultaneously, the sign of the 
soldering $\sigma_{a}^{{}_{A'A}}$ of the spin algebra to the spacetime vectors is reversed.) 
The projective scaling subgroup $\C^{{}^{\times}}$ of $\G$ is defined to act on 
the fields as 
${\big(\sigma_{a}^{{}_{A'A}},\,\Psi,\,\nabla_{b}\big)\,\mapsto\,\big(|z|\,\sigma_{a}^{{}_{A'A}},\,z\,\Psi,\,z\nabla_{b}z^{-1}\big)}$ 
for a projective scaling field $z$, being a section of the ${\C^{{}^{\times}}}$ valued line bundle. 
Thus at this point, the action of the structure group $\G$ and the global charge conjugation group ${\{I,\,\overline{(\cdot)}\}\equiv\Z_{2}}$ on the fields ${\big(\sigma_{a}^{{}_{A'A}},\,\Psi,\,\nabla_{b}\big)}$ is fully specified.

The actual Lagrangian shall be a real volume form valued pointwise vector bundle mapping
\begin{equation} \label{eq:20}
\begin{array}{l}
 (o,\,\sigma_{a}^{{}_{A'A}},\,\Psi,\,\nabla_{a}\Psi,\,P(\nabla)_{ab}) \;\longmapsto\; \mathrm{L}(o,\,\sigma_{a}^{{}_{A'A}},\,\Psi,\,\nabla_{a}\Psi,\,P(\nabla)_{ab}) \cr
\end{array}
\end{equation}
with the requirement of being invariant to the vector bundle automorphisms 
of $A(\M)$ compatible with the structure group $\G$. 
The symbol $P(\nabla)_{ab}$ denotes the curvature tensor of a covariant derivation $\nabla_{a}$, 
and $o$ denotes the spacetime orientation sign (${o=\pm1}$). 
The action functional is, as usually, defined to be local integrals of the 
pertinent volume form over compact regions of the spacetime $\M$. 
We also require the action functional of the theory to be 
invariant to the change of the spacetime orientation $o$, which implies that 
the Lagrangian should flip sign when changing the spacetime orientation 
$o$ to opposite. 
This explicitely forbids Chern--Simons-like terms in the model. 
In addition to these quite conventional gauge-theory-like symmetry prescriptions, 
we require the Lagrangian to be invariant to a shift 
transformation of the gauge-covariant derivation according to 
${\nabla_{a}\mapsto\nabla_{a}+C_{a}}$ in the manner of Section~\ref{secdirac}, 
where $C_{a}$ denotes a smooth covector field taking its values in the 
ideal of $\G$, corresponding to the 
${\C^{{}^{\times}}\times\big(\mathrm{H}_{3}(\C)\rtimes\D(1)\big)}$ part.\footnote{For 
the sectors carrying faithful representation of the structure group $\G$, such 
as the matter field sector, it is enough to demand a connection shift invariance 
with respect to the ${\C^{{}^{\times}}\times\D(1)}$ part of $\G$. Under that 
requirement, due to the local $\G$-invariance of the Lagrangian, connection 
shift invariance with respect to 
${\Ad_{\,\G}\big(\C^{{}^{\times}}\times\D(1)\big)}={\C^{{}^{\times}}\times\big(\mathrm{H}_{3}(\C)\rtimes\D(1)\big)}$ 
follows.} 
This requirement means that the Lagrangian should not depend on all the 
$\G$-connection fields, but only on modes with $\U(1)$ or $\SL(2,\C)$ charges. 
The search for all such invariant volume form valued expressions in principle can be addressed by the 
\texttt{LieAlgebras} Maple package \cite{Anderson2016}. However, due to the 
relatively large dimension of the total pointwise degrees of freedom, the 
pertinent library was not able to answer this question in its full generality. 
We were able to find, though, all the invariant terms, with certain fixed polynomial 
degree in $P(\nabla)_{ab}$ and in $\nabla_{a}\Psi$. There is strong 
evidence that these are all the invariants. The pertinent invariant terms 
are enumerated in the following, listed according to their 
polynomial degree in $P(\nabla)_{ab}$ and $\nabla_{a}\Psi$.

\textbf{Yang--Mills-like term.} The tensor field 
$\bv(o,\sigma)\,\bg(\sigma)^{ab}\,\bg(\sigma)^{cd}$ 
only depends on the orientation $o$ and the soldering form $\sigma$, and it is 
$\G$-invariant. Due to the structure of the group $\G$, the curvature 
$P(\nabla)_{ab}$ of a covariant derivation $\nabla_{a}$ does have 
a canonical action not only as a $\Lin(A)$-valued two-form, but also as a 
$\Lin(S^{*})$-valued two-form with the usual identification 
${S^{*}\equiv M(\A)/M^{2}(\A)\cong\Lambda_{\bar{0}1}}$. Therefore, its 
restricted trace ${\left.\Tr\right|_{\Lambda_{\bar{0}1}}P(\nabla)_{ab}}$ is 
meaningful, and is a $\G$-gauge covariant expression. 
With the introduced quantities, it does not come as a surprise 
that the only invariant Lagrangian bilinear in the curvature 
$P(\nabla)_{ab}$ and satisfying positive energy density condition for gauge fields is:
\begin{equation}
\begin{array}{l}
 \mathrm{L}_{{}_{\mathrm{YM}}}(o,\,\sigma_{a}^{{}_{A'A}},\,\Psi,\,\nabla_{a}\Psi,\,P(\nabla)_{ab}) \;:=\; \cr
 \qquad \Big. \bv(o,\sigma)\,\bg(\sigma)^{ac}\,\bg(\sigma)^{bd}
 \,\Imag\big(\left.\Tr\right|_{\Lambda_{\bar{0}1}}P(\nabla^{\R})_{ab}\big)\,\Imag\big(\left.\Tr\right|_{\Lambda_{\bar{0}1}}P(\nabla^{\R})_{cd}\big).
\end{array}
\end{equation}
This is nothing but literally the Maxwell Lagrangian, as expressed in our field variables. 
It is remarkable that only the $\U(1)$ part of the connection gives contribution, 
while the expression being $\G$-covariant.

\textbf{Einstein--Hilbert-like term.} The tensor field 
$\bv(o,\sigma)\,\bg(\sigma)^{ab}\,\bL(\overline{\Psi},\Psi)$ is $\G$-invariant. 
Thus, it is not surprising that the only invariant Lagrangian linear in the curvature 
$P(\nabla)_{ab}$ is:
\begin{equation}
\begin{array}{l}
 \mathrm{L}_{{}_{\mathrm{EH}}}(o,\,\sigma_{a}^{{}_{A'A}},\,\Psi,\,\nabla_{a}\Psi,\,P(\nabla)_{ab}) \;:=\; \cr
 \qquad \Big. \bv(o,\sigma)\,\bg(\sigma)^{ab}\,\bL(\overline{\Psi},\Psi) 
 \,\Real\big(\left.\Tr\right|_{\Lambda_{\bar{0}1}}\big(\I\Sigma(\sigma){}_{a}{}^{c}\,P(\nabla^{\R})_{cb}\big)\big).
\end{array}
\label{eqEH}
\end{equation}
This is nothing but a rather straightforward generalization of the Einstein--Hilbert Lagrangian, 
as expressed in spinorial variables. The only difference is that the prefactor of 
the scalar curvature is the field $\bL(\overline{\Psi},\Psi)$ instead 
of the constant\footnote{See e.g.\ \cite{1988NuPhB.302..645W} for a discussion of the impact of a dynamical Newton's constant in cosmology. Such mechanism is invoked in Brans-Dicke-type theories.} $\text{(Planck length)}^{{}_{-2}}$. It is remarkable that only 
the $\SL(2,\C)$ part of the connection gives contribution while the full expression 
being $\G$-covariant. An interesting feature of this Lagrangian term is that 
it is invariant to the shift of the top-form component of $\Psi$, according 
to the transformation ${\Psi\mapsto\Psi+b(\Psi)\,\I\,\bomega}$ with any 
$\Real\big(M^{4}(A)\big)$ valued field $\bomega$.

\textbf{Klein--Gordon-like term} is not allowed. The field 
$\bv(o,\sigma)\,\bg(\sigma)^{ab}\,\bL\big(\overline{(\cdot)},\cdot\big)$ is 
$\G$-invariant, and therefore the expression
\begin{eqnarray}
 \mathrm{L}_{{}_{\mathrm{KG}}}(o,\,\sigma_{a}^{{}_{A'A}},\,\Psi,\,\nabla_{a}\Psi,\,P(\nabla)_{ab}) \;:=\;
 \bv(o,\sigma)\,\bg(\sigma)^{ab}\,\bL\big(\,\overline{\,\I\nabla_{a}^{\R}(\Psi)},\,\I\nabla_{b}^{\R}(\Psi)\big)
\end{eqnarray}
is $\G$-invariant. However, it is not invariant to the shift symmetry 
${\nabla_{a}\mapsto\nabla_{a}+C_{a}}$ with $C_{a}$ being smooth covector field 
taking its values in the ideal of $\G$, corresponding to the 
${\C^{{}^{\times}}\times\big(\mathrm{H}_{3}(\C)\rtimes\D(1)\big)}$. Thus, a Klein--Gordon-like 
second order term in $\nabla_{a}\Psi$ is disallowed by the shift 
symmetry requirement on the connection.

\begin{figure}[!h]
\begin{center}
\includegraphics[width=5cm]{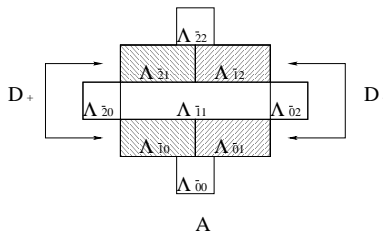}
\end{center}
\vspace*{-6mm}
\caption{Illustration of the fact that whenever a fixed $\Zs$-grading of the 
spin algebra $A$ is taken, then the $\pm1$ $\U(1)$ charge subspaces 
${D_{+}:=\Lambda_{\bar{1}0}{\oplus}\Lambda_{\bar{2}1}}$ and 
${D_{-}:=\Lambda_{\bar{0}1}{\oplus}\Lambda_{\bar{1}2}}$ can be considered as 
embedded Dirac bispinor spaces in $A$. Conversely: the spin algebra $A$ 
can be considered as a generalization of the Dirac bispinor / Clifford algebra 
concept.}
\label{figDirac}
\end{figure}

\textbf{Dirac-like term.} Here the calculations have to rely more intensively 
on the symbolic Maple calculation. It turns out that the $\G$-gauge-covariance, 
the diffeomorphism covariance, along with the CPT covariance singles out 
13 linearly independent Lagrangians, which are first order in $\nabla_{a}\Psi$. 
However, the requirement of connection shift invariance mentioned above 
singles out 1 unique invariant combination of these, resembling to a 
generalization of a Dirac term. It reads:
\begin{equation}
\begin{array}{l}
 \mathrm{L}_{{}_{\mathrm{D}}}(o,\,\sigma_{a}^{{}_{A'A}},\,\Psi,\,\nabla_{a}\Psi,\,P(\nabla)_{ab}) \;:=\; \cr
 \qquad \Bigg. \bv(o,\sigma)\,\tfrac{1}{\vert b(\Psi)\vert}\; 
  \tfrac{1}{\sqrt{2}}\Real\Big(\bL\big(\overline{\Psi},\;\bgamma(\sigma,\overline{\Psi},\Psi)^{a}\;\text{\footnotesize$b(\Psi)$}\,\I\nabla_{a}^{\R}(\frac{1}{b(\Psi)}\Psi)\big)\Big) \cr
\end{array}
\label{eqDiracLike}
\end{equation}
where one defines the map $\bgamma(\sigma,\overline{\Psi},\Psi')^{a}$ 
as a ${T^{*}\rightarrow\Lin(A)\otimes{\Real\big(M^{4}(A)\big)^{*}}}$ 
pointwise linear vector bundle mapping according to the formula
\begin{equation}
\begin{array}{l}
 \bgamma(\sigma,\overline{\Psi},\Psi')^{a}(\cdot) \;:=\; 
  \tfrac{1}{\sqrt{2}}\; \sigma^{a}_{{}^{A'A}}\; \Big(\, (R_{\bdelta^{A}}\,\overline{\Psi}\,)\;\bL\big(R_{\bar{\bdelta}^{A'}}\,\Psi',\,\cdot\big) 
    \;+\; (R_{\bar{\bdelta}^{A'}}\,\overline{\Psi}\,)\;\bL\big(R_{\bdelta^{A}}\,\Psi',\,\cdot\big) \,\Big). \cr
\end{array}
\end{equation}
Here, the notation $R_{\bdelta^{A}}$ and $R_{\bar{\bdelta}^{A'}}$ 
denote the pointwise injections ${S^{*}\rightarrow R_{\A}}$ and ${\bar{S}^{*}\rightarrow R_{\bar{\A}}}$, 
defined previously in Eq.(\ref{eqR}). 
This Lagrangian is a kind of generalization of the Dirac kinetic term in the 
following sense. Introduce a fixed $\Zs$-grading of $A$, and take 
the $\U(1)$ charged subspaces with charge $\pm1$ which are 
${D_{+}:=\Lambda_{\bar{1}0}{\oplus}\Lambda_{\bar{2}1}}$ and 
${D_{-}:=\Lambda_{\bar{0}1}{\oplus}\Lambda_{\bar{1}2}}$, respectively. Then, 
consider a background field $\Psi_{0}$ which takes its value in the spin-free subspace, i.e. in the 
center $Z(A)$ of the spin algebra $A$. With these conditions, the tensor 
$\bgamma(\sigma,\overline{\Psi}_{0},\Psi_{0})^{a}$ can be seen 
to admit the Clifford property against $\bg(\sigma)^{ab}$, 
over the subspaces $D_{+}$ and $D_{-}$ of $A$. 
In this sense, $\bgamma(\sigma,\overline{\Psi},\Psi)^{a}$ 
can be considered as a kind of modified vertex function, in field theory speak. 
Also, one can show that the nondegenerate sesquilinear invariant form 
$\bL\big(\overline{(\cdot)},\cdot\big)$, when restricted 
to $D_{+}$ or $D_{-}$, corresponds to the one generated by the Dirac adjoint 
in ordinary Dirac bispinor formalism. This generalization scheme is illustrated in Figure~\ref{figDirac}. 
It is remarkable, that the Dirac-like Lagrangian term Eq.(\ref{eqDiracLike}) is only meaningful 
for invertible fields, i.e.\ for matter fields $\Psi$ which have 
$b(\Psi)\neq 0$ (non-vanishing scalar component). A further remarkable property 
of Eq.(\ref{eqDiracLike}) is that it does not depend on the top-form subspace, 
i.e.\ the expression is invariant to a shift ${\Psi\mapsto\Psi+\bomega}$ 
by any $M^{4}(A)$ valued field $\bomega$.

\textbf{Fourth order self-interaction potential.} Relying on the symbolic Maple calculation it 
turns out that there are 5 linearly independent self-interaction terms, merely dependent 
on $\Psi$ and $\sigma_{a}^{{}_{A'A}}$. These are all combinatorial variants 
of the $\G_{s}$-invariant form field 
${\bv(o,\sigma)\,\bL(\cdot,\cdot)\,\bL(\cdot,\cdot)}$. 
The number of 5 invariants can also be understood by taking the representation Eq.(\ref{eqLdef}) 
of the multilinear form ${\bL(\cdot,\cdot)\otimes \bL(\cdot,\cdot)}$, 
which reads as
\begin{eqnarray*}
 \bar{\blambda}(\cdot,\cdot)\otimes\blambda(\cdot,\cdot)\otimes\bar{\blambda}(\cdot,\cdot)\otimes\blambda(\cdot,\cdot),
\end{eqnarray*}
and by subsequent enumeration of its linearly independent combinatorial 
contractions with ${\overline{\Psi}\otimes\overline{\Psi}\otimes\Psi\otimes\Psi}$, 
understood as a tensor of
\begin{eqnarray*}
 A\otimes A\otimes A\otimes A \equiv \bar{\A}\otimes\A\otimes\bar{\A}\otimes\A\otimes\bar{\A}\otimes\A\otimes\bar{\A}\otimes\A.
\end{eqnarray*}
Apart from invariance requirements, there is a clear guideline to select 
physically relevant combinations from the 5 invariant potential terms:
the requirement of non-negativity of the potential. 
Two of the five invariants, based on 
${\bL\big(\overline{\Psi},\Psi\big)\,\bL\big(\overline{\Psi},\Psi\big)}$ and on 
${\bL\big(\overline{\Psi},\overline{\Psi}\big)\,\bL\big(\Psi,\Psi\big)}$, 
are easily seen to be positive semidefinite. It is not easy to judge whether 
the remaining three invariants can be cast to a positive semidefinite form: 
generally, it is known not to be a simple problem to automatically deduce if a quartic form 
is positive semidefinite, unless it obviously can be written as sums of 
squares. A possible guideline to select a preferred combination of the 5 invariant 
potentials could be that one requires the symmetries as all the other invariant 
Lagrangian terms do obey, in particular that Eq.(\ref{eqEH}) obeys. 
Namely, that the Lagrangian should be invariant to the shift of the top-form 
component according to ${\Psi\mapsto\Psi+b(\Psi)\,\I\,\bomega}$ with $\bomega$ being 
a $\Real\big(M^{4}(A)\big)$ valued field. 
That requirement can be shown to uniquely select the invariant based on 
${\bL\big(\overline{\Psi},\Psi\big)\,\bL\big(\overline{\Psi},\Psi\big)}$, namely:
\begin{equation}
\begin{array}{l}
 \mathrm{L}_{{}_{\mathrm{V}}}(o,\,\sigma_{a}^{{}_{A'A}},\,\Psi,\,\nabla_{a}\Psi,\,P(\nabla)_{ab}) \;:=\; 
  \bv(o,\sigma)\, \bL\big(\overline{\Psi},\Psi\big)\,\bL\big(\overline{\Psi},\Psi\big). \cr
\end{array}
\label{eqVterm}
\end{equation}
In Section~\ref{secNFreeParams}, we shall present a 
further symmetry argument, which also suggests that the above invariant 
Lagrangian is a preferred unique combination for a self-interaction potential 
term.

As mentioned before, due to the high dimensionality of the problem we were not able 
to formally prove that the above invariants exhaust the set of all linearly independent 
invariant Lagrangians, but there is strong evidence that these are all. In 
any case, the linear combination of the known invariants
\begin{equation}
\begin{array}{l}
 \mathrm{L}_{A_{\mathrm{YM}},\,A_{\mathrm{EH}},\,A_{\mathrm{D}},\,A_{\mathrm{V}}} \;:=\; 
   A_{\mathrm{YM}}\,\mathrm{L}_{{}_{\mathrm{YM}}} \;+\;
   A_{\mathrm{EH}}\,\mathrm{L}_{{}_{\mathrm{EH}}} \;+\;
   A_{\mathrm{D}}\,\mathrm{L}_{{}_{\mathrm{D}}} \;+\;
   A_{\mathrm{V}}\,\mathrm{L}_{{}_{\mathrm{V}}} \cr
\end{array}
\end{equation}
with real coupling constants 
${A_{\mathrm{YM}},A_{\mathrm{EH}},A_{\mathrm{D}},A_{\mathrm{V}}}$ provides also 
an invariant Lagrangian. The question naturally arises: 
to what degree the behavior of such a theory depends on these 
coupling constants? We address this question in the following.

\section{The number of truly independent couplings in the toy model}
\label{secNFreeParams}

In this section we show that at the classical level, 3 of the 4 independent 
coupling constants ${A_{\mathrm{YM}},A_{\mathrm{EH}},A_{\mathrm{D}},A_{\mathrm{V}}}$ 
can be eliminated by field redefinition transformations. 
Thus, there remains only one independent coupling constant, which can be attributed 
e.g.\ to the strength of the gravitational interaction in the model.

In order to address the question of how many of the coupling factors 
of the toy model are truly independent, one first needs to establish 
the notion of equivalence of two instances of the theory. An instance
\begin{eqnarray*}
\Big(\M',\,A'(\M'),\,\G'(\M'),\,(A'_{\mathrm{YM}},A'_{\mathrm{EH}},A'_{\mathrm{D}},A'_{\mathrm{V}}),\,\mathrm{L}'_{A'_{\mathrm{YM}},\,A'_{\mathrm{EH}},\,A'_{\mathrm{D}},\,A'_{\mathrm{V}}}\Big)
\end{eqnarray*}
of the theory is defined to be equivalent to an other instance
\begin{eqnarray*}
\Big(\M,\,A(\M),\,\G(\M),\,(A_{\mathrm{YM}},A_{\mathrm{EH}},A_{\mathrm{D}},A_{\mathrm{V}}),\,\mathrm{L}_{A_{\mathrm{YM}},\,A_{\mathrm{EH}},\,A_{\mathrm{D}},\,A_{\mathrm{V}}}\Big)
\end{eqnarray*}
if and only if there exists a principal bundle isomorphism ${\G'(\M')\rightarrow\G(\M)}$ 
with underlying vector bundle isomorphism ${A'(\M')\rightarrow A(\M)}$ and 
underlying diffeomorphism ${\M'\rightarrow\M}$, such that 
$\mathrm{L}_{A_{\mathrm{YM}},A_{\mathrm{EH}},A_{\mathrm{D}},A_{\mathrm{V}}}$
is pulled back to
$\mathrm{L}'_{A'_{\mathrm{YM}},A'_{\mathrm{EH}},A'_{\mathrm{D}},A'_{\mathrm{V}}}$, 
up to a nonzero real multiplier. The overall normalization can be disregarded 
for a classical field theory, since the Euler--Lagrange equations do not depend on the 
absolute normalization of the Lagrange form, and the relative hierarchy 
of the Noether charges is also independent of that. 
Assume that we have one instance of the theory with all the coupling constants 
${A_{\mathrm{YM}},A_{\mathrm{EH}},A_{\mathrm{D}},A_{\mathrm{V}}}$ 
being nonzero. Then, by means of the above definition, all such theories are equivalent 
to an instance with coupling factors ${1,\frac{A_{\mathrm{EH}}}{A_{\mathrm{YM}}},\frac{A_{\mathrm{D}}}{A_{\mathrm{YM}}},\frac{A_{\mathrm{V}}}{A_{\mathrm{YM}}}}$, 
i.e.\ when the Yang--Mills coupling factor is fixed to $1$, by convention. 
Thus, it is enough to study theories with coupling factors 
${1,A_{\mathrm{EH}},A_{\mathrm{D}},A_{\mathrm{V}}}$.

We now address the question whether some of the remaining couplings 
${1,A_{\mathrm{EH}},A_{\mathrm{D}},A_{\mathrm{V}}}$ can be eliminated by field 
redefinition transformations. 
By counting the homogeneity degree of the terms of 
${\mathrm{L}_{{}_{\mathrm{YM}}},\,\mathrm{L}_{{}_{\mathrm{EH}}},\,\mathrm{L}_{{}_{\mathrm{D}}},\,\mathrm{L}_{{}_{\mathrm{V}}}}$, 
we establish the fact that some further coupling factors can be 
eliminated, using a kind of ``classical renormalization'', 
while keeping the invariant observables intact. 
In order to make the argument more exact, we need to formally introduce 
the deformation theory of Lagrangians, and thus of action functionals.

Let 
${S:\R\times\mathcal{F}\rightarrow\mathcal{R},\,(a,f)\mapsto S(a,f)}$ be 
some continuously differentiable functional, with $\mathcal{F}$ being some 
topological affine space with a norm type topology, and $S$ 
taking its values in some topological vector space $\mathcal{R}$. 
Let us denote the underlying vector space of $\mathcal{F}$ by $\delta{\mathcal{F}}$. 
For instance, $S$ may be the action functional from the space of a deformation parameter ($\R$), and the 
field configuration space ($\mathcal{F}$) over some compact region of spacetime, 
mapping onto the real numbers (the space $\mathcal{R}$ is $\R$ in that case), and the underlying vector space 
$\delta{\mathcal{F}}$ of $\mathcal{F}$ is then the space of field variations with the uniquely 
and naturally defined $C^{k}$ norm topology. Over such spaces, the ordinary 
Fr\'echet differentiability, i.e.\ the usual notion of differentiability 
based on the ordo functions, is meaningful and uniquely defined. 
We call such a map ${(a,f)\mapsto S(a,f)}$ a \emph{deformation family of the 
functional} ${f\mapsto S(1,f)}$.

Recall that given ${a\in\R}$ and ${f\in\mathcal{F}}$, the 
partial Fr\'echet derivative in the second variable 
${D_{2}S(\cdot,\cdot)\big\vert_{(a,f)}}$ is a continuous linear map 
${\delta{\mathcal{F}}\rightarrow\mathcal{R}}$. 
Given a closed subspace $\delta^{\circ}\!\mathcal{F}$ of $\delta{\mathcal{F}}$, 
one may consider the restriction of the above linear map to that subspace, 
denoted by ${D_{2}^{\delta^{\circ}\!\mathcal{F}}S(\cdot,\cdot)\big\vert_{(a,f)}}$. 
For instance, $\delta^{\circ}\!\mathcal{F}$ may be the subspace of field 
variations $\delta{\mathcal{F}}$ which vanish on the boundary of the compact region over which the 
action functional is considered. With that example, the equation 
${D_{2}^{\delta^{\circ}\!\mathcal{F}}S(\cdot,\cdot)\big\vert_{(1,f)}=0}$ would 
be equivalent to the Euler--Lagrange equation of the action $S(1,\cdot)$, 
at a field configuration $f\in\mathcal{F}$ (variation with fixed boundary values). 
Let then ${F:\R\times\mathcal{F}\rightarrow\mathcal{F},(a,f)\mapsto F(a,f)}$ be a continuously 
differentiable map such that for all parameters ${a\in\R}$ the mapping 
${f\mapsto F(a,f)}$ is one-to-one and onto, and that ${F(1,\cdot)}$ is the identity of $\mathcal{F}$. 
We call ${(a,f)\mapsto F(a,f)}$ a \emph{deformation family of the space} $\mathcal{F}$. 
We then say that a deformation family 
${(a,f)\mapsto S(a,f)}$ of a functional and a deformation family 
${(a,f)\mapsto F(a,f)}$ of its configuration space $\mathcal{F}$ are \emph{compatible} if for all parameters 
${a\in\R}$ one has that ${S(a,F(a,\cdot))=S(1,\cdot)}$. 
This, for the case of an action functional, would mean that the deformation 
of the action is compensated by a counter-deformation of the field configuration space.

Assuming that $S$ and $F$ are compatible, one has that 
${D\left(S(a,F(a,\cdot))\right)\big\vert_{f}=D_{2}\left(S(\cdot,\cdot)\right)\big\vert_{(1,f)}}$ 
for all ${a\in\R}$ and ${f\in\mathcal{F}}$. The left hand side of that equation can 
be reformulated via the chain rule of differentiation, thus one infers 
${D_{2}(S(\cdot,\cdot))\big\vert_{(a,F(a,f))}\;D_{2}(F(\cdot,\cdot))\vert_{(a,f)}}={D_{2}\left(S(\cdot,\cdot)\right)\big\vert_{(1,f)}}$. 
We call a deformation family ${(a,f)\mapsto F(a,f)}$ of the space $\mathcal{F}$ 
to be \emph{regular}, whenever for all parameters ${a\in\R}$ and configurations 
${f\in\mathcal{F}}$ the ${\delta{\mathcal{F}}\rightarrow\delta{\mathcal{F}}}$ 
linear map ${D_{2}(F(\cdot,\cdot))\big\vert_{(a,f)}}$ is onto. Moreover, 
we call it \emph{regular over a closed subspace} $\delta^{\circ}\!\mathcal{F}$ 
of $\delta{\mathcal{F}}$, whenever for all parameters ${a\in\R}$ and configurations 
${f\in\mathcal{F}}$ the ${\delta{\mathcal{F}}\rightarrow\delta{\mathcal{F}}}$ 
linear map ${D_{2}(F(\cdot,\cdot))\big\vert_{(a,f)}}$ can be restricted 
as a ${\delta^{\circ}\!\mathcal{F}\rightarrow\delta^{\circ}\!\mathcal{F}}$ linear 
map which is onto. 
If $S$ and $F$ are compatible, and $F$ is regular over 
$\delta^{\circ}\!\mathcal{F}$, then from the above chain rule argument one infers that
\begin{eqnarray}
{D_{2}^{\delta^{\circ}\!\mathcal{F}}\!\left(S(\cdot,\cdot)\right)\big\vert_{(1,f)}=0} &\;\Longrightarrow\;& {D_{2}^{\delta^{\circ}\!\mathcal{F}}\!(S(\cdot,\cdot))\big\vert_{(a,F(a,f))}=0}.
\end{eqnarray}
Applying this identity to 
our specific case, $S$ being the action functional, this means that under such conditions, 
taking a field configuration ${f\in\mathcal{F}}$ which solves the Euler--Lagrange equation 
${D_{2}^{\delta^{\circ}\!\mathcal{F}}(S(\cdot,\cdot))\big\vert_{(1,f)}=0}$, then for all ${a\in\R}$ its 
deformed version ${F(a,f)\in\mathcal{F}}$ is also a solution of the deformed 
Euler--Lagrange equation ${D_{2}^{\delta^{\circ}\!\mathcal{F}}(S(\cdot,\cdot))\big\vert_{(a,F(a,f))}=0}$. 
One can make a rather evident observation that whenever the deformation 
family of the field configurations ${a\mapsto F(a,\cdot)}$ is spacetime 
pointwise, then it is regular over $\delta^{\circ}\!\mathcal{F}$ 
(and over the entire ${\delta\mathcal{F}}$) 
if and only if for all ${a\in\R}$ the spacetime pointwise derivative of $F(a,\cdot)$ 
against the field configurations is onto. 
That is an easily testable (finite dimensional) condition, which we will use.

With the above arguments we have shown that under appropriate conditions, 
one can generate a flow of corresponding solutions ${a\mapsto F(a,f)}$ for 
a flow ${a\mapsto S(a,\cdot)}$ of theories, from a single instance of 
the theory ${S(1,\cdot)}$ and its solution $f$. 
For different parameters ${a\in\R}$, however, the deformed solution ${F(a,f)}$ 
of the deformed theory ${S(a,\cdot)}$ might eventually describe physically 
different configurations. 
For instance, from the above first principles, there is no guarantee that 
the physically relevant invariants, such as some relevant Noether charges 
of the solutions, are the same throughout the deformation family. 
In the following we investigate that under what additional conditions the Noether 
charges are constant throughout the deformation flow ${a\mapsto F(a,f)}$ 
of a solution.

Consider a spacetime pointwise deformation family of a Lagrangian 
${a\mapsto\mathrm{L}(a,\cdot)}$ together with a spacetime pointwise regular and 
compatible counter-deformation ${a\mapsto F(a,\cdot)}$ of the field configuration space. 
We assume a Palatini-like variational principle, i.e.\ for 
a fixed ${a\in\R}$, the field configuration $f$ is a pair ${(v,\nabla)}$, $v$ being 
section of a vector bundle $V(\M)$ (matter fields), and $\nabla$ being covariant 
derivative on $V(\M)$ (combined gauge and gravitational connection). 
The Lagrangian is assumed to depend on three field 
variables, the matter fields ($v$), the matter field covariant derivatives 
($\nabla_{b}v$), and the curvature ($P(\nabla)_{cd}$), and hence it is a mapping 
${\big(a,\,(v,\,\nabla_{b}v,\,P(\nabla)_{cd})\big)}\mapsto{\mathrm{L}\big(a,\,(v,\,\nabla_{b}v,\,P(\nabla)_{cd})\big)}$. 
We shall check under what conditions the Noether current densities along the 
deformation family ${a\mapsto\mathrm{L}(a,\cdot)}$ are constant 
with respect to the deformation parameter ${a\in\R}$. 
As a shorthand notation, we will use ${\big(\,{}^{{}^{(a)}}\!\!v,{}^{{}^{(a)}}\!\nabla\,\big)}$ for the 
image of a field configuration ${(v,\nabla)}$ by the field deformation map 
${F(a,\cdot)}$. Let $\mathsterling$ be a first order 
differential operator over the sections of $V(\M)$ which generates a 
local vector bundle automorphism over $V(\M)$. Let 
$u^{b}$ be the tangent vector field of the base manifold $\M$, subordinate 
to $\mathsterling$, describing its corresponding flow on the base manifold. Assume 
that for all ${a\in\R}$ the symmetry generator $\mathsterling$ leaves the Lagrangian 
${\mathrm{L}(a,\cdot)}$ intact, i.e.\ that the Lagrangian ${\mathrm{L}(a,\cdot)}$ is 
$\mathsterling$-covariant. Assuming that ${(v,\nabla)}$ is an Euler--Lagrange 
solution of ${\mathrm{L}(1,\cdot)}$, then because of the above conditions, 
${\big(\,{}^{{}^{(a)}}\!\!v,{}^{{}^{(a)}}\!\nabla\,\big)}$ shall also be an 
Euler--Lagrange solution of ${\mathrm{L}(a,\cdot)}$ for all ${a\in\R}$. 
Moreover for any fixed ${a\in\R}$, the volume form valued vector field
\begin{equation}
\begin{array}{l}
 \mathrm{J}^{b}_{\mathsterling}\left(a,\,\big({}^{{}^{(a)}}\!\!v,{}^{{}^{(a)}}\!\nabla({}^{{}^{(a)}}\!\!v),P({}^{{}^{(a)}}\!\nabla)\big)\right) := \cr
 \Bigg.\qquad D_{2,2}^{b}\mathrm{L}\left(a,\,\big({}^{{}^{(a)}}\!\!v,{}^{{}^{(a)}}\!\nabla({}^{{}^{(a)}}\!\!v),P({}^{{}^{(a)}}\!\nabla)\big)\right)\,\mathsterling\big({}^{{}^{(a)}}\!\!v\big) \cr
 \Bigg.\qquad + 2\,D_{2,3}^{[bc]}\mathrm{L}\left(a,\,\big({}^{{}^{(a)}}\!\!v,{}^{{}^{(a)}}\!\nabla({}^{{}^{(a)}}\!\!v),P({}^{{}^{(a)}}\!\nabla)\big)\right)\,\left[\mathsterling,{}^{{}^{(a)}}\!\nabla_{c}\right] \cr
 \qquad       \,- \,\mathrm{L}\left(a,\,\big({}^{{}^{(a)}}\!\!v,{}^{{}^{(a)}}\!\nabla({}^{{}^{(a)}}\!\!v),P({}^{{}^{(a)}}\!\nabla)\big)\right)\,u^{b} \cr
\end{array}
\label{eqNoether}
\end{equation}
will be the corresponding $\mathsterling$-Noether current density, which is divergence free. 
(A vector density, i.e.\ a volume form valued vector field has a naturally 
defined divergence operator.) The symbols ${D_{2,1}\mathrm{L}}$, 
${D_{2,2}\mathrm{L}}$, ${D_{2,3}\mathrm{L}}$ denote the spacetime pointwise 
partial derivative of $\mathrm{L}$ against its 2,1-th, 2,2-th and 2,3-th 
variable, respectively, i.e.\ against the matter fields, against the 
matter field covariant derivatives, and against the curvature. (In 
the above notation, the 1-st variable is reserved for the deformation 
parameter ${a\in\R}$ itself.) From the formula Eq.(\ref{eqNoether}) of the 
Noether current density, one can directly read off a rather evident 
sufficient condition for 
${a\mapsto\mathrm{J}^{b}_{\mathsterling}\left(a,\,\big({}^{{}^{(a)}}\!\!v,{}^{{}^{(a)}}\!\nabla({}^{{}^{(a)}}\!\!v),P({}^{{}^{(a)}}\!\nabla)\big)\right)}$ 
to be constant along the deformation parameter ${a\in\R}$. Assume that the following conditions hold:
\begin{enumerate}[(i)]
 \item \label{condi} ${\big(a,\,(v,\,\nabla{v},\,P(\nabla))\big)\mapsto\mathrm{L}\big(a,\,(v,\,\nabla{v},\,P(\nabla))\big)}$ 
is a spacetime pointwise deformation family of the Lagrangian.
 \item ${\big(a,\,(v,\,\nabla{v},\,P(\nabla))\big)\mapsto F\big(a,\,(v,\,\nabla{v},\,P(\nabla))\big)}$ 
is a spacetime pointwise deformation family of the field configurations which is regular.
 \item \label{condiii} the deformation family ${a\mapsto\mathrm{L}(a,\cdot)}$ and ${a\mapsto F(a,\cdot)}$ are compatible, i.e.\ for any field configuration ${(v,\nabla)}$ one has
${\mathrm{L}\left(a,\,\big({}^{{}^{(a)}}\!\!v,{}^{{}^{(a)}}\!\nabla({}^{{}^{(a)}}\!\!v),P({}^{{}^{(a)}}\!\nabla)\big)\right)=\text{\emph{const}}}$ along ${a\in\R}$.
 \item \label{condiv} for all deformation parameters $a\in\R$ the symmetry generator $\mathsterling$ leaves the Lagrangian ${\mathrm{L}(a,\cdot)}$ invariant, i.e.\ the Lagrangian ${\mathrm{L}(a,\cdot)}$ is $\mathsterling$-covariant.
 \item for all deformation parameters ${a\in\R}$ the symmetry generator $\mathsterling$ leaves the deformation mapping ${F(a,\cdot)}$ invariant, i.e.\ ${F(a,\cdot)}$ is $\mathsterling$-covariant.
 \item \label{condvi} one has the compatibility condition that for any field configurations ${(v,\nabla)}$ and ${(v',\nabla')}$
\begin{equation*}
\begin{array}{rl}
  D_{2,2}^{b}\mathrm{L}\left(a,\,\big({}^{{}^{(a)}}\!\!v,{}^{{}^{(a)}}\!\nabla({}^{{}^{(a)}}\!\!v),P({}^{{}^{(a)}}\!\nabla)\big)\right)\,\big({}^{{}^{(a)}}\!\!v'{-}{}^{{}^{(a)}}\!\!v\big) & \cr
+\,\Big. 2\,D_{2,3}^{[bc]}\mathrm{L}\left(a,\,\big({}^{{}^{(a)}}\!\!v,{}^{{}^{(a)}}\!\nabla({}^{{}^{(a)}}\!\!v),P({}^{{}^{(a)}}\!\nabla)\big)\right)\,\big({}^{{}^{(a)}}\!\nabla'_{c}{-}\!{}^{{}^{(a)}}\!\nabla_{c}\big) & 
\;=\; \text{\emph{const}} \cr
\end{array}
\end{equation*}
 holds along ${a\in\R}$.
\end{enumerate}
Then, for all the Euler--Lagrange solutions ${(v,\nabla)}$ of ${\mathrm{L}(1,\cdot)}$ 
the corresponding deformed field configuration ${\big(\,{}^{{}^{(a)}}\!\!v,{}^{{}^{(a)}}\!\nabla\,\big)}$ 
is an Euler--Lagrange solution of the deformed Lagrangian ${\mathrm{L}(a,\cdot)}$ 
for any ${a\in\R}$, moreover the Noether current density 
${\mathrm{J}^{b}_{\mathsterling}\left(a,\,\big({}^{{}^{(a)}}\!\!v,{}^{{}^{(a)}}\!\nabla({}^{{}^{(a)}}\!\!v),P({}^{{}^{(a)}}\!\nabla)\big)\right)}$ 
is constant along the deformation parameter ${a\in\R}$.

Using the above formalism for the deformation of a theory, one can generalize 
the notion of equivalence of instances of a theory with different coupling 
factors. We define instances of a theory within a deformation family 
to be \emph{equivalent in the generalized sense}, if there exists a regular 
compatible counter-deformation map of the field configurations, for which 
also the $\mathsterling$-Noether current density is constant along the 
deformation family, for all the vector bundle automorphism generators 
$\mathsterling$, respecting the structure group. 
The rationale behind this notion of equivalence is that one can 
generate the corresponding solutions of the deformed instances of the theory from 
each-other, moreover these corresponding solutions will have the same 
Noether charges for all the fundamental symmetry generators. Therefore, it 
seems to be rational to regard such a deformation family as describing the 
same physics throughout the flow of the deformation parameter. 
The sufficient conditions \mbox{(\ref{condi})--(\ref{condvi})} outlined above, 
are useful tools for recognizing such generalized equivalence of 
theory instances, and will be applied in the following to our toy model 
in order to eliminate some of the coupling factors.

Using the above notions, one can see that an instance of our 
toy model with nonvanishing coupling coefficients 
${1,\,A_{\mathrm{EH}},\,A_{\mathrm{D}},\,A_{\mathrm{V}}}$ is equivalent in 
the generalized sense to an instance with couplings 
${1,\,A_{\mathrm{EH}}/a^{2},\,A_{\mathrm{D}}/a^{3},\,A_{\mathrm{V}}/a^{4}}$. 
That is simply seen by observing the homogeneity degree of the invariant 
Lagrangians in the soldering form $\sigma_{b}^{{}_{B'B}}$, from which 
one infers that a deformation 
${a\mapsto\mathrm{L}_{A_{\mathrm{YM}},\,A_{\mathrm{EH}}/a^{2},\,A_{\mathrm{D}}/a^{3},\,A_{\mathrm{V}}/a^{4}}}$ 
of the Lagrangian may be compensated by a compatible counter-deformation 
${(\sigma_{b}^{{}_{B'B}},\,\Psi,\,\nabla_{c})\mapsto(a\,\sigma_{b}^{{}_{B'B}},\,\Psi,\,\nabla_{c})}$, 
which satisfies \mbox{(\ref{condi})--(\ref{condvi})} for all ${a\neq 0}$. 
Choosing specifically ${a=A_{\mathrm{D}}^{{}_{1/3}}}$, we arrive at the conclusion that such 
an instance of the theory is equivalent to the instance
${1,\,A_{\mathrm{EH}}/A_{\mathrm{D}}^{{}_{2/3}},\,1,\,A_{\mathrm{V}}/A_{\mathrm{D}}^{{}_{4/3}}}$. 
It is thus enough to study instances of the theory with couplings 
$1,\,A_{\mathrm{EH}},\,1,\,A_{\mathrm{V}}$.

Further applying the above deformation theory, one can eliminate another 
independent coupling constant from the set ${1,\,A_{\mathrm{EH}},\,1,\,A_{\mathrm{V}}}$. 
For that, one needs to use the fact that both the Yang--Mills-like term, 
the Dirac-like term and the Einstein--Hilbert-like term is invariant to 
an affine shift transformation ${\Psi\mapsto\Psi+b(\Psi)\,\I\,\bomega}$ with 
$\bomega$ being a ${\Real\big(M^{4}(A)\big)}$ valued field. 
In the previous section we suggested 
this symmetry requirement to be imposed also on the self-interaction 
term $\mathrm{L}_{{}_{\mathrm{V}}}$, in which case the only surviving potential 
term can be Eq.(\ref{eqVterm}). We now suggest a further symmetry argument 
to support that requirement. One may observe that the field deformation family 
\begin{equation}
\begin{array}{l}
 \Big(\sigma_{b}^{{}_{B'B}},\,\Psi,\,\nabla_{c}\Big) 
  \;\longmapsto\; \Big({}^{{}^{(a)}}\!\sigma_{b}^{{}_{B'B}},\,{}^{{}^{(a)}}\!\!\Psi,\,{}^{{}^{(a)}}\!\nabla_{c}\Big) := 
    \Big(\sigma_{b}^{{}_{B'B}},\, \big(\Psi+(a{-}1)\tfrac{1}{2\,b(\overline{\Psi})}\bL(\overline{\Psi},\Psi)\big), \,\nabla_{c}\Big)
\end{array}
\end{equation}
satisfies \mbox{(\ref{condi})--(\ref{condvi}}) for all ${a\neq 0}$, 
moreover it leaves the Yang--Mills-like term as well as the 
Dirac-like term invariant, whereas it acts as a scaling on the quadratic expression 
${\Psi\mapsto \bL\big(\overline{\Psi},\Psi\big)}$: one has the identity 
${\bL\big(\overline{{}^{{}^{(a)}}\!\!\Psi},{}^{{}^{(a)}}\!\!\Psi\big)=a\,\bL(\overline{\Psi},\Psi)}$. 
Therefore, it acts on the Einstein--Hilbert-like term as a scaling transformation 
by $a$. On the potential term Eq.(\ref{eqVterm}), which is proportional to 
$\bL(\overline{\Psi},\Psi)^{2}$, the pertinent field deformation acts 
as a scaling by $a^{2}$. With these conditions, the instance of the theory 
with couplings ${1,\,A_{\mathrm{EH}},\,1,\,A_{\mathrm{V}}}$ is equivalent 
in the generalized sense to an instance 
${1,\,A_{\mathrm{EH}}/a,\,1,\,A_{\mathrm{V}}/a^{2}}$, specifically to 
${1,\,1,\,1,\,A_{\mathrm{V}}/A_{\mathrm{EH}}^{2}}$. Thus, it is enough 
to study the theory with couplings 
${1,\,1,\,1,\,A_{\mathrm{V}}}$, i.e.\ with a single free coupling factor 
$A_{\mathrm{V}}$, describing the intensity of the self-interaction. Alternatively, 
one may transform this one free parameter to the pre-factor of the Einstein--Hilbert-like 
action, in which case it is enough to consider couplings 
${1,\,A_{\mathrm{EH}},\,1,\,\pm1}$. Since only a single coupling is left as a 
free parameter, such a toy model can be considered as unified.

The flat spacetime limit of the toy model can be deduced from the Lagrangian, 
when the instance of the theory with couplings ${1,\,A_{\mathrm{EH}},\,1,\,\pm1}$ 
is considered at the limit ${A_{\mathrm{EH}}\rightarrow\infty}$. 
(Here, the coupling $A_{\mathrm{EH}}$ plays the role of scaling ${(\text{Planck length})^{-2}}$, 
so in order to switch off the gravity, that has to go to infinity.) 
It is seen, that in the flat spacetime limit, the theory is left with no freely 
adjustable coupling constants.

\begin{remark}
Since the model turns out to have a single independent coupling, it is
quite natural to ask the question about the remaining degrees of freedom
of the matter field sector after a gauge fixing. Initially, a section $\Psi$ of the spin algebra bundle $A(\M)\equiv{\Lambda\big(\bar{S}^{*}(\M)\big)\otimes\Lambda\big(S^{*}(\M)\big)}$ can be represented by a tuple of spinor-tensor fields
\begin{equation*}
\begin{array}{l}
\Psi\equiv\Big(\phi,\,\bar{\xi}_{{}_{(+)A'}},\,\xi_{{}_{(-)A}},\,\bar{\varepsilon}_{{}_{(+)A'B'}},\,v{}_{{}_{A'B}},\,\varepsilon_{{}_{(-)AB}},\,\bar{\chi}_{{}_{(+)A'B'C}},\,\chi_{{}_{(-)C'AB}},\,\omega_{{}_{A'B'AB}}\Big). \cr
\end{array}
\end{equation*}
By a gauge transformation with the component $\C^{{}^{\times}}$ of $\G$, 
one can fix a gauge such that $b(\Psi)=1$, i.e.\ ${\phi=1}$ in the above representation. Then, by the $\mathrm{H}_{3}(\C)$ part of $\G$, one can choose a gauge that for instance ${\xi_{{}_{(-)A}}=0}$ and ${\varepsilon_{{}_{(-)AB}}=0}$ holds (``Dirac sea gauge'', i.e.\ only the net fermion content is present in the description). One can then use the affine symmetry ${\Psi\mapsto\Psi+b(\Psi)\,\I\,\bomega}$ with
$\bomega$ being a ${\Real\big(M^{4}(A)\big)}$ valued field. That can make sure that in the above gauge, the top form $\omega_{{}_{A'B'AB}}$ is hermitian. Finally, the $\D(1)$ component of $\G$ can be used to fix the scale of the top form, so that ${\omega_{{}_{A'B'AB}}=\bomega_{{}_{A'B'AB}}}$, where 
$\bomega_{{}_{A'B'AB}}$ is a fixed prescribed (non-dynamical) hermitian top form. (In a GR-like formalism, on would write ${\bomega_{{}_{A'B'AB}}=\pm\,\bar{\bepsilon}_{{}_{A'B'}}{\otimes}\,\bepsilon_{{}_{AB}}}$ with $\bepsilon_{{}_{AB}}$ fixed.) In the gauge field sector, due to the affine shift symmetry ${\nabla_{b}\mapsto\nabla_{b}+C_{b}}$, with $C_{b}$ being a ${\C^{{}^{\times}}\times\big(\mathrm{H}_{3}(\C)\rtimes\D(1)\big)}$ charged gauge potential, only the $\U(1)$ and $\SL(2,\C)$ sector of the connection gives contribution.
\end{remark}

\begin{remark}                                                 
The fact that the Lagrangian admits an internal symmetry implies the  existence of a corresponding Noether current, which is conserved on-shell, i.e.\ for fields satisfying the
Euler--Lagrange equations.
If in addition to the internal symmetry, the Lagrangian is invariant to shifts
of the connection $\nabla_{d}\mapsto\nabla_{d}+C_{d}$ with $C_{d}$ valued in an ideal $\mathfrak i$ of the internal symmetry Lie algebra,
then one can show that Noether currents corresponding to symmetries $\mathfrak i$
vanish on-shell. In our model, this implies the (on-shell) vanishing of the
${\C^{{}^{\times}}\times\big(\mathrm{H}_{3}(\C)\rtimes\D(1)\big)}$ Noether current.
For the case of the ordinary Dirac Lagrangian, discussed in Section~\ref{secdirac}, this implies
the (on-shell) vanishing of the Noether current associated to the dilatation charges. Despite the fact that nilpotent symmetries do not provide additional nonzero conserved charges, they do constrain the matter content and the set of allowed coupling constants.
\end{remark}

\section{Concluding remarks}
\label{secConclusion}

In this paper a toy model of a unified general relativistic gauge theory is 
constructed which exhibits a curious behavior: 
not all its local internal symmetry generators, 
which act locally and faithfully on the matter fields, 
are accompanied by corresponding gauge boson fields. 
As an introductory example it was shown that already the ordinary Dirac kinetic 
Lagrangian exhibits an extremely simplified version for such behavior: the 
gauge boson field corresponding to an internal dilatation symmetry does not 
give rise to any physically observable fields. 
In other words: the Lagrangian has a hidden affine symmetry, 
namely it is invariant with respect to an affine shift of the dilatation gauge connection. 
We showed that such behavior can also be exhibited by more complicated 
internal symmetry groups, and even by indecomposable (unified) ones. The necessary 
condition, however, is that these ``exotic'' symmetry generators, whose 
gauge boson fields can be transformed out, span an $\ad$-invariant sub-Lie algebra of the 
internal symmetries. Due to a general structural theorem of 
Lie algebras (Levi--Mal'cev decomposition), 
this implies that only theories having some nilpotent internal symmetry 
generators besides the usual compact ones can show such behavior. 
We have constructed a Lagrangian that exhibits these properties. 
The symmetries of the constructed theory, to linearized order, has the 
structure of a unified group, with compact, Poincar\'e and nilpotent components, 
the latter part acting as a ``glue'' in the unification.

Heuristically speaking, the constructed model describes the 
field equations of a classical field, which spacetime pointwise has degrees 
of freedom similar to a second quantized fermionic theory, i.e.\ with pointwise 
degrees of freedom obeying Pauli principle. 
As such, it may be a kind of semiclassical limit of a QFT-like model. In this QFT heuristic 
picture, besides the usual compact gauge, Lorentz and dilatation symmetries, 
the theory is symmetric to the transformation when equal amount of fermions 
and charge conjugate fermions are injected into a configuration spacetime pointwise, 
and this happens to be isomorphic to a pointwise $\mathrm{H}_{3}(\C)$ Heisenberg 
internal group action. It also turns out that the ``exotic'', 
$\mathrm{H}_{3}(\C)$ gauge fields can be completely transformed out from the 
theory due to the extra affine shift symmetry on the connection, which is 
a symmetry similar to what ordinary Dirac equation exhibits against the 
dilatation gauge fields. Thus, the nilpotent symmetries 
$\mathrm{H}_{3}(\C)$, necessary for the unification, do act locally and 
faithfully on the matter fields, without being accompanied by physical gauge 
boson fields.

\appendix
\addcontentsline{toc}{section}{Appendix}
\section*{Appendix}

\section{On the structure of generic Lie groups and Lie algebras}
\label{appliegroups}

As it is well known, the universal covering group of a connected Lie group 
is uniquely characterized by its Lie algebra, which can be studied by 
purely algebraic methods. Thus, for studying Lie groups it is important to 
first understand the structure of Lie algebras. 
In the following we shall recall some general known facts concerning 
the structure of finite dimensional real Lie algebras. Not all of these 
are well known in the folklore of gauge theory literature for model building, 
since in the traditional model building, only semisimple or reductive Lie 
algebras are considered.

\subsection{Ideal, semi-direct sum, direct sum}

A subspace $\mathfrak{i}$ of a Lie algebra $\mathfrak{e}$ is said to be an 
\emph{ideal} if $[x,y]$ belongs to $\mathfrak i$ for all $x\in\mathfrak{e}$ 
and $y\in\mathfrak{i}$. Notice that this condition is strictly more restrictive 
than the requirement of $\mathfrak{i}$ being a sub-Lie algebra of $\mathfrak{e}$. 
An example for an ideal is the translation generator sub-Lie algebra inside 
the Lie algebra of the Poincar\'e group, while the Lorentz generator sub-Lie 
algebra is not an ideal within, merely a sub-Lie algebra. 
The notion of ideal is arguably the most important concept in the theory of 
Lie algebras. The usual notation $\ad_{x}y:=[x,y]$ shall 
occasionally be applied ($x,y\in\mathfrak{e}$) in the following.

For every ideal $\mathfrak{i}\subset\mathfrak{e}$ one can always find a 
(non-unique) complementary linear subspace, i.e.\ a linear subspace 
$C\subset\mathfrak{e}$ such that $\mathfrak{i}\cap C=\{0\}$ and 
$\mathfrak{i}+C=\mathfrak{e}$. In the Lie algebra theory literature, such 
disjoint linear sum, being simply the vector space sense direct sum, is 
often denoted as $\mathfrak{e}=\mathfrak{i}\dotplus C$. 
Given an ideal $\mathfrak{i}$, in general there need not exist a complementary subspace which 
is also a sub-Lie algebra of $\mathfrak{e}$. Whenever such a complementary 
sub-Lie algebra $\mathfrak{c}$ does exist, we say that $\mathfrak{e}$ is a 
\emph{semi-direct sum} of $\mathfrak{i}$ with $\mathfrak{c}$, 
and denote it by $\mathfrak{e}=\mathfrak{i}\opluslhrim\mathfrak{c}$. 
For instance, the Poincar\'e Lie algebra is a semi-direct sum of the translation 
and of the Lorentz Lie algebra. If the complementing sub-Lie algebra 
$\mathfrak{c}$ is also an ideal, then elements of $\mathfrak{i}$ commute with 
elements of $\mathfrak{c}$, and $\mathfrak{e}$ is said to be a 
\emph{direct sum} of $\mathfrak{i}$ and $\mathfrak{c}$, denoted by 
$\mathfrak{e}=\mathfrak{i}\oplus\mathfrak{c}$. For instance, the 
Standard Model (SM) internal Lie algebra is a direct sum 
$\ua(1)\oplus\su(2)\oplus\su(3)$. On the other hand, for instance the 
Poincar\'e Lie algebra is a semi-direct sum, but not a direct sum of the 
translation and of the Lorentz Lie algebra. When a Lie algebra is not a direct 
sum of other smaller Lie algebras, it is called \emph{direct-indecomposable}, 
or simply \emph{indecomposable}, or in physics it is called \emph{unified}. 
The GUT strategy aims at finding a field theoretical description of particle 
physics with a unified internal symmetry group.

\subsection{A measure of non-commutativity: abelian, nilpotent, solvable, and semisimple Lie algebras}

It is natural to cathegorize Lie algebras according to the degree of their 
non-commutativity. Quite naturally, the least non-commutative Lie algebras 
$\mathfrak{e}$ are the \emph{abelian} ones, i.e.\ the ones satisfying 
$[\mathfrak{e},\mathfrak{e}]=\{0\}$, or equivalently, which satisfy 
$\ad_{x}=0$ for all $x\in\mathfrak{e}$. A next, slightly less commutative class of 
Lie algebras is the class of \emph{nilpotent} Lie algebras. Their defining 
property is that the so-called lower central series terminates in a finite
number of steps: with the definition $\mathfrak{e}^{1}:=\mathfrak{e}$, 
$\;\mathfrak{e}^{k+1}:=[\mathfrak{e},\mathfrak{e}^{k}]$, 
one has that $\mathfrak{e}^{k}=\{0\}$ for some finite non-negative 
integer $k$. It is known (Engel's theorem) \cite{Snobl2014,Onishchik1990,IseTakeuchi1991,Jacobson1962} 
that this condition is equivalent to the property that operator $\ad_{x}$
is nilpotent for every $x\in\mathfrak{e}$, hence the name. 
Such Lie algebras play a role in physics, for instance in SUSY.
An even less commutative class of Lie algebras is the class 
of \emph{solvable} Lie algebras, which satisfy the property that their 
so-called derived series vanish in finite steps: 
with the definition $\mathfrak{e}^{(0)}:=\mathfrak{e}$, 
$\;\mathfrak{e}^{(k+1)}:=[\mathfrak{e}^{(k)},\mathfrak{e}^{(k)}]$ 
one has that $\mathfrak{e}^{(k)}=\{0\}$ for some finite non-negative integer $k$.
The structure of solvable Lie algebras is slightly more complex than that 
of nilpotent ones. One could say, that the least commutative Lie algebras 
are the \emph{semisimple} ones, which are defined by the property that they 
do not contain solvable ideals other than the trivial $\{ 0 \}$.
Usually in gauge theory only semisimple Lie algebras, 
e.g.\ direct sums of $\su(N)$, are considered, along with abelian ones, 
which are always direct sums of some copies of the $\ua(1)$. Typically, general 
Lie algebras, possibly containing nilpotent or solvable component, are not 
used for field theory model building. In the present paper we address this 
more general possibility, and also discuss the rationale behind the traditional 
approach in gauge theory, while pointing out possible loopholes within.

\subsection{Structure of general Lie algebras: the Levi--Mal'cev decomposition theorem}

In every Lie algebra $\mathfrak{e}$ there exists a very distinguished ideal: 
the solvable ideal of the largest possible dimension, which is called 
the \emph{radical} of $\mathfrak{e}$ and is denoted by $\rad(\mathfrak{e})$. 
A further distinguished ideal is the largest dimensional nilpotent ideal, called the 
\emph{nilradical} of $\mathfrak{e}$ and is denoted by $\nil(\mathfrak{e})$. 
By construction, the radical and nilradical are unique, and one always has 
$\nil(\mathfrak{e})\subset\rad(\mathfrak{e})$. 
One of the foundational results about Lie algebras is the Levi--Mal'cev 
decomposition theorem \cite{Snobl2014,Onishchik1990,IseTakeuchi1991,Jacobson1962}, 
which states that the radical does admit 
a complementary sub-Lie algebra $\mathfrak{l}$, called \emph{Levi factor}. 
That is, one has the semi-direct sum splitting Eq.(\ref{eqLevi}), 
where the Levi factor $\mathfrak{l}$ is semisimple and isomorphic 
to the quotient Lie algebra $\frac{\mathfrak{e}}{\rad(\mathfrak{e})}$. 
As such, the Levi factors are isomorphic to each other, 
but they are not a uniquely determined embedded sub-Lie 
algebra in $\mathfrak{e}$. 
However, the choice of a Levi factor is unique 
up to an inner automorphism, defined by the conjugation by the exponential 
of $\ad_{z}$ for some element $z\in\nil(\mathfrak{e})$.
In this sense Levi factors are essentially unique.
Also, a side result of the Levi--Mal'cev 
theorem is that any semisimple sub-Lie algebra of $\mathfrak{e}$ must 
be contained within a Levi factor, i.e.\ a Levi factor is the maximal 
semisimple sub-Lie algebra with respect to the inclusion relation. 
An enlightening example of Levi--Mal'cev decomposition is provided by 
the Lie algebra of the Poincar\'e group Eq.(\ref{eqPoincare}), 
in which case, the radical coincides with the nilradical, and it is abelian. 
As outlined in \cite{Laszlo2017}, the Lie algebra of the super-Poincar\'e group can 
also be considered as an example to the Levi--Mal'cev decomposition, 
with a non-abelian, but two-step nilpotent radical.

Results above indicate that constructive characterization 
of the radical, nilradical and Levi factor is quite important. That 
can be done via Cartan's criterion 
\cite{Snobl2014,Onishchik1990,IseTakeuchi1991,Jacobson1962}, which employs the well known 
notion of Killing form. The \emph{Killing form} 
$K(x,y):=\Tr\left(\ad_{x}\ad_{y}\right)$ for $x,y\in\mathfrak{e}$ is
an invariant symmetric bilinear form, 
i.e.\ is a naturally given scalar product on $\mathfrak{e}$ 
(possibly of indefinite signature and possibly degenerate). 
The statement of Cartan's criterion can be formulated as:
\emph{(i)} the radical $\rad(\mathfrak{e})$ is 
the subspace within $\mathfrak{e}$ which is orthogonal to 
$[\mathfrak{e},\mathfrak{e}]$ with respect to the Killing form, moreover 
\emph{(ii)} Levi factor $\mathfrak{l}$ of $\mathfrak{e}$ is a maximal dimensional 
sub-Lie algebra on which the Killing form is nondegenerate.

Another important property of semisimple Lie algebras, and hence of the 
Levi factor of every Lie algebra, is the Weyl's theorem on complete 
reducibility \cite{Snobl2014,Onishchik1990,IseTakeuchi1991,Jacobson1962}. 
Its consequence is that every ideal of a semisimple Lie algebra has a 
complementing ideal, and therefore any semisimple Lie algebra is a direct 
sum of \emph{simple} Lie algebras: these are Lie algebras which do not 
have any ideals apart from the trivial ones, i.e.\ apart from the 
zero and the entire Lie algebra. 
Knowing the above properties, one can draw the following ``big picture'' 
of the structure of general Lie algebras:
\begin{eqnarray}
 \underbrace{\mathfrak{e}}_{\substack{\text{arbitrary}\\\text{Lie algebra}}} & \;=\; & \underbrace{\rad(\mathfrak{e})}_{\substack{\text{maximal}\\\text{solvable ideal,}\\\text{Killing form}\\\text{is degenerate}\\\text{(\emph{radical})}}} \opluslhrim\; \underbrace{\overbrace{\mathfrak{l}_{1}}^{\substack{\text{no ideals}\\\text{inside}\\\text{(\emph{simple})}}}\!\!{\oplus}{\dots}{\oplus}\!\!\overbrace{\mathfrak{l}_{n}}^{\substack{\text{no ideals}\\\text{inside}\\\text{(\emph{simple})}}}}_{\substack{\text{maximal}\\\text{semisimple sub-Lie algebra,}\\\text{Killing form}\\\text{is nondegenerate}\\\text{(\emph{Levi factor}})}}.
\label{eqLevi2}
\end{eqnarray}
The structure of simple Lie algebras is rather thoroughly explored: they are 
classified by the Dynkin diagrams. In physics literature by the 
standard theory of Lie algebras, mostly the theory of simple Lie 
algebras is meant. If nontrivial radicals are also allowed, the classification 
theory of simple Lie algebras is not enough, and one needs to look at the 
possible structure of solvable Lie algebras as well.

\subsection{Structure of radicals of Lie algebras}
\label{secrad}

The classification of all finite dimensional real or complex Lie algebras 
with nonvanishing radical is unresolved, moreover is known to be a 
``wild problem'' in mathematics. Complete classification exists only 
for low dimensional Lie algebras. There are however, some results on the 
generalities of the possible structure of such Lie algebras. 
For completeness, we recall some of these results, mostly from 
\cite{Snobl2010,Snobl2014}.

Let us consider a finite dimensional real Lie algebra with Levi--Mal'cev 
decomposition Eq.(\ref{eqLevi}). 
The identities $\nil(\mathfrak{e})\subset\rad(\mathfrak{e})$, 
$[\mathfrak{e},\rad(\mathfrak{e})]\subset\nil(\mathfrak{e})$, 
$[\mathfrak{l},\mathfrak{l}]=\mathfrak{l}$ are well known. 
If $\mathfrak{e}$ is indecomposable, i.e.\ not a direct sum of smaller 
Lie algebras, then the representation of $\mathfrak{l}$ by $\ad$ on 
$\rad(\mathfrak{e})$ is known to be faithful \cite{Snobl2014}. 
From now on, assume that $\mathfrak{e}$ is indecomposable. 
Then, one has the result by Turkowski, recalled in \cite{Snobl2010,Snobl2014}, 
that there exists a (non unique) subspace $q$ within $\rad(\mathfrak{e})$ 
complementing the ideal $\nil(\mathfrak{e})$, 
i.e.\ $\rad(\mathfrak{e})=\nil(\mathfrak{e})\dotplus q$, 
such that the action of $\mathfrak{l}$ by the $\ad$ on $q$ vanishes. 
The subspace $q$, however, may not always be a sub-Lie algebra, i.e.\ the 
preceding $\dotplus$ may not be a semi-direct sum $\opluslhrim$. 
Whenever the subspace $q$ is sub-Lie algebra, then it is necessarily 
abelian: $[q,q]=\{0\}$. 
The structure of the nilradical can be characterized by results of \v{S}nobl \cite{Snobl2010,Snobl2014}: 
there exists a (non unique) tuple of complementing subspaces 
$m_{1},\dots,m_{k}$ within $\nil(\mathfrak{e})$, such that 
$\nil(\mathfrak{e})=m_{k}\dotplus\dots\dotplus m_{1}$, with 
$\nil(\mathfrak{e})^{j}=m_{j}\dotplus\nil(\mathfrak{e})^{j+1}$, and 
$m_{j+1}\subset[m_{1},m_{j}]$, and 
$\ad_{\mathfrak{l}} m_{j}\subset m_{j}$ ($j=1,{\dots},k$), 
moreover $\mathfrak{l}$ acts by $\ad$ on $m_{1}$ faithfully. 
All this can be summarized in a ``big picture'' of the structure of 
indecomposable Lie algebras:
\begin{equation}
\begin{array}{c}
 \qquad\qquad\text{\tiny(arrows: nonvanishing, faithful, adjoint action)}\cr
 \cr
 \underbrace{\mathfrak{e}}_{\substack{\text{arbitrary}\\\text{indecomp.}\\\text{Lie algebra}}} =\; \underbrace{\!\!\underbrace{\overset{{}^{\tikzmark{mk}}}{m_{k}}{\dotplus}{\dots}{\dotplus}\overset{{}^{\tikzmark{m1}}}{m_{1}}}_{\substack{\nil(\mathfrak{e})\text{, the maximal}\\\text{nilpotent ideal,}\\\text{Killing form is zero}\\\text{(\emph{nilradical})}}}\!\!\!\!\dotplus\, \overset{{}^{\tikzmark{q}}}{q}}_{\substack{\rad(\mathfrak{e})\text{, the maximal}\\\text{solvable ideal,}\\\text{Killing form}\\\text{is degenerate}\\\text{(\emph{radical})}}} \;\;\opluslhrim\!\! \underbrace{\underbrace{\overset{\tikzmark{l1}}{\mathfrak{l}_{1}}}_{\substack{\text{no ideals}\\\text{inside}\\\text{(\emph{simple})}}}\!\!\!\!{\oplus}{\dots}{\oplus}\!\!\!\!\underbrace{\overset{\tikzmark{ln}}{\mathfrak{l}_{n}}}_{\substack{\text{no ideals}\\\text{inside}\\\text{(\emph{simple})}}}}_{\substack{\mathfrak{l}\text{, a maximal}\\\text{semisimple sub-Lie algebra,}\\\text{Killing form}\\\text{is nondegenerate}\\\text{(\emph{Levi factor}})}}. \cr
\end{array}
\tikz[remember picture,overlay] \draw[<-] (mk.north) |- +(0ex,2.5ex) -| (ln.north);
\tikz[remember picture,overlay] \draw[<-] (m1.north) |- +(0ex,2.5ex) -| (ln.north);
\tikz[remember picture,overlay] \draw[<-] (mk.north) |- +(0ex,2.5ex) -| (l1.north);
\tikz[remember picture,overlay] \draw[<-] (m1.north) |- +(0ex,2.5ex) -| (l1.north);
\tikz[remember picture,overlay] \draw[<-] (mk.north) |- +(0ex,2.5ex) -| (q.north);
\tikz[remember picture,overlay] \draw[<-] (m1.north) |- +(0ex,2.5ex) -| (q.north);
\label{eqRad}
\end{equation}

\begin{remark}
A further constraint on the structure of radical is a theorem of 
\v{S}nobl (2010) \cite{Snobl2010}: if $\mathfrak{e}$ is an indecomposable Lie 
algebra over $\C$, and its Levi factor $\mathfrak{l}$ acts irreducibly by $\ad$ 
on the top subspace $m_{1}$ of $\nil(\mathfrak{e})$, then the complementing 
subspace $q$ to the nilradical $\nil(\mathfrak{e})$ within the radical 
$\rad(\mathfrak{e})$ is $0$ or $1$ complex dimensional. 
In the latter case, one has that $q\cong\da(1){\oplus}\,\ua(1)$, i.e.\ $q$ 
closes as an (abelian) sub-Lie algebra. Also, it is seen that under such 
conditions, there can be maximum one copy of the $\ua(1)$ component within. 
(This might remind us about the structure of the Standard Model Lie 
algebra, which also has merely one copy of $\ua(1)$, and thus well may be 
the factor $\frac{\mathfrak{e}}{\nil(\mathfrak{e})}$ of some larger 
indecomposable Lie algebra $\mathfrak{e}$.)
\end{remark}

\begin{remark}
In the case when $\mathfrak{e}$ is the Lie algebra of a real linear 
algebraic group, there are some further constraints on 
the structure of $\rad(\mathfrak{e})$. Such constraints are implied by Mostow's 
decomposition theorem of linear algebraic groups \cite{borel1991}: 
a connected real linear algebraic group can be decomposed as a 
semi-direct product of an idempotent normal subgroup and of a so-called 
reductive subgroup.
\end{remark}

\subsection{Lie algebras in traditional model building: quadratic, reductive and compact Lie algebras}
\label{subsecquadratic}

As outlined, every Lie algebra has an $\ad$-invariant, but possibly 
indefinite and possibly degenerate scalar product: the Killing form. 
It is often of interest to consider Lie algebras with a 
\emph{non}degenerate (possibly indefinite) invariant scalar product. 
Such Lie algebras are called \emph{quadratic}. Quadratic Lie algebras 
play a natural role as internal Lie algebras in gauge theory, since the 
nondegeneracy of the invariant scalar product would ensure that all gauge 
fields would propagate. Not all possible quadratic Lie algebras are 
fully classified as of now.

An important class of quadratic Lie algebras are called \emph{reductive}. 
These can be defined by the following equivalent properties: 
\emph{(i)} its adjoint representation is completely reducible (direct sum of irreducible ones), 
\emph{(ii)} it admits a faithful finite dimensional completely reducible representation, 
\emph{(iii)} its radical coincides with its center, 
\emph{(iv)} it is a direct sum of an abelian ideal and of a semisimple Lie algebra. 
As such, a reductive Lie algebra $\mathfrak{e}$ has the structure:
$\mathfrak{e}=\ua(1){\oplus}{\dots}{\oplus}\,\ua(1){\oplus}\,\mathfrak{l}_{1}{\oplus}{\dots}{\oplus}\,\mathfrak{l}_{n}$, 
where the components $\mathfrak{l}_{1},\dots,\mathfrak{l}_{n}$ are simple. 
Clearly, a reductive Lie algebra is quadratic: the semisimple part 
$\mathfrak{l}_{1}{\oplus}{\dots}{\oplus}\,\mathfrak{l}_{n}$ has the nondegenerate 
Killing form, whereas $\ua(1)$ has its invariant scalar product by its 
identification with the imaginary numbers $\I\,\R$. 
It is instructive to note that for every Lie algebra $\mathfrak{e}$ the 
quotient by the nilradical $\frac{\mathfrak{e}}{\nil(\mathfrak{e})}$ is reductive \cite{Bourbaki1975}. 
Usually, in field theory 
model building, the most general Lie algebras appearing are the 
reductive ones. For example, the vector bundle of fermion fields in the 
Standard Model having electromagnetic, weak and strong charges will have 
the reductive Lie algebra $\ua(1){\oplus}\,\su(2){\oplus}\,\su(3)\oplus\sla(2,\C)$ 
as the Lie algebra of their structure group. In case of a generic Lie algebra 
$\mathfrak{e}$, one could say that $\nil(\mathfrak{e})$ is responsible for 
the deviation from reductivity, as seen from Eq.(\ref{eqRad}).

A quadratic Lie algebra, whose invariant scalar product is positive 
definite is called \emph{compact}. These are always isomorphic to the Lie 
algebra of some compact Lie group, and conversely, the Lie algebra of 
every compact Lie group is compact in this sense, hence the name. Compact Lie algebras 
are always reductive, therefore they admit decomposition of the form
$\mathfrak{e}=\ua(1){\oplus}{\dots}{\oplus}\,\ua(1){\oplus}\,\mathfrak{l}_{1}{\oplus}{\dots}{\oplus}\,\mathfrak{l}_{n}$, 
where now the components $\mathfrak{l}_{1},\dots,\mathfrak{l}_{n}$ are compact simple. 
The internal symmetries in a traditional gauge theory are encoded by 
compact Lie algebras. The rationale of this requirement is that the Yang--Mills 
kinetic energy density contains this internal scalar product, and that is 
required to be positive definite. Quite naturally, the Standard Model internal 
Lie algebra $\ua(1){\oplus}\,\su(2){\oplus}\,\su(3)$ is compact.

\subsection{Constraints on symmetry unification patterns by the Levi--Mal'cev decomposition}
\label{secconseq}

If one studies the possible enlargements of Lie groups, the Levi--Mal'cev 
theorem gives important constraints: the Lie algebra enlargement 
must respect the Levi--Mal'cev decomposition Eq.(\ref{eqLevi2}). 
In particular, their Lie algebras must obey the following rule: 
the embedded image of a Levi factor of the smaller Lie algebra, being semisimple, 
must sit in some Levi factor of the larger Lie algebra. 
In particular it has to intersect trivially with the radical
of the larger algebra.
Moreover, the embedded image every simple component of the Levi factor of the 
smaller Lie algebra has intersection with precisely one 
simple component of the Levi factor of the larger one. 
From this observation, 
O'Raifeartaigh developed a classification theorem \cite{LOR1965a,LOR1965b} of the 
finite dimensional real Lie algebra extensions of the Poincar\'e Lie algebra, 
as recalled in Section~\ref{secliegroups}.
The O'Raifeartaigh theorem can be illustrated as follows:
\begin{eqnarray}
\begin{array}{lcl}
\begin{array}{l}
\text{case (A) and (B):} \cr
 \cr
\begin{array}{cclcl}
 \quad{}_{\tikzmark{a}}\!\mathfrak{e} & = & \quad{}_{\tikzmark{r}}\!\!\rad(\mathfrak{e}) & {\opluslhrim} & \;\;{}_{\tikzmark{l1}}\!\mathfrak{l}_{1}{\oplus}{...}{\oplus}\mathfrak{l}_{n} \cr
 & & & & \cr
 \tikzmark{b}\!\mathfrak{p}           & = & \;\;\tikzmark{t}\!\mathfrak{t}               & {\opluslhrim} & \tikzmark{l}\!\boldell \cr
\end{array}
\end{array} & \qquad\qquad & 
\begin{array}{l}
\text{case (C):} \cr
 \cr
\begin{array}{cclcl}
 \quad{}_{\tikzmark{sa}}\!\mathfrak{e} & = & \quad\rad(\mathfrak{e})         & {\opluslhrim} & \;\;{}_{\tikzmark{sl1}}\!\mathfrak{l}_{1}{\oplus}{...}{\oplus}\mathfrak{l}_{n} \cr
 & & & & \cr
 \tikzmark{sb}\!\mathfrak{p}           & = & \;\;\mathfrak{t}\tikzmark{st} & {\opluslhrim} & \tikzmark{sl}\!\boldell
\end{array} \cr
\end{array} \cr
\end{array}
\tikz[remember picture,overlay] \draw[line width=0.5mm, ->, >=stealth] (b) to [out=130,in=220] (a);
\tikz[remember picture,overlay] \draw[line width=0.5mm, ->, >=stealth] (t) to [out=130,in=220] (r);
\tikz[remember picture,overlay] \draw[line width=0.5mm, ->, >=stealth] (l) to [out=130,in=220] (l1);
\tikz[remember picture,overlay] \draw[line width=0.5mm, ->, >=stealth] (sb) to [out=130,in=220] (sa);
\tikz[remember picture,overlay] \draw[line width=0.5mm, ->, >=stealth] (st) to [out=35,in=220] (sl1);
\tikz[remember picture,overlay] \draw[line width=0.5mm, ->, >=stealth] (sl) to [out=130,in=220] (sl1);
\label{eqLOR}
\end{eqnarray}

%The O'Raifeartaigh theorem is illustrated in Figure~\ref{figLOR}.

%\begin{figure}[!h]
%\caption{Illustration of the O'Raifeartaigh classification theorem of finite 
%dimensional Lie algebra extensions of the Poincar\'e Lie algebra. The are 
%three disjoint cases: 
%case (A) is the direct sum (trivial) extension, 
%case (B) is the non-direct sum extension 
%via extended radical, and 
%case (C) stands for embedding into a simple Lie algebra.}
%\label{figLOR}
%\end{figure}

\subsection{Levi--Mal'cev decomposition and the Lie algebra of the super-Poincar\'e group}

Although the SUSY algebra is usually presented as a super-Lie algebra, 
but via choosing appropriate variables, it can be cast into a real Lie algebra form, 
as recalled e.g.\ in \cite{Laszlo2017}. 
It is the Lie algebra of a concrete finite dimensional real Lie group, 
called to be the super-Poincar\'e group. 
The Lie algebra of the super-Poincar\'e group is of the form
\begin{equation}
\begin{array}{c}
 \text{\tiny(arrows: nonvanishing adjoint action)}\cr
 \cr
 \cr
\underbrace{\Big(\underbrace{\overset{\tikzmark{a}}{\mathfrak{t}}}_{\substack{\text{translation}\\\text{generators}}}\dotplus\underbrace{\overset{\tikzmark{c}}{t_{s}}}_{\substack{\text{supertransl.}\\\text{generators}}}\Big) \opluslhrim \underbrace{\overset{\tikzmark{e}\;\;\tikzmark{b}}{\boldell}}_{\substack{\text{Lorentz}\\\text{generators}}}}_{\text{Lie algebra of the super-Poincar\'e group}}
\tikz[remember picture,overlay] \draw[<-] (a.north) |- +(0ex,4.5ex) -| (b.north);
\tikz[remember picture,overlay] \draw[<-] (c.north) |- +(0ex,2.5ex) -| (e.north);
\end{array}
\label{eqsuper}
\end{equation}
It has a two-step nilradical, consisting of ${\mathfrak{t}\dotplus t_{s}}$, 
and its Levi factor is $\boldell$. The super-Poincar\'e Lie algebra has 
extended versions, being of the form
\begin{equation}
\begin{array}{c}
 \text{\tiny(arrows: nonvanishing adjoint action)}\cr
 \cr
 \cr
\underbrace{\Big(\big(\!\!\!\!\!\underbrace{\overset{\tikzmark{a}}{\mathfrak{t}}}_{\substack{\text{translation}\\\text{generators}}}\dotplus\underbrace{\overset{\tikzmark{c}}{t_{s}^{{}_{\mathrm{ext}}}}}_{\substack{\text{extended}\\\text{supertransl.}\\\text{generators}}}\!\!\!\!\!\!\big)\opluslhrim\!\!\underbrace{\overset{{}^{\tikzmark{f}}}{q}}_{\substack{\text{compact}\\\text{abelian}\\\text{internal}\\\text{generators}}}\!\!\Big) \opluslhrim \Big(\underbrace{\overset{\tikzmark{d}}{\mathfrak{l}_{1}{\oplus}{...}{\oplus}\,\mathfrak{l}_{n}}}_{\substack{\text{compact}\\\text{non-abelian}\\\text{internal}\\\text{generators}}}\oplus\!\!\!\!\underbrace{\overset{\tikzmark{e}\;\;\tikzmark{b}}{\boldell}}_{\substack{\text{Lorentz}\\\text{generators}}}\!\!\!\!\Big)}_{\text{Lie algebra of the extended super-Poincar\'e group}}
\tikz[remember picture,overlay] \draw[<-] (a.north) |- +(0ex,4.5ex) -| (b.north);
\tikz[remember picture,overlay] \draw[<-] (c.north) |- +(0ex,2.5ex) -| (d.north);
\tikz[remember picture,overlay] \draw[<-] (c.north) |- +(0ex,2.5ex) -| (e.north);
\tikz[remember picture,overlay] \draw[<-] (c.north) |- +(0ex,2.5ex) -| (f.north);
\end{array}
\label{eqsuperext}
\end{equation}
It is instructive to compare its structure to that of the generic Lie 
algebras Eq.(\ref{eqRad}) and to the scheme of the O'Raifeartaigh theorem 
Eq.(\ref{eqLOR}). 
The (extended) super-Poincar\'e group demonstrates the case (B) of the O'Raifeartaigh theorem.

\subsection{Conservative extensions of the Poincar\'e group}

The conservative extensions of the Poincar\'e Lie algebra was defined 
via the requirement Eq.(\ref{eqcons}). Due to O'Raifeartaigh theorem, 
if it is indecomposable, then it must be O'Raifeartaigh case (B), similar 
to the (extended) super-Poincar\'e. For a conservative Poincar\'e extension 
$\mathfrak{e}$, one has 
$\frac{\mathfrak{e}}{\nil(\mathfrak{e})}={\ua(1){\oplus}{\dots}{\oplus}\,\ua(1){\oplus}\,\mathfrak{l}_{1}{\oplus}{\dots}{\oplus}\,\mathfrak{l}_{n}{\oplus}\,\boldell}$, 
with $\mathfrak{l}_{1},\dots,\mathfrak{l}_{n}$ being simple, and 
$\boldell\equiv\sla(2,\C)$ being the Lorentz Lie algebra. In a gauge theory like 
setting, it is natural to require that the non-Lorentz part of 
$\frac{\mathfrak{e}}{\nil(\mathfrak{e})}$ is compact, i.e.\ that 
$\frac{\mathfrak{e}}{\nil(\mathfrak{e})}/\boldell$ is compact. As discussed in 
\cite{Laszlo2017,Laszlo2018}, in that case the conservative Poincar\'e 
Lie algebra extensions have the structure
\begin{equation}
\begin{array}{c}
 \text{\tiny(arrows: nonvanishing adjoint action)}\cr
 \cr
 \cr
\underbrace{\Big(\!\!\!\!\underbrace{\overset{\tikzmark{a}}{\mathfrak{t}}}_{\substack{\text{translation}\\\text{generators}}}\oplus\underbrace{\underbrace{\big(\!\!\!\!\underbrace{\overset{\tikzmark{c}}{\mathfrak{n}}}_{\substack{\text{nilpotent}\\\text{internal}\\\text{generators}}}\!\!\dotplus\!\!\underbrace{\overset{{}^{\tikzmark{f}}}{q}}_{\substack{\text{compact}\\\text{abelian}\\\text{internal}\\\text{generators}}}\!\!\!\!\big)}_{\substack{\text{solvable}\\\text{internal}\\\text{generators}}}\!\!\Big) \opluslhrim \Big(\underbrace{\overset{\tikzmark{d}}{\mathfrak{l}_{1}{\oplus}{...}{\oplus}\,\mathfrak{l}_{n}}}_{\substack{\text{compact}\\\text{non-abelian}\\\text{internal}\\\text{generators}}}}_{\text{all internal (gauge) symmetry generators}}\oplus\!\!\!\!\underbrace{\overset{\tikzmark{e}\;\;\tikzmark{b}}{\boldell}}_{\substack{\text{Lorentz}\\\text{generators}}}\!\!\!\!\Big)}_{\text{conservative Poincar\'e extension generators, acting on matter fields}}
\tikz[remember picture,overlay] \draw[<-] (a.north) |- +(0ex,4.5ex) -| (b.north);
\tikz[remember picture,overlay] \draw[<-] (c.north) |- +(0ex,2.5ex) -| (d.north);
\tikz[remember picture,overlay] \draw[<-] (c.north) |- +(0ex,2.5ex) -| (e.north);
\tikz[remember picture,overlay] \draw[<-] (c.north) |- +(0ex,2.5ex) -| (f.north);
\end{array}
\label{eqConservative}
\end{equation}
It is istructive to compare this structure to that of the generic Lie 
algebras Eq.(\ref{eqRad}) and to the scheme of the O'Raifeartaigh theorem 
Eq.(\ref{eqLOR}).

In a conservative Poincar\'e extension, 
all the non-Standard-Model-like symmetry generators are expelled into 
the ideal of nilpotent internal symmetries $\mathfrak{n}$. 
The unification happens because $\mathfrak{n}$ 
carries both compact and Lorentz charges, similarly to the case of SUSY. 
An important property of the conservative unification pattern is that despite of 
the indecomposable (unified) structure Eq.(\ref{eqConservative}), 
there is a forgetful homomorphism back onto the usual direct sum of the 
Poincar\'e symmetries and the compact internal symmetries
${(\mathfrak{t}\opluslhrim\boldell)\oplus q \oplus \mathfrak{l}_{1}{\oplus}{\dots}{\oplus}\,\mathfrak{l}_{n}}$. 
That is, one could think of a theory in which a unified symmetry concept 
like Eq.(\ref{eqConservative}) acts on the fundamental field degrees of freedom, 
whereas the usual Poincar\'e plus Standard Model compact gauge symmetries 
act on some derived field quantities, which are functions of the fundamental 
field degrees of freedom. One could call such a mechanism ``symmetry hiding'', 
in contrast to symmetry breaking.

\acknowledgments
This research was supported in part by Perimeter Institute for Theoretical Physics, 
which is financed by the Government of Canada through the Department of 
Innovation, Science and Economic Development and by the Province of Ontario 
through the Ministry of Research and Innovation.
Part of this work was done while L.~Andersson was in residence at Institut 
Mittag-Leffler in Djursholm, Sweden during the fall of 2019, supported by 
the Swedish Research Council under grant no.~2016-06596. 
The work of A.~L\'aszl\'o was supported in part by the Hungarian Scientific Research Fund 
(NKFIH 123842-123959). 
B.~Ruba was supported by the Faculty of Physics,
Astronomy and Applied Computer Science grant MSN 2019 (N17/MNS/000040)
for young scientists and PhD students.

%% References:
\bibliographystyle{JHEP}
\bibliography{newlocal}

\end{document}